\documentclass[a4paper,11pt]{article}
\pdfoutput=1 

\usepackage{jheppub} 

\usepackage{amsmath}
\usepackage{amsfonts}
\usepackage{amssymb}
\usepackage{enumerate}
\usepackage{braket}
\usepackage{xcolor}
\usepackage{caption}
\usepackage{subcaption}
\usepackage{hyperref}

\DeclareMathOperator{\tr}{tr}

\renewcommand{\d}{{\rm d}}
\renewcommand{\i}{{\rm i}}
\newcommand{\e}{{\mathrm e}}
\newcommand\paper{paper}
\newcommand\half{{\tfrac{1}{2}}}
\newcommand\itwo{{I_2}}
\newcommand\ithree{{I_3}}

\newcommand\inine{{I_9}}
\newcommand\nhat{{{\hat{\bf n}}}}

\newcommand\lambdad{\lambda^{(d)}}
\newcommand\lambdathree{\lambda^{(3)}}
\newcommand\mintertboundsq{c^2_{\rm MB}}
\renewcommand{\j}{{\mathsf{j}}}
\newcommand{\spinstate}{{{\mathsf m}}}

\newcommand\WignerQ{\Phi^{Q}}
\newcommand\WignerQp{\Phi^{Q+}}
\newcommand\WignerQm{\Phi^{Q-}}
\newcommand\WignerQpm{\Phi^{Q\pm}}
\newcommand\WignerP{\Phi^{P}}
\newcommand\WignerPp{\Phi^{P+}}
\newcommand\WignerPm{\Phi^{P-}}
\newcommand\WignerPpm{\Phi^{P\pm}}

\newcommand\WignerNonProjectiveQ{\widetilde\Phi^{Q}}
\newcommand\WignerNonProjectiveP{\widetilde\Phi^{P}}

\newcommand*{\rhomixWW}{\rho_\mathrm{H/WW}}
\newcommand*{\rhomixZZ}{\rho_\mathrm{H/ZZ}}

\title{\boldmath Quantum state tomography, entanglement detection and Bell violation prospects in weak decays of massive particles}

\author[a]{Rachel Ashby-Pickering,}
\author[a,b]{Alan J. Barr,}
\author[a,b]{Agnieszka Wierzchucka}

\affiliation[a]{Department of Physics, Keble Road, University of Oxford, OX1 3RH}
\affiliation[b]{Merton College, Merton Street, Oxford, OX1 4JD}
\emailAdd{alan.barr@physics.ox.ac.uk}

\abstract{A rather general method for determining the spin density matrix of a multi-particle system from angular decay data is presented.
The method is based on a Bloch parameterisation of the $d$-dimensional generalised Gell-Mann representation of $\rho$ and exploits the associated Wigner- and Weyl-transforms on the sphere.
Each parameter of a (possibly multipartite) spin density matrix can be measured from a simple average over an appropriate set of experimental angular decay distributions.
The general procedures for both projective and non-projective decays are described, and the  Wigner $P$ and $Q$ symbols calculated for the cases of spin-half, spin-one, and spin-3/2 systems. 
The methods are used to examine Monte Carlo simulations of $pp$  collisions for bipartite systems:  $pp\rightarrow W^+W^-$, $pp\rightarrow ZZ$, $pp\rightarrow ZW^+$, $pp\rightarrow W^+\bar{t}$, $t\bar{t}$, and those from the Higgs boson decays $H\rightarrow WW^{*}$ and $H\rightarrow ZZ^*$.
Measurements are proposed for entanglement detection and Bell inequality violation in bipartite systems.
}


\begin{document}
\maketitle
\flushbottom

\section{Introduction}
\label{sec:intro}

In quantum mechanical systems the most general characterisation one may make of a system is knowledge of its density matrix $\rho$. 
Knowing $\rho$, one may determine the expectation value of any measurement from the value $\braket{A} = \tr(\rho A)$ of its Hilbert-space operator $A$. 
With knowledge of $\rho$ one can therefore go on to calculate any quantity one might desire, be it a spin expectation along another arbitrary axis, a measurement of entanglement, a Bell-operator expectation value, or any other observable of choice. 

However, since quantum measurements generally involve incompatible operators it is not possible in general to determine $\rho$ from measurements of a single instance of the state. 
Nevertheless, with an ensemble of similarly-prepared states one may perform multiple measurements of incompatible measurements. 
Determining the density matrix from an ensemble of measurements is known within the quantum computing literature as `quantum state tomography'~\cite{PhysRevLett.83.3103,PhysRevA.64.052312,Thew-qudit-tomography-2002}.

Though historically not often described in these terms within the particle physics literature, 
for the special case of spin-half fermions measuring the spin polarisation is equivalent to performing quantum tomography on that state. 
This equivalence follows since the spin-density matrix of a $\j=\half$ system, a `qubit', is fully characterised by the three parameters of its Bloch vector or polarisation vector, hence measurement of those polarisation parameters is equivalent to performing full quantum state tomography for those particles. 

Spin density matrix tomography of systems of pairs of spin-half particles has been considered previously in particle physics. For example a method has been proposed~\cite{Bernreuther_2015,Afik_2021} 
for determining the spin density matrix of pairs of top and anti-top quarks at the LHC. 
The method has been used to study in simulation how one could measure the entanglement of the pair~\cite{Fabbrichesi:2021npl,Afik_2021,Severi:2021cnj}. 
Experimental tests of entanglement in high-energy systems have been proposed in systems of pairs of $\Lambda$ baryons~\cite{Gong:2021bcp}, top-quarks~\cite{Severi:2021cnj,Afik:2022kwm,Aoude:2022imd,Fabbrichesi:2022ovb,Afik:2022dgh}, $\tau$ leptons~\cite{Fabbrichesi:2022ovb} and photons~\cite{Fabbrichesi:2022ovb}, and the quantum discord and steering in $t\bar{t}$ systems has also been investigated~\cite{Afik:2022dgh}.
In bipartite systems of spin-half particles, experimental measurements of the spin density matrix have been made for the $t\bar{t}$ system by the ATLAS and CMS collaborations~\cite{ATLAS:2016bac,CMS:2019nrx}, demonstrating spin-correlations in those systems. 

Experimental tests have shown violation of Bell inequalities 
for pairs of qubits using
photons~\cite{PhysRevLett.28.938,PhysRevLett.49.1804}, ions~\cite{Rowe}, superconducting systems~\cite{josephson} and nitrogen vacancy centres
~\cite{pfaff}, 
and in pairs of three-outcome measurements using photons~\cite{PhysRevLett.89.240401}.
So-called ``loophole-free'' Bell inequality tests were performed by three groups in 2015~\cite{Hensen:2015ccp,Giustina_2015,Shalm_2015}. 
Proposals have also been made to test Bell inequalities in $e^+e^-$ collisions~\cite{Tornqvist}, charmonium decays~\cite{Baranov_2008,10.1093/ptep/ptt032}, positronium decays~\cite{Ac_n_2001}, and Higgs boson decays to $WW^{*}$~\cite{BARR2022136866}, and in top quark~\cite{Severi:2021cnj,Fabbrichesi:2022ovb} and tau leptons~\cite{Fabbrichesi:2022ovb}.

In this \paper{} we are concerned with determining the spin density matrix of a system of one or more particles, including those of higher spin i.e. with $\j\geq \half$. 
These are `qudits' of dimension $d=2\j+1$ in the quantum computing terminology. 
Unlike in the spin-half case, when one considers $j>\half$ the spin-polarisation vector no longer fully characterises the density matrix, and so a more general procedure is required to recover $\rho$ from data. 
We develop such a procedure, and use it to examine various states including those containing one or more $W^\pm$ or $Z^0$ bosons, which present interesting test cases due to having both $d>2$ and chiral decays. 

\subsection{Theoretical and experimental background}\label{sec:background}

The theory behind the relationship between spin polarisations and angular distributions dates back to Wigner's work~\cite{WignerGroupTheory}. Jacob~\cite{Jacob:1958ply} showed how two-body decays are related to angular distributions for particular helicity, and  
Jacob and Wick~\cite{Jacob:1959at} described how transitions between states may be represented in terms of helicity amplitudes.  

A general introduction to quantum state tomography~\cite{Thew-qudit-tomography-2002} using Gell-Mann based Bloch vectors~\cite{Bertlmann-Bloch-2008} has been described for non-orthogonal bases, which guides our own study. 
Since the $W^\pm$ and $Z^0$ are massive vector ($\j=1$) bosons their spins states are in a Hilbert space of $d=2\j+1=3$, making them `qutrits' the language of quantum computing.

Quantum state tomography of qutrits has previously been explored in the context of nuclear magnetic resonance (NMR) experiments~\cite{1703.06102}.
Experimental measurements of pairs of entangled qutrits have been performed for example in photonic systems~\cite{Langford:2003zp}. 
Unlike the photons, the weak vector bosons are ``their own polarimeters''~\cite{Tornqvist}. The chiral nature of the weak interaction means that the directions of the emitted fermions are correlated with their parent's spin, meaning that the bosons effectively make a measurement of their own spin during their decay process\footnote{This sensitivity of weak decays to spin underlies proposals to measure light quark polarisation at the LHC~\cite{Kats2015light}, $c$-quark polarisation in $W+c$ samples at the LHC~\cite{Kats2016Wc}, and to use heavy baryons as polarimeters at colliders~\cite{Galanti2015}.}.

The angular decay distribution of pairs of leptons from single gauge boson production has previously been parameterised in terms of angular coefficients~\cite{Collins:1977iv,Lam:1980uc,Mirkes:1994eb,Mirkes:1994dp,Hagiwara:2006qe,Bern:2011ie,Stirling:2012zt,Rahaman:2021fcz}. One of these paper showed how one can obtain five parity-even parameters characterising $\rho$ from expectation values~\cite{Mirkes:1994dp}.
The parity-odd terms were later included~\cite{Hagiwara:2006qe}, completing the eight real parameters required to characterise a single $\j=1$ boson. Expressions for the eight parameters in terms of expectation values, for the case of massless daughter leptons have also been calculated~\cite{Bern:2011ie}.

The angular distribution resulting from decays of various multipartite states is described in terms of Cartesian vectors and tensors in Refs.~\cite{Martens:2017cvj,Rahaman:2021fcz}. Quantum tomographic methods of obtaining the density matrix parameters either by multi-variable fitting
\cite{Martens:2017cvj} or from asymmetry observables~\cite{Rahaman:2021fcz} have been proposed.
It has also been shown~\cite{Rahaman:2021fcz} that knowledge of the full bipartite density matrix can provide a significant improvement over just determination of polarisations in gaining sensitivity to anomalous couplings.

Experimentally, polarisations of $W^\pm$ and $Z^0$ boson have been the subject of considerable experimental study. 
The UA1~\cite{UA1:1988rck} and UA2~\cite{UA2:1987qqc} collaborations used the angular distributions to determine the spin and couplings of the $W^\pm$ and $Z^0$ bosons.
The polarisation of $W$ bosons was measured in $ep$ collisions by the H1 Collaboration~\cite{H1:2009rfg},
and measured in the decay of the top quark 
by the CDF and D\O~\cite{CDF:2012dup,CDF:2010cnn,D0:2010zpi} collaborations and at the LHC by the ATLAS~\cite{ATLAS:2012nhi} and 
CMS~\cite{CMS:2016asd} collaborations. Measurements of $W^\pm$ polarisation in events containing jets have also been performed by ATLAS~\cite{ATLAS:2012au} and CMS~\cite{CMS:2011kaj}. 
Polarisation and angular measurements of a single $Z$ boson were published by the CDF~\cite{CDF:2011ksg}, CMS~\cite{CMS:2015cyj} and ATLAS~\cite{ATLAS:2012au} collaborations. 
Diboson measurements of $W^\pm$ polarisations were performed by L3~\cite{L3:2002ygr} and OPAL~\cite{OPAL:2003uzt}. DELPHI~\cite{DELPHI:2008uqu} used similar measurements to set limits on anomalous triple gauge couplings.
Polarisations of pairs of gauge bosons have also been measured by ATLAS and CMS~\cite{ATLAS:2019bsc,CMS:2020etf,CMS:2021icx}, including the recent observation of joint-polarisation states~\cite{ATLAS-CONF-2022-053}.
The simultaneous production of three charged weak bosons, $WWW$, has recently been observed by ATLAS~\cite{ATLAS:2022xnu}.

We extend the previous literature by exploiting the properties of the generalised Gell-Mann (GGM) basis for the density matrix, by relating the tomography process of particle decays to the  Wigner and Weyl transformations and to the Kraus and measurement operators of quantum information theory, and by extending the tomography method to arbitrary numbers of parents of arbitrary spin $\j$. 
We illustrate how the method allows one to address a programme of questions relevant to the foundations of quantum mechanics, such as in entanglement detection and Bell inequality violation in $WW$, $WZ$ and $ZZ$ diboson systems.\footnote{During the final states of preparing this paper a preprint was posted examining Bell inequalities and entanglement in $H\rightarrow ZZ$ systems~\cite{Aguilar-Saavedra:2022wam}.}


\section{Generalised Gell-Mann parameterisations of a multipartite density matrix}\label{sec:rho}

The density matrix~\cite{Fano-RevModPhys.29.74} captures our knowledge about a partially-understood quantum system for which the complete state 
$\ket{\psi}$ is not completely known. It is defined by the convex sum
\begin{equation}
\rho=\sum_i p_i \ket{\psi_i}\bra{\psi_i},
\end{equation}
where the $p_i$ can be interpreted (in a non-unique way) as a set of classical probabilities for the states within an ensemble. 
The density matrix is a positive semi-definite operator meaning that
\begin{equation}\label{eq:rhopositive}
\bra{\phi}\rho\ket{\phi}\geq0 \qquad \forall \ket{\phi} \in \mathcal{H}, 
\end{equation}
and is constrained by the definition of probability to have $\sum_i p_i=1$, a constraint that corresponds to the requirement that $\rho$ must have unit trace.

The spin density matrix $\rho$ for a single spin-$\j$ particle is a Hermitian operator on the $d=2\j+1$ dimensional Hilbert space of the state. 
Taking into account that it is Hermitian and the trace constraint it has $d^2-1$ real parameters. 
The density matrix for a multi-particle system is defined in the Hilbert space 
$\mathcal H_1\otimes \mathcal H_2\otimes \ldots \mathcal H_n$
which is of dimension $d=(2\j_1+1)(2\j_2+1)\ldots(2\j_n+1)$, 
and which again has $d^2-1$ real parameters.
The task of quantum state tomography is to determine each of those density matrix parameters for a single or multi-particle density matrix from experimental measurements. 

In order to reconstruct the density matrix experimentally an explicit parameterisation of it must be chosen.
We find advantages\footnote{The parameterisation is orthogonal, complete, non-redundant and symmetric, facilitating calculations such as the bound of the concurrence \eqref{eq:mintertinGM} in Section~\ref{sec:entanglementintro}.} to using a parameterisation based on the generalised $d$-dimensional Gell-Mann (GGM) operators 
$\lambdad$~\cite{kimura2003bloch,Bertlmann-Bloch-2008}.
These operators are the $d^2-1$ traceless Hermitian  generators of the group $\mathrm{SU}(d)$.
The lower-dimensional versions are familiar in particle physics: for $d=2$ they are the three Pauli matrices $\lambda^{(2)}_i=\sigma_i$, while for $d=3$ they are Gell-Mann's eight matrices~\cite{PhysRev.125.1067}, which we will denote here $\lambda^{(3)}_i$. 

The $\lambdad_i$ satisfy the trace orthogonality\footnote{Since the generalised Gell-Mann matrices are Hermitian this trace orthogonality is equivalent to orthogonality under the Hilbert-Schmidt inner product $\braket{A,B}_{\rm HS} = \tr(A^\dag B)$.} relations
\begin{equation}\label{eq:GMtrace}
\tr(\lambdad_j \lambdad_k) = 2\delta_{jk},\qquad \tr(\lambdad_j)=0.
\end{equation}
This mutual trace orthogonality, orthogonality with the identity, and their linearly independence, means that they form an ideal basis for $\rho$. 
For the single-particle $d$-dimensional spin density matrix we therefore express the density matrix in the form
\begin{equation}\label{eq:rhosingleGM}
\rho^{(d)} = \frac{1}{d}  I_d+ \sum_{i=1}^{d^2-1} a_i \lambdad_i ,
\end{equation}
where $I_d$ is the $d$-dimensional identity matrix and the $a_j$ are real parameters\footnote{The parameters are constrained, since 
we must enforce the positivity condition \eqref{eq:rhopositive}. 
For the case of $d=2$ that constraint on the $a_i$ is 
\[
\sum_{i=1}^3 a_i^2 \leq \frac{1}{4} \qquad\qquad {\rm for}~d=2,
\]
where the limiting case corresponds to density matrices representing pure states. The conditions for $d>2$ are more complicated~\cite{kimura2003bloch,Byrd_2003,Kryszewski_2006}.
} which form a $(d^2-1)$-dimensional generalised Bloch vector. 
Explicit forms for generalised Gell-Mann operators and of the corresponding density matrices may be found in Appendix~\ref{sec:ggm}.

Our choice of GGM matrices which satisfy the orthogonality relations \eqref{eq:GMtrace} 
has the desired consequence  that each of the real coefficients 
$a_j$ can be obtained from the expectation values of corresponding operators. For any Hermitian operator $A$
its expectation value is given by
\begin{equation}
\braket{A} = \tr(\rho A),
\end{equation}
so the expectation values of the operators corresponding to the Gell-Mann matrices yield
\begin{align}
\braket{\lambdad_i} & = \tr(\rho \lambdad_i), \nonumber\\
                            & = 2 a_i \label{eq:GMextractparam}.
\end{align}
Using \eqref{eq:GMextractparam} we can determine the $d^2-1$ parameters $a_i$ of the density matrix
\eqref{eq:rhosingleGM}, provided that we can find the expectation values of the $\lambdad_i$ operators from data. 
In what follows we will described a general procedure for doing so. 

For multi-particle systems we can generalise this parameterisation. 
For example for a two-particle state of particles of spin $\j_1$ and $\j_2$, we can parameterise the bipartite density matrix
\begin{equation}\label{eq:doubleGM}
\rho^{( d ) } = \tfrac{1}{d} I_{d} 
 + \sum_{i=1}^{d_1^2-1} a_i \lambda^{(d_1)}_i \otimes \tfrac{1}{d_2} I_{d_2}
 + \sum_{j=1}^{d_2^2-1} \tfrac{1}{d_1}I_{d_1} \otimes  b_j \lambda^{(d_2)}_j 
 + \sum_{i=1}^{d_1^2-1} \sum_{j=1}^{d_2^2-1} c_{ij} \lambda^{(d_1)}_i \otimes \lambda^{(d_2)}_j \,,
\end{equation}
in terms of the generalised Gell-Mann matrices and their tensor products, 
where $d_i=(2\j_i+1)$, and the dimension of the full Hilbert space is $d=d_1d_2$. The $a_i$ and $b_j$ are the Bloch vectors for the individual particles and the $c_{ij}$ parameterise correlations between the two particles. 
\par
Systems with larger numbers of distinguishable parents can be parameterised in an analogous way, using sums over appropriate tensor products up to $\lambda^{(d_1)}_{i_1}\otimes\lambda^{(d_2)}_{i_2} \otimes\ldots\otimes\lambda^{(d_n)}_{i_n}$ of the appropriate $d_i$-dimensional generalised Gell-Mann matrices for each particle.

\subsection{Converting operators between parameterisations}\label{sec:lambda_0}

We have noted that once the density matrix (for single or multi-particle systems) 
is known in the GGM parameterisation it can be calculated in any other by use of the trace orthogonality relations \eqref{eq:GMtrace}. 
Alternatively, one may also calculate the representation of any desired Hermitian operator $A$ in the GGM basis 
using these same orthogonality conditions. 
To do so for a general Hermitian operator we need to supplement the generalised $d$-dimensional matrices with an additional matrix
\begin{equation}\label{eq:lambda_0}
\lambdad_0 = \sqrt{\frac{2}{d}}I_d,
\end{equation}
to allow for operators with non-zero traces. 
The normalisation of $\lambdad_0$ has been chosen such that it also obeys a generalisation of \eqref{eq:GMtrace}.
We can then express any Hermitian operator 
\begin{equation}\nonumber
A = \sum_{i=0}^{d^2-1} a_i \lambdad_i 
\end{equation}
in terms of the $d^2$ parameters
\begin{equation}\label{eq:operatortoGM}
a_i = \half \tr (\lambdad_i A) \qquad i\in \{0,\,\ldots,\, (d^2-1)\}.
\end{equation}
One particularly useful application is the mapping between the spin and the GGM operators, which is calculated in Appendix~\ref{sec:spinmatrices}, and which is used in the examination of the states of the example systems in Section~\ref{sec:examples}.


\section{The Weyl-Wigner formalism for spin}

Theoretical physicists are very used to describing the process whereby particles with particular spin density matrices decay according to angular distributions, and they know how to calculate those distributions. 
That process (operators $\rightarrow$ angular distributions) is frequently simulated in the field of particle physics both using analytical calculations and Monte Carlo simulations.\footnote{We make use of spin correlations in Monte Carlo simulations~\cite{Richardson2001,Frixione:2007zp}, a technique automated in \cite{Artoisenet:2012st} using the methods first proposed by Collins~\cite{Collins:1987cp} and Knowles~\cite{Knowles:1987cu,Knowles:1988hu,Knowles:1988vs}, 
and later used in Madspin~\cite{Alwall_2014}.}

Experimentalists analysing decay data wish to invert that process. Their goal is to analyse a set of angular decay distributions, and from those distributions obtain the spin density matrix of the originating quantum system. 
A possibility one can consider is to fit the parameters, perhaps based on likelihood functions and angular decay data.
The strategy employed here is instead to find analytically the inverse mapping. 
This has the advantage that one can then find analytical expressions for quantities derived from the density matrix, such as those related to Bell inequalities and entanglement detection. 
The analytical approach also allows one rapidly to explore density matrices in simulated Monte Carlo data sets, and hence to find regions where interesting quantum mechanical measurements might be made. 

An ideal formalism for undertaking this inverse task is the Wigner-Weyl transformation, the invertible mapping between functions in the quantum phase space and Hilbert space operators. We are interested in the case the phase space is the space of angular distributions, and the operators of interest are the spin density operators. 

For the forward mapping (`operators $\rightarrow$ angles') we will make use of Wigner $Q$ symbols which map the bounded operators $A:\mathcal{H}\rightarrow \mathcal{H}$ in the Hilbert space $\mathcal H$ of the spin states onto the space of functions on the sphere $S^2$. They are defined by~\cite{LiBraunGarg2013}
\begin{equation}\label{eq:wignerQgeneral}
\Phi_A^Q(\nhat) = \braket{\nhat | A | \nhat},
\end{equation}
where $\nhat$ is a unit vector in $\mathcal R^3$. 
These capture the forward mapping from operators to functions on the sphere.

The inverse mapping from the space of functions on $S^2$ to the space of bounded operators $A:\mathcal H\rightarrow \mathcal H$ is achieved by using the Wigner $P$ symbols, the defining property of which is~\cite{LiBraunGarg2013}
\begin{equation}\label{eq:wignerPgeneral}
A = \frac{2\j+1}{4\pi} \int \d\Omega_{\nhat}\,\ket{\nhat}\Phi_A^P(\nhat)\bra{\nhat},
\end{equation}
where the angular integral is over directions of unit vector $\nhat$. 

We can then perform quantum state tomography for projective measurements as follows.
First one parameterises the spin density matrix into a generalised Bloch-vector form \eqref{eq:rhosingleGM}, using a generalised Gell-Mann operator basis for a multi-paritite system. 
Then one finds the Wigner $Q$ symbols for each of those Gell-Mann operators, which provide us with the mapping from operators to angular distributions. From the Wigner $Q$ symbols the corresponding set of Wigner $P$ symbols are then obtained. 
Experimentally measured expectation values of those Wigner $P$ symbols will then provide us with each of the real parameters of the spin density matrix. 

We will also describe how this method can be generalised to non-projective measurements, and to particles of arbitrary angular momentum quantum number $\j$. We provide examples for single particle and multipartite systems with $\j=\half$ and $\j=1$. 

\section{\boldmath Projective spin-1 procedure ($W^\pm$ boson tomography)}\label{sec:singleW}

We start with a particularly illustrative example -- that of the spin-1, massive, $W^+$ or $W^-$ charged weak boson. 
The $W$ vector boson is of particular interest since (a) its spin state is of dimension $d=3$, and unlike lower-spin cases is an example for which the forms of the Wigner $P$ and $Q$ symbols are not proportional and 
(b) decays of the $W$ violates parity maximally, leading to projective (von-Neumann) measurements during decay, which make it a good introductory example the Wigner-Weyl procedures. 

Our task is to find the eight Bloch vector parameters $a_i$ of the $W^+$ or $W^-$ boson's $d=3$ spin density matrix\footnote{The explicit form of this matrix for $d=3$ is shown as \eqref{eq:rhothree} in Appendix~\ref{sec:ggm}.} from angular decay data. We will work in the (massive) vector boson's rest frame where spin is most naturally represented. 

\subsection{\boldmath Wigner $Q$ symbols $(\j=1)$}\label{sec:tomographyWQ}
\label{sec:density}

In its decays the $W^\pm$ couples only to left-chiral spin-half fermions
and to right-chiral spin-half anti-fermions~\cite{ParticleDataGroup:2020ssz}. 
In the excellent approximation $m_\ell \ll m_W^\pm$
in which the lepton masses can be neglected, the
$W^+\rightarrow \ell^+\nu$ decay produces a $\ell^+$ in a positive helicity state
and a $\nu$ in a negative helicity state\footnote{For the more general situation with non-negligible daughter masses the reader is directed to Appendix~\ref{sec:Wnonprojective}.}
so the spin of the $W^+$ is effectively measured
to be $+1$ in the momentum direction of the $\ell^+$.
Similarly, the $W^-\rightarrow \ell^-\bar\nu$ decay produces a $\ell^-$ in a negative helicity state
and a $\bar\nu$ in a positive helicity state. The spin of the $W^-$ is therefore measured
to be $-1$ in the momentum direction of the $\ell^-$.

By measuring the decay direction of the outgoing leptons in the $W^\pm$ boson rest frame we effectively make a projective measurement, i.e. a von Neumann measurement, of the $W^\pm$ boson spin in that frame. Hence with access to an ensemble of such decays we can reconstruct the $W$ boson spin density matrix.

More formally, the probability density function for a $W^\pm$ boson with 
spin density matrix given by \eqref{eq:rhosingleGM} (with $d=3$) to emit a charged lepton $\ell^\pm$ 
into infinitesimal solid angle $\d\Omega$ along the direction $\nhat(\theta,\phi)$ is~\cite{Jacob:1958ply}
\begin{equation}\label{eq:definepdf}
p(\ell^\pm_\nhat; \rho) = \tfrac{d}{4\pi} \tr(\rho \Pi_{\pm,\nhat}) \,,
\end{equation}
with $d=3$ for the $W^\pm$ bosons.
The spin-1 projection operator  $\Pi_{+,\nhat}\equiv \ket{+}_{\nhat}\bra{+}_{\nhat}$  selects a positive helicity $\ell^+$  in the direction $\nhat$ accompanied by a negative helicity $\nu$ in the direction $-\nhat$. 
Similarly $\Pi_{-,\nhat}\equiv \ket{-}_{\nhat}\bra{-}_{\nhat}$
selects a negative helicity $\ell^-$  in the direction $\nhat$ accompanied by a positive helicity $\bar\nu$ in the direction $-\nhat$ for the $W^-$ case. 
The normalisation of~\eqref{eq:definepdf} is such that  
\begin{equation}
\int\d\Omega_\nhat \, p(\ell^\pm_{\nhat}; \rho) = 1.
\end{equation}

In forming \eqref{eq:definepdf} we have made the approximation that the leptons are measured to be ``in the $\nhat$ direction''. More precisely we are assuming that their probability density functions $|\psi(\Omega)|^2$ are concentrated in a small range around $\nhat$, by which we mean a range across which the Wigner symbols change only slightly. 
This approximation is valid provided that both the de Broglie wavelength of the particle and the angular resolution of the detector are sufficiently small compared to the range over which the Wigner functions on the sphere change. 
This requirement is easily satisfied for our experiments of interest, e.g. when measuring the multi-GeV momentum decay products of $W$ bosons at LHC experiments~\cite{ATLAS:2008xda,CMS:2008xjf} which have angular resolutions for leptons $\sigma_{\theta,\phi}\ll\pi/\j$.

The explicit form of the projection operator $\Pi_{\pm,\nhat}$ can be found by starting with the states $\ket{\pm}_z\bra{\pm}_z$ 
which project the pure state with spin $\pm1$ in the $z$ direction, and then rotating each to point along an arbitrary direction $\nhat$ defined by the polar angle $\theta$ and the azimuth $\phi$:
\begin{align} 
\Pi_{\pm,\nhat}   & = \ket{\pm\nhat}\bra{\pm \nhat} \nonumber \\
                           &= U_{\theta,\phi} \ket{\pm}_z\bra{\pm}_z U_{\theta,\phi}^\dag.
\end{align}
The unitary rotation operator $U_{\theta,\phi}$ is given by the product 
\begin{equation}\label{eq:dmatrix}
U_{\theta,\phi} = \exp(- \i S_z \phi) \exp(- \i S_y \theta) ,
\end{equation}
where the $S_i$ are the appropriate spin operators, the representations of which are shown for $d=3$ in Appendix~\ref{sec:spinmatrices}. 
The unitary operation \eqref{eq:dmatrix} is a Wigner $D^\j(\alpha,\beta,\gamma)$ matrix~\cite{WignerGroupTheory} for $\j=1$, with $\alpha=\phi$, $\beta=\theta$, $\gamma=0$.
Explicitly, the positive projector $\Pi_{+,\nhat}$ is given for $d=3$ by the matrix
\begin{equation} \label{eq:pnhat} 
\Pi_{+,\nhat} = \left(
\begin{array}{ccc}
 \cos ^4 \tfrac{\theta }{2} & \frac{1}{2\sqrt{2}}e^{-\i \phi } \sin \theta (1+\cos\theta) & \frac{1}{4} \e^{-2 \i \phi } \sin ^2 \theta  \\
 \frac{1}{2\sqrt{2}} \e^{\i \phi } \sin \theta (1+\cos \theta) & \frac{1}{2} \sin ^2 \theta & \frac{1}{2\sqrt{2}} \e^{-\i \phi } \sin \theta (1-\cos\theta) \\
 \frac{1}{4} \e^{2 \i \phi } \sin ^2\theta  & \frac{1}{2\sqrt{2}}\e^{\i \phi } \sin \theta (1-\cos\theta) & \sin ^4 \frac{\theta }{2} \\
\end{array}
\right).
\end{equation}
The negative projector $\Pi_{-,\nhat}$ is obtained from \eqref{eq:pnhat} with the replacement 
$\theta\rightarrow\pi-\theta$, $\phi\rightarrow\phi+\pi$. 

In terms of the parameters of the Gell-Mann basis for $\rho$, the full expressions for the angular probability density functions~\eqref{eq:definepdf} are
\begin{equation}
p(\ell^\pm_{\nhat}; \rho)  = \frac{3}{4\pi} \left(\frac{1}{3} + \sum_{i=1}^8 \WignerQpm_j a_j \right) \label{eq:pGMbasis}.
\end{equation}
The eight functions $\WignerQp_j(\nhat)$ for the $W^+$ boson (and the eight $\WignerQm_j(\nhat)$ for the $W^-$ boson) corresponding to each of the Gell-Mann operators are calculable from \eqref{eq:definepdf}, using $\rho$ from \eqref{eq:rhosingleGM} and $\Pi_{\pm,\nhat} $ from  \eqref{eq:pnhat} and are given by
\begin{align} 
\WignerQpm_1 &= \tfrac{1}{\sqrt 2} \sin \theta ( \cos \theta \pm 1 ) \cos \phi  &
\WignerQpm_5 &= \tfrac{1}{2} \sin ^2\theta \sin 2 \phi \nonumber\\
\WignerQpm_2 &= \tfrac{1}{\sqrt{2}} \sin \theta (\cos \theta \pm 1) \sin \phi &
\WignerQpm_6 &= \tfrac{1}{\sqrt 2} \sin \theta ( - \cos \theta \pm 1 ) \cos \phi \nonumber\\
\WignerQpm_3 &= \tfrac{1}{8} (\pm 4 \cos \theta+3 \cos 2 \theta +1) &
\WignerQpm_7 &= \tfrac{1}{\sqrt{2}} \sin \theta (- \cos \theta \pm 1) \sin \phi \nonumber\\
\WignerQpm_4 &=  \tfrac{1}{2} \sin ^2\theta \cos 2\phi & 
\WignerQpm_8 &=  \tfrac{1}{8 \sqrt{3}}\left(\pm12 \cos \theta-3 \cos 2 \theta -1\right) .\label{eq:WignerQGM}
\end{align}

\begin{figure}
\begin{center} 
\includegraphics[width=0.24\linewidth]{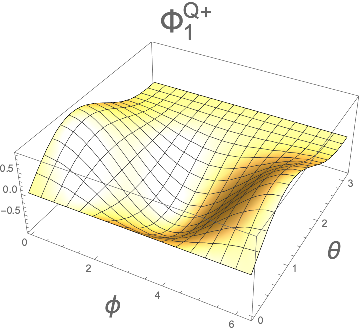}
\includegraphics[width=0.24\linewidth]{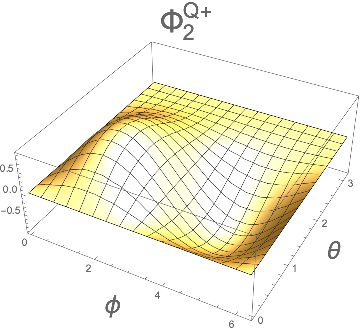}
\includegraphics[width=0.24\linewidth]{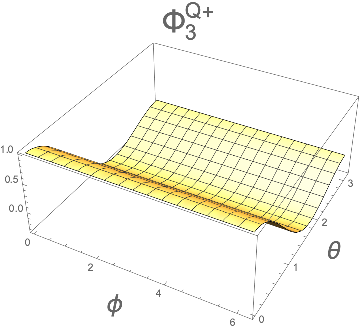}
\includegraphics[width=0.24\linewidth]{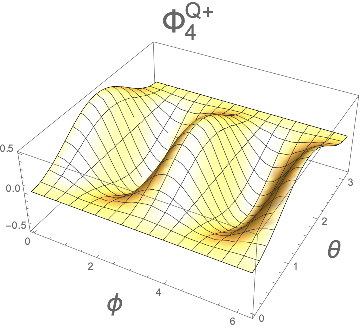}
\includegraphics[width=0.24\linewidth]{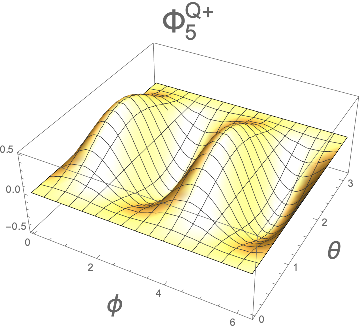}
\includegraphics[width=0.24\linewidth]{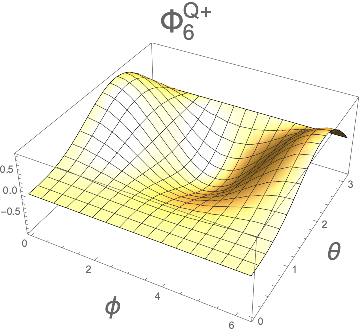}
\includegraphics[width=0.24\linewidth]{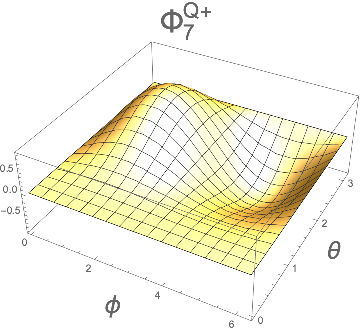}
\includegraphics[width=0.24\linewidth]{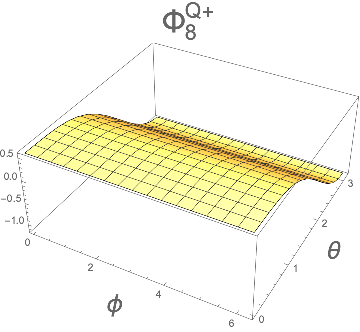}
\caption{\label{fig:angularGM}
Plots showing the form of the eight Wigner $Q$ symbols $\WignerQp_j$ corresponding to each of the Gell-Mann operators $\lambdathree_j$. These functions show the contribution \eqref{eq:pGMbasis} of the corresponding density matrix parameters to the angular probability density function for lepton emission in a $W^+$ boson decay. 
}
\end{center}
\end{figure}

The symbols  $\WignerQpm_j$ that we have assigned to these functions are suggestive that they might be Wigner $Q$ symbols for the Gell-Mann operators. To see that they are indeed functions of that type it suffices to note that
\begin{align}\label{eq:proveQ}
\WignerQpm_j &= \braket{\pm\nhat | \lambdathree_j | \pm\nhat} \nonumber\\
                        &= \tr\left({\lambdathree_j \Pi_{\pm,\nhat}}\right),
\end{align}
so the $\WignerQp_j$ do indeed satisfy the requirements \eqref{eq:wignerQgeneral} of a Wigner $Q$ symbol, as do (separately) the  $\WignerQm_j$.
To illustrate the effect of each Gell-Mann operator's contribution to the angular PDF, 
the eight Wigner $\WignerQp_j$ functions are shown in Figure~\ref{fig:angularGM}.
Thus we have completed the Wigner transform --- the mapping in the forward direction from the density matrix to the angular distributions. 

Before we move on to obtain the corresponding Wigner $P$ symbols required for the inverse (Weyl) transform, we make an observation. Leaving aside the normalisation constant of $\frac{3}{4\pi}$, the probability density functions  for lepton emission distributions \eqref{eq:definepdf} are themselves also Wigner transforms 
\begin{align}
 \tr(\rho \Pi_{\pm,\nhat}) 
	&= \braket{\nhat | \rho | \nhat} \nonumber\\
 	&= \Phi^{Q,\pm}_\rho, 
\end{align}
this time of the density matrix --- so the lepton angular pdfs are examples Wigner functions. 
This is a result worth reflecting on: the seemingly abstract Wigner function for the spin density matrix has, for the $W^\pm$ bosons, a very physical meaning: it is (up to a normalisation constant) the probability density function for charged lepton emission.\footnote{%
The Wigner pseudo-probability function can in general take negative values, however our `Wigner function' does not take negative values, since its obtained from a non-negative map operating on a non-negative function. This can be considered to be a result of our earlier assumption that the lepton was emitted within a small range of solid angle close to $\nhat$. Before making the Wigner transform we had effectively already integrated the probability density over that narrow range of solid angles.}
\par

\subsection{\boldmath Wigner $P$ symbols $(\j=1)$}\label{sec:tomographyWP}

To reconstruct the density matrix from the data one can either fit the GGM parameters of $\rho$, or analytically perform the inverse transform. 
We take the analytical path --- calculating and then performing the inverse mapping provided by the Weyl transform. 
We therefore seek a set of eight Wigner $P$ symbols (or rather two such sets, one each for $W^+$ and $W^-$), to complement the $Q$ symbols we have already found, that will satisfy the defining property \eqref{eq:wignerPgeneral}.

To help us find these Wigner $P$ symbols, we note that 
\eqref{eq:wignerPgeneral}  implies that the product of any two generalised Gell-Mann operators (for general $d$) is given by
\[
\lambdad_i \lambdad_j = \lambdad_i \left(\frac{d}{4\pi} \int \d \Omega_{\nhat} \WignerPp_j(\nhat) \ket{\nhat}\bra{\nhat}\right)
\]
and so the trace over them yields (again for general $d$),
\begin{align} 
2\delta_{ij} &= \tr( \lambdad_i \lambdad_j) \nonumber\\
 		&= \sum_{\alpha}  \bra{\alpha} \left( \lambdad_i \frac{d}{4\pi} \int \d \Omega_{\nhat}\, \WignerPp_j(\nhat)  \ket{\nhat}\bra{\nhat} \right) \ket{\alpha} \nonumber\\
		&= \frac{d}{4\pi} \int \d \Omega_{\nhat} \, 
			\WignerPp_j(\nhat) \sum_{\alpha} \braket{\nhat|\alpha} \braket{\alpha  | \lambdad_i | \nhat}  \nonumber\\
		&= \frac{d}{4\pi} \int \d \Omega_{\nhat}~ \WignerPp_j(\nhat) \WignerQp_i(\nhat)\,, \label{eq:PQorthogonalW}
\end{align}
where in the last step we have used the completeness relation $\sum_\alpha\ket{\alpha}\bra{\alpha}=I_d$ for the states $\ket{\alpha}$, and the definition \eqref{eq:wignerQgeneral} of the Wigner $Q$ symbol $\WignerQp_i$ for $\lambdad_i$.

We seek two sets of eight functions $\WignerPpm_j(\nhat)$ which are orthogonal to the $\WignerQpm_j(\nhat)$ in the sense of \eqref{eq:PQorthogonalW}.
Since the Gell-Mann operators are traceless a similar calculation to \eqref{eq:PQorthogonalW} shows that the Wigner $P$ symbols must also maintain orthogonality to the identity
\begin{equation}\label{eq:WignerPOrthogonalIdentityW}
0=\int \d \Omega_\nhat \, \WignerPpm_k(\nhat).
\end{equation}
\par
\begin{figure}
\begin{center}
\includegraphics[width=0.24\linewidth]{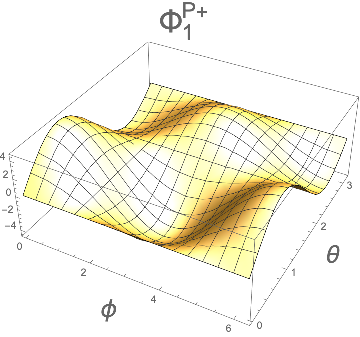}
\includegraphics[width=0.24\linewidth]{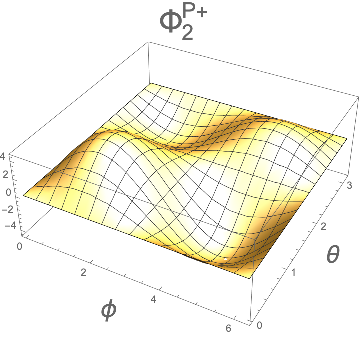}
\includegraphics[width=0.24\linewidth]{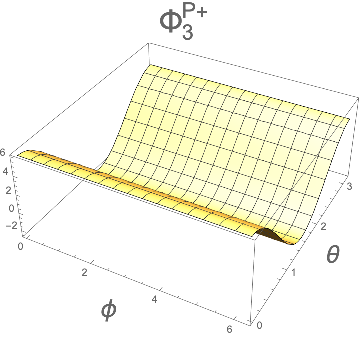}
\includegraphics[width=0.24\linewidth]{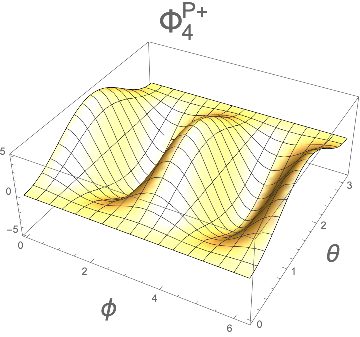}
\includegraphics[width=0.24\linewidth]{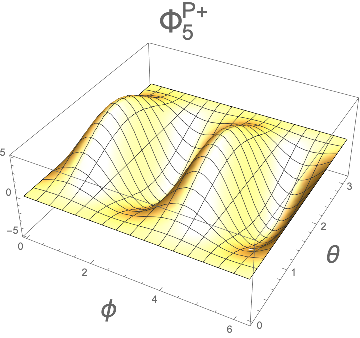}
\includegraphics[width=0.24\linewidth]{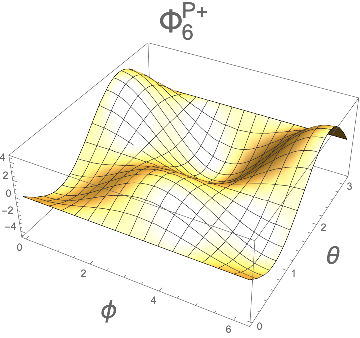}
\includegraphics[width=0.24\linewidth]{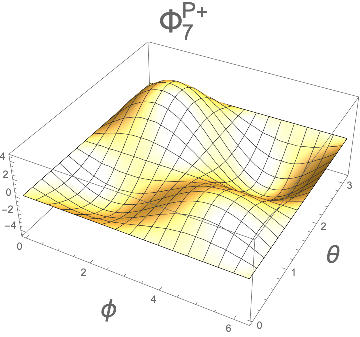}
\includegraphics[width=0.24\linewidth]{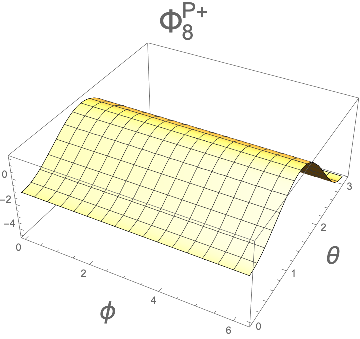}
\caption{\label{fig:WignerPGMs}
Plots showing the form of the eight Wigner $P$ symbols $\WignerPp_i$ 
corresponding to each of the Gell-Mann operators $\lambdathree_i$. These functions are used in recovering the density matrix parameters from the lepton angular probability distribution according to~\eqref{eq:extractGMW}. 
}
\end{center}
\end{figure}

Eight such symbols $\WignerPp_j$ for the $d=3$ Gell-Mann operators (and their $\WignerPm_j$ equivalents for $W^-$ bosons) can be constructed from the $\WignerQp_j$ (respectively $\WignerQm_j$)
\begin{equation}\label{eq:pfromq}
\WignerPpm_i (\nhat) = [M^{-1}]_{ij} \WignerQpm_j (\nhat)
\end{equation}
where the elements of the real symmetric matrix $M$ (which is the same for the $\pm$ cases) are given by the inner products
\begin{eqnarray}
M_{ij} &=& \frac{d}{2}\left< \WignerQpm_i \WignerQpm_j \right>\\
 &=& \frac{d}{2} \frac{1}{4\pi}\int\d\Omega_\nhat\,\WignerQpm_i  (\nhat) \WignerQpm_j(\nhat)
\end{eqnarray}
of the $Q$ symbols. 
The corresponding $\WignerPpm$ for $d=3$ are\label{sec:WignerPfunctions}
\renewcommand\arraystretch{1.5}
\begin{align}
\WignerPpm_1 &=  \sqrt{2} (5 \cos \theta \pm 1) \sin \theta  \cos \phi  \nonumber &
\WignerPpm_5 &=   5 \sin ^2\theta \sin 2 \phi  \nonumber\\
\WignerPpm_2 &=   \sqrt{2} (5 \cos \theta \pm 1) \sin \theta  \sin \phi \nonumber&
\WignerPpm_6 &=   \sqrt{2} (\pm 1-5 \cos \theta)\sin \theta  \cos \phi \\
\WignerPpm_3 &=  \tfrac{1}{4} (\pm 4 \cos \theta+15 \cos 2 \theta +5) \nonumber&
\WignerPpm_7 &=   \sqrt{2} (\pm 1-5 \cos \theta)\sin \theta  \sin \phi \\
\WignerPpm_4 &=  5 \sin ^2\theta \cos 2 \phi &
\WignerPpm_8 &=   \tfrac{1}{4 \sqrt{3} } \left( \pm12 \cos \theta-15 \cos 2 \theta  - 5 \right). \label{eq:WignerPGM}
\end{align}
We observe that the Wigner $P$ and $Q$ symbols for a particular $\lambda_i$ are not generally proportional to one another.
For illustrative purposes the $\WignerPp_j$ are shown graphically in Figure~\ref{fig:WignerPGMs}. 
\par
Using these Wigner $P$ symbols, together with the orthogonality relationships \eqref{eq:PQorthogonalW} and \eqref{eq:WignerPOrthogonalIdentityW}, and the probability density function \eqref{eq:pGMbasis}, each of the eight Gell-Mann density matrix parameters $a_j$ for any $W^+$ or $W^-$ boson may separately be extracted from the lepton angular emission data. The result is the remarkably simple expression
\begin{equation}\label{eq:extractGMW}
a_j = \frac{1}{2} \int \d \Omega_\nhat\, p(\ell^\pm_{\nhat}; \rho)\,\WignerPpm_j .
\end{equation}

With this result one may determine the eight real parameters $a_i$ of $\rho$ from data.
The experimental measurement $\hat a_i$ of the density matrix parameter $a_i$ is obtained by a simple angular average 
\begin{equation}\label{eq:extractGMWexpt}
\hat a_i = \frac{1}{2} \left\langle \WignerPpm_i \right\rangle_{\rm av}
\end{equation}
over the ensemble of decays.
By performing eight such averages (one each for $i\in {1,\ldots,8}$) one obtains each of the parameters of the $W$ boson spin density matrix~\eqref{eq:rhosingleGM}.
By this method we achieve our initial goal of performing quantum state tomography for the spin of a single $W^+$ or $W^-$ boson.

Examples of the application of this procedure for ensembles of simulated $W^+W^-$ pairs from different simulated sources can be found in Section~\ref{sec:tomographyWW}. 

\section{General, non-projective measurements}\label{sec:nonprojective}

The $W^\pm$ boson example described in Section~\ref{sec:singleW} was a particularly convenient introduction to the Wigner-Weyl method because, unusually, the decays of the $W^\pm$ for $m_\ell\ll m_W$ 
are equivalent to \textit{projective} measurements (via operators $\Pi_\nhat$) along  the emitted lepton direction $\nhat$.

In the more general case the angular information is consistent with more than one spin hypothesis along that direction and so will be equivalent to a \emph{non-projective} measurement along $\nhat$. 
We now, therefore, turn to the process of determining the spin density matrix of parents that undergo these more general types of decay. 
We will restrict ourselves initially to the single-particle case; the generalisation to a multipartite case follows in  Section~\ref{sec:nonprojective:multi}.

\subsection{Non-projective single particle decay}\label{sec:nonprojective:single}

We will continue to assume that we can identify daughters with parents (at least in cases where parents are not identical particles) and that we can measure the momentum ${\bf p}$ of one of the daughter particles in the corresponding parent rest frame. 
We shall further assume that we are not able to obtain additional spin measurement information other than obtained from the direction $\nhat = {\bf p}/{|{\bf p}|}$ (and in some cases the speed) of that daughter in that frame; for example we cannot also make a further measurement of the daughter's spin after its production.\footnote{In cases where further information can be obtained from other methods, or for tomography of parents-of-parents that information can be transferred around the system using techniques similar to those described for Monte Carlo simulations in Ref.~\cite{Richardson2001}.}

Let the decay matrix element of a parent with eigenvalue $\spinstate$ of the spin projection operator
to a daughter state $f$ be
\begin{equation}
\mathcal{M}(i_{\spinstate}\rightarrow f).
\end{equation}
From the matrix element we may define a Kraus operator~\cite{nielsen_chuang_2010}  $K_l$
for the decay process, which, when the spin measurement axis is aligned with $\nhat$, and given our assumptions about the known information, is diagonal in the parent's rest frame. In that spin-aligned frame the Kraus operator elements are given by\footnote{This equation employs a non-standard normalization in which the Kraus operators are scaled by 
$D_f = \sum_m |M(i_m \!\to\! f)|^2$. 
Although this normalization does not, in general, satisfy the more standard completeness condition $\sum_f K_f^\dagger K_f = I$, in the examples considered in this work $D_f$ is uniform across all final states. 
Consequently, the overall factor cancels in the construction of the  generalized Wigner $Q$ and $P$ symbols, leaving all tomographic and numerical results unchanged. 
We note, however, that this equation should not be interpreted as a general trace-preserving Kraus decomposition when $D_f$ varies with~$f$.}
\begin{equation}
[K_l]_{\mathsf{m}\mathsf{m}^\prime} =  \frac{\mathcal{M}(i_{\spinstate}\rightarrow f)}{\sqrt{\sum_\mathsf{m}\mathcal{M}^*(i_{\spinstate}\rightarrow f)\mathcal{M}(i_{\spinstate}\rightarrow f)}} \, \delta_{\mathsf{m}\mathsf{m}^\prime}.
\end{equation}
We next define a set of quantum measurement operators~\cite{nielsen_chuang_2010},
\begin{equation}\label{eq:defineFKraus}
F_l = K_l^*K_l,
\end{equation}
which, in the daughter-aligned frame, have diagonal elements
\begin{equation}
[F_l]_{\mathsf{m}\mathsf{m}^\prime} =  \frac{\Gamma(i_{\spinstate}\rightarrow f)}{\sum_\mathsf{m}\Gamma(i_{\spinstate}\rightarrow f)} \delta_{\mathsf{m}\mathsf{m}^\prime}.
\end{equation}
\par
The $F_l$ of the decay in the Standard Model is one of set of possible decay measurement operators $\{F_l\}$ that together would form a positive operator-valued measure (POVM),
that is, a set of positive semi-definite Hermitian operators $\{F_i\}$ on the Hilbert space of the parent and which together would sum to the identity
\begin{equation}\label{eq:povmidentity}
\sum_{l=1}^d F_l = I_d.
\end{equation}
One special case of a POVM is a set of orthogonal projectors $\{\Pi_i\}$ 
\[
\sum_{l=1}^d \Pi_l = I_d, \qquad \Pi_l\Pi_m = \delta_{lm} \Pi_l.
\]
The projective operators correspond to von-Neumann measurements, while the more general POVM operators $\{F_i\}$ correspond to outcomes of a more general non-projective measurements~\cite{nielsen_chuang_2010}.

After the particular $F_l$ corresponding to the amplitudes for the decay is calculated in the frame aligned with the daughter direction we define a rotated version
\begin{equation}\label{eq:rotateF}
F_{l,\nhat}  = U_{\theta,\phi} \, F_l \, U_{\theta,\phi}^\dag,
\end{equation}
where $U_{\theta,\phi}$ is the Wigner matrix for the rotation. 
For a fixed coordinate system the probability density function for the emission of the daughter in direction $\nhat$ is then given by 
\begin{eqnarray}\label{eq:decayangularF}
p(\nhat;\rho,F_l) &=& \frac{d}{4\pi}\tr( F_{l,\nhat}\,\rho) .
\end{eqnarray}

To proceed with the tomographic process we define a non-projective generalisation 
\begin{eqnarray}\label{eq:nonprojectiveQ}
\WignerNonProjectiveQ_{F_l,j} (\nhat) &=& \tr\left({\lambdad_j F_{l,\nhat}}\right) 
\end{eqnarray}
of the Wigner $Q$ symbols~\eqref{eq:wignerQgeneral}.
The tilde, and the measurement operator subscript distinguishes this new symbol from the projective symbols $\WignerQ_i$. Since the measurement operator is a linear combination of projectors by linearity we may now expand the angular emission probability density function
\begin{eqnarray}\label{eq:generalpdfQ}
p(\nhat;\rho,F_l) = \frac{d}{4\pi}\left( \frac{1}{d} + \sum_{j=1}^{d^2-1} a_j \WignerNonProjectiveQ_{F_l,j} (\nhat) \right).
\end{eqnarray}
in terms of these new $\WignerNonProjectiveQ_{F_l,j}$ symbols.
The physical meaning of \eqref{eq:generalpdfQ} is that each term in the sum gives the contribution to the angular emission probability of a daughter emitted near the $\nhat$ direction provided by the $a_j$ component of the $d$-dimensional spin density matrix of the form \eqref{eq:rhosingleGM} for a decay characterised along the decay axis by the diagonal measurement matrix $F_{l}$.

To determine the density matrix from data we must again find the inverse mapping, that from the space of functions over the angles of $\nhat$ to the space of operators. 
We therefore wish to find for our particular operator $F_l$ a set of $(d^2-1)$ non-projective equivalents 
$\WignerNonProjectiveP_{F_k,l}(\nhat)$ to the Wigner $P$ symbols which will
satisfy orthogonality relationships equivalent to \eqref{eq:PQorthogonalW} and \eqref{eq:WignerPOrthogonalIdentityW}. 
That is, the $\WignerNonProjectiveP_{F_l,i}(\nhat)$ should be orthogonal to the generalised $Q$ symbols
\begin{equation}\label{eq:PQorthogonalNonprojective}
2\delta_{ij} = \frac{d}{4\pi} \int \d \Omega_{\nhat}~ \WignerNonProjectiveP_{F_l,i}(\nhat) \WignerNonProjectiveQ_{F_l,j}(\nhat)\,, 
\end{equation}
and also orthogonal to the identity,
\begin{equation}\label{eq:WignerPNonProjectiveOrthogonalIdentity}
0 = \int \d \Omega\, \WignerNonProjectiveP_{F_l,i}(\nhat).
\end{equation}

These generalised $P$ symbols can be found from a method similar to that described for projective measures around \eqref{eq:pfromq}. We form the real symmetric matrix $M$ with elements
\begin{equation}\label{eq:QQinner}
M_{ij} = \frac{1}{2}\frac{d}{4\pi}\int \d\Omega_\nhat\, \WignerNonProjectiveQ_{F_l,i} \WignerNonProjectiveQ_{F_l,j}.
\end{equation}
Provided that $M$ is invertible the $P$ functions are given by
\begin{equation}\label{eq:pfromqgen}
\WignerNonProjectiveP_{F_l,i} = [M^{-1}]_{ij} \WignerNonProjectiveQ_{F_l,j}.
\end{equation}
The GGM Bloch vector parameters of the density matrix \eqref{eq:rhosingleGM} are then given by
\begin{equation}\label{eq:extractGM}
a_j = \frac{1}{2} \int \d \Omega_\nhat\, p(\nhat; \rho)\,\WignerNonProjectiveP_{F_l,j} .
\end{equation}
The experimental measurement of each density matrix parameter can be obtained from the average over an ensemble of decays
\begin{equation}\label{eq:generalexptGM}
\hat{a}_j = \tfrac{1}{2} \left< \WignerNonProjectiveP_{F_l,j} \right>_{\rm av}.
\end{equation}

Unlike in the projective case, the matrix $M$ is no longer guaranteed to have an inverse. For example, in the extreme case where  in which the decay directions are independent of the parent's spin the Kraus operator is $F_l=\tfrac{1}{d}I_d$, meaning that the $\WignerNonProjectiveQ$ each vanish due to \eqref{eq:nonprojectiveQ} and the traceless nature of the $\lambdad_j$. Hence $M$ vanishes, and the $\WignerNonProjectiveP_{F_l,\lambdad_i}$ cannot be defined. So we find the logical result that in cases where the decay directions are independent of the spin, then it is not possible to reconstruct the spin density matrix from an ensemble of such decays.

\subsection{General multipartite measurements}\label{sec:nonprojective:multi}

We now consider our most general case -- that of the density matrix of a general multipartite system with measurements which are not necessarily projective. We continue to assume that we can determine the event-by-event information about the decay directions of a daughter in each respective parents' rest frame, and that each of those decay distributions depends on the respective parent's density matrix. 

The salient points can be found in a bipartite system of two parents, with Hilbert space of dimension $d_1\times d_2$. 
The joint probability density function (PDF) for the emissions of the respective daughters is given by the effect of two quantum operations \eqref{eq:rotateF} --- $F^{(A)}_{l,\nhat_1}$  associated with the decay of parent A and $F^{(B)}_{m,\nhat_2}$ associated with that of parent B. 

Defining the combined bipartite Kraus operator
\begin{equation}\label{eq:krausproduct}
K_{l,m,\nhat_1,\nhat2} = K^{(A)}_{l,\nhat_1} \otimes K^{(B)}_{m,\nhat_2}
\end{equation}
the probability density function over the pair of emission directions is given by
\begin{equation}\label{eq:definepdfgeneralKraus}
p({\nhat_1}, {\nhat_2}; \rho) = \frac{d_1d_2}{(4\pi)^2} \tr\left(K_{l,m,\nhat_1,\nhat2}~\rho~K_{l,m,\nhat_1,\nhat2}^\dag \right).
\end{equation}
Using the cyclical nature of the trace, the definition \eqref{eq:defineFKraus} of the measurement operator and the factorisation \eqref{eq:krausproduct} of the Kraus operators we may write this in the form
\begin{equation}\label{eq:definepdfgeneral}
p({\nhat_1}, {\nhat_2}; \rho) = \frac{d_1d_2}{(4\pi)^2} \tr\left(\rho\, F^{(A)}_{l,\nhat_1} \otimes F^{(B)}_{m,\nhat_2}\right) ,
\end{equation}
This PDF is a map from the space of density operators of the bipartite system $\rho:\mathcal H_{AB} \rightarrow \mathcal H_{AB}$ to functions on $S^2\otimes S^2$. To reconstruct the two-particle density matrix we therefore wish to invert that map.

Provided that the inner-product matrices \eqref{eq:QQinner} for each single-particle decay are invertible we may define generalised $P$ symbols $\WignerNonProjectiveP$ symbols for each according to \eqref{eq:pfromqgen}. 
The $a_i$ parameters can be obtained from data using an average~\eqref{eq:generalexptGM} over the appropriate single-particle PDF.
The $b_j$ parameters are found from corresponding averages of the angular distributions of the daughter directions $\nhat_2$ from parent $B$, averaging over the $\WignerNonProjectiveP$ for that decay.

The $c_{ij}$ parameters can be extracted (under the same conditions) from the double integral over the pair of daughter emission directions,
\begin{equation}\label{eq:extractcgeneral}
c_{ij} = \left(\tfrac{1}{2}\right)^2 \iint \d \Omega_{\nhat_1} \d \Omega_{\nhat_2} \, p(\ell^+_{\nhat_1}, \ell^-_{\nhat_2}; \rho) ~ \WignerNonProjectiveP_{F^{(A)}_l,i}(\nhat_1)~\WignerNonProjectiveP_{F^{(B)}_m,j}(\nhat_2)\, ,
\end{equation}
where the angular integrals are each performed in the respective parents' rest frames. The parameters can be measured experimentally from the averages
\begin{equation}\label{eq:measurecgeneral}
\hat c_{ij} = \left(\tfrac{1}{2}\right)^2  \left<\WignerNonProjectiveP_{F^{(A)}_l,i}(\nhat_1) \,\WignerNonProjectiveP_{F^{(B)}_m,j}(\nhat_2)\, \right>_{\rm av},
\end{equation}
of products of generalised Wigner $P$ symbols.
Thus, provided that the matrices \eqref{eq:QQinner} for each decay are non-singular we can measure all of the parameters $a_i$, $b_j$ and $c_{ij}$ of the 
bipartite spin density matrix \eqref{eq:doubleGM}, providing full quantum state tomography of the bipartite system. The procedure generalises in a similar manner to multipartite systems of more than two distinguishable parents. 

When two or more parents are indistinguishable the density matrix must be constructed so as to have the appropriate symmetry under exchange of the parent particles' labels. 
For example, if the parents are a pair of identical bosons then the state vector must be symmetric under interchange of the labels of the members of the pair. The bipartite density matrix must then also respect the same symmetry, so that:
\begin{equation}\label{eq:measureaidentical}
\hat{a}_i = \hat{b}_i = \frac{1}{4} \left< 
{\WignerNonProjectiveP_{F_l^{(1)},i}}(\nhat_1) + {\WignerNonProjectiveP_{F_m^{(2)},i}}(\nhat_2) 
\right>_{\rm av}.
\end{equation}
and
\begin{equation}\label{eq:measurecidentical}
\hat c_{ij} = \hat c_{ji} = \frac{1}{8}  \left<
\WignerNonProjectiveP_{F_l^{(1)},i}(\nhat_1) \,\WignerNonProjectiveP_{F_m^{(2)},j}(\nhat_2) 
~~+ ~~i \leftrightarrow j
\right>_{\rm av},
\end{equation}
where by defining separate measurement operators $F_l^{(1)}$ and $F_m^{(2)}$ we leave open the possibility that the measurement operators for the daughters differ.
\par

\section{Example decays}\label{sec:Fexamples}

In Section~\ref{sec:singleW} we described the tomographic procedure for the case of the decay of a single $W^+$ or $W^-$ decay, 
which had the simplifying factor of being represented by projective decay operators. 
Having developed the more general formalism for non-projective decays and for multi-particle tomography in Section~\ref{sec:nonprojective}, we now describe several illustrative examples for parents with $\j=\half$ or $\j=1$.

\subsection{Top quark decay/spin-half fermion decay}\label{sec:generalspinhalf}

As a first example let us take the case of leptonic decay of a top quark. 
The differential decay rate is given by
\begin{equation}\label{eq:spinanalysing}
\frac{1}{\Gamma}\frac{\d\Gamma}{\d \cos\theta} = 
\tfrac{1}{2}\left( 1+|{\vec{\mathcal P}}|\kappa \cos\theta \right),
\end{equation}
where ${\vec{\mathcal P}}$ is the polarisation vector of length $0\leq|{\vec{\mathcal P}}|\leq 1$, and $\theta$ is the decay direction of the lepton relative to the polarisation direction~\cite{Brandenburg:2002xr}.
The value of the `spin analysing power' $\kappa$ varies in the range $-1\leq\kappa\leq 1$. 
In leptonic top decays $\kappa_{\ell^+}\approx +1.0$, while
the $b$-quark daughter in top decay has spin analysing power of about $\kappa_{b}\approx -0.41$~\cite{Brandenburg:2002xr}.

Consider now a two-dimensional POVM $\{F_+,F_-\}$ where
\begin{equation}\label{eq:povmhalf}
F_\pm = \half( I_2 \pm \kappa \sigma_3).
\end{equation}
The corresponding measurement operators are projective when $\kappa=1$ for which $F_\pm\rightarrow\Pi_{\pm}$ (and when $\kappa=-1$ for which $F_\pm\rightarrow\Pi_{\mp}$), and are non-projective for intermediate values $-1<\kappa<1$.
The probability density function \eqref{eq:decayangularF} for daughter emission in direction $\nhat$, and with density 
matrix\footnote{Our general parameterisation \eqref{eq:rhosingleGM} of $\rho$ is related to that in \eqref{eq:rhospinhalf} from Ref.~\cite{Brandenburg:2002xr} by $a_i=\alpha_i/2$.}
\begin{equation}\label{eq:rhospinhalf}
\rho_2 = \tfrac{1}{2} \left( \itwo + {\boldsymbol \alpha} \cdot \boldsymbol\sigma \right)
\end{equation}
is given by
\begin{eqnarray}
p(\nhat; \rho) &=& \frac{2}{4\pi} \tr \left( F_{+,\nhat}\, \rho \right)\nonumber\\
               &=& \frac{1}{4\pi} (1+ \kappa\, {\boldsymbol\alpha}\cdot\nhat).
\end{eqnarray}
This corresponds with \eqref{eq:spinanalysing} for polarisation vector $\vec{\mathcal P}={\boldsymbol\alpha}$. Thus the POVM \eqref{eq:povmhalf} reproduces the decay dynamics. 

For this spin-half case the generalised $Q$ symbols can be calculated using the procedures of Section~\ref{sec:nonprojective:single} and are
\begin{eqnarray}\label{eq:fermionQ}
\WignerNonProjectiveQ_{F_+,\sigma_1} &=& \kappa \sin\theta \cos\phi \nonumber\\
\WignerNonProjectiveQ_{F_+,\sigma_2} &=& \kappa \sin\theta \sin\phi  \nonumber\\
\WignerNonProjectiveQ_{F_+,\sigma_3} &=& \kappa \cos\theta.
\end{eqnarray}
The matrix \eqref{eq:QQinner} of their inner products is invertible except when $\kappa=0$. 
Hence for tomography to be possible it is necessary and sufficient that the decay have non-zero spin analyzing power. Provided that condition is satisfied the corresponding generalised $P$ symbols are
\begin{eqnarray}\label{eq:fermionP}
\WignerNonProjectiveP_{F_+,\sigma_1} &=& \frac{3}{\kappa} \sin\theta \cos\phi \nonumber\\
\WignerNonProjectiveP_{F_+,\sigma_2} &=& \frac{3}{\kappa} \sin\theta \sin\phi  \nonumber\\
\WignerNonProjectiveP_{F_+,\sigma_3} &=& \frac{3}{\kappa} \cos\theta.
\end{eqnarray}
\par
The special cases of $F$ corresponding to projective measurements $\Pi_\pm$ can be found by setting $\kappa=1$ or $-1$. For these cases \eqref{eq:fermionQ} and \eqref{eq:fermionP} become the expressions for the $d=2$ Wigner symbols $\WignerQpm_{\sigma i}$ and $\WignerPpm_{\sigma i}$ respectively. 
\par
We note that for each GGM index $(i=1,\,2,\,3)$ the (generalised) Wigner $P$ and $Q$ symbols are proportional to one another, a property that is particular to $d=2$.
We also note that the measurement operators \eqref{eq:povmhalf} are the most general diagonal POVM for a $d=2$ system, and so the methodology of this section applies equally to decays of any other spin-half system.

Examples of applying these $\j=\half$ symbols in simulated data may be found in Appendix~\ref{sec:qubittomography}.

\subsection{\boldmath $Z\rightarrow \ell^+\ell^-$ decay}\label{sec:Zdecay}

As a further example let us consider the case $Z\rightarrow \ell^+\ell^-$. 
The $Z$ boson coupling to fermions is given by~\cite{thomson_2013}
\begin{equation}
-\i \frac{g_{\rm W}}{\cos\theta_{\rm W}} \gamma^\mu \half (c_V-c_A \gamma^5) ,
\end{equation}
where $\theta_{\rm W}$ is the weak mixing angle, 
the vector and axial couplings are $c_V=c_L+c_R$ and $c_A=c_L-c_R$, and the right- and left-chiral couplings are $c_R$ and $c_L$ respectively.
For decays $Z\rightarrow\ell^+\ell^-$ to the charged leptons the axial and vector couplings are $c_A=-0.5064$ and $c_V=-0.0398$ respectively~\cite{ParticleDataGroup:2020ssz}, and so $c_L = -0.273$ and $c_R=+0.233$. 
Neglecting the mass of the charged leptons, higher-order EW effects, and any photon propagator contribution, 
the $\ell^+$-aligned Kraus operator is
\begin{equation}
K_{Z\rightarrow\ell\ell} = \frac{\i}{\sqrt{|c_R|^2+|c_L|^2}}~\mathrm{diag}\left(c_R,\,0,\,c_L\right).
\end{equation}
The diagonal measurement operator \eqref{eq:defineFKraus} is given by
\begin{eqnarray}
F_{Z\rightarrow\ell\ell} &=& K^\dag_{Z\rightarrow\ell\ell} K_{Z\rightarrow\ell\ell}\\
 &=& \frac{1}{|c_R|^2+|c_L|^2}~\mathrm{diag}\left(|c_R|^2,\,0,\,|c_L|^2\right).
\end{eqnarray}
The generalised $Q$ symbols are calculated from \eqref{eq:nonprojectiveQ} and can be expressed as a linear combinations 
\begin{equation}
    \WignerNonProjectiveQ_{F_{Z\rightarrow\ell\ell},j} = 
    \frac{1}{|c_R|^2+|c_L|^2} \left( |c_R|^2\WignerQp_{\lambda_j} 
    + |c_L|^2 \WignerQm_{\lambda_j} \right)
\end{equation}
of the projective Wigner $Q$ symbols.
The matrix \eqref{eq:QQinner} of their inner products is invertible provided that $|c_L|\ne |c_R|$ (a condition satisfied for $Z\rightarrow \ell^+\ell^-$, but not for e.g. the electromagnetic process $\gamma\rightarrow \ell^+\ell^-$). Under such conditions the generalised $P$ functions are given by 
\begin{equation}\label{eq:ZP}
    \WignerNonProjectiveP_{F_{Z\rightarrow\ell\ell},j} = A_{jk} \WignerPp_{\lambda_k} 
\end{equation}
where the matrix $A$ is 
\[\renewcommand\arraystretch{1.05}
\frac{1}{|c_R|^2-|c_L|^2}\left(
\begin{array}{cccccccc}
 |c_R|^2 & 0 & 0 & 0 & 0 & |c_L|^2 & 0 & 0 \\
 0 & |c_R|^2 & 0 & 0 & 0 & 0 & |c_L|^2 & 0 \\
 0 & 0 & |c_R|^2-\frac{1}{2}|c_L|^2 & 0 & 0 & 0 & 0 & \frac{\sqrt{3}}{2}|c_L|^2 \\
 0 & 0 & 0 & |c_R|^2-|c_L|^2 & 0 & 0 & 0 & 0 \\
 0 & 0 & 0 & 0 & |c_R|^2-|c_L|^2 & 0 & 0 & 0 \\
 |c_L|^2 & 0 & 0 & 0 & 0 & |c_R|^2 & 0 & 0 \\
 0 & |c_L|^2 & 0 & 0 & 0 & 0 & |c_R|^2 & 0 \\
 0 & 0 & \frac{\sqrt{3}}{2}|c_L|^2 & 0 & 0 & 0 & 0 & \frac{1}{2}|c_L|^2+|c_R|^2 \\
\end{array}
\right).
\]
When $|c_R|=1$ and $|c_L|=0$ the matrix $A$ becomes the identity, and the generalised $P$ symbols become the Wigner $P$ symbols \eqref{eq:WignerPGM} found previously for the $W^+$ boson decay in the limit of massless fermion daughters.

\section{Example quantum applications}\label{sec:applications}

The ability to reconstruct the multipartite spin density matrix for multipartite systems of arbitrary spin $\j$ allows us to undertake a broad programme of studies of quantum properties using the experimental data from colliders. In what follows we provide some illustrative examples of possible applications.

\subsection{Entanglement detection for qubit and qutrit pairs}\label{sec:entanglementintro}

Entanglement is a fundamental property of quantum  systems\footnote{For introductions see for example Refs.~\cite{Horodecki:2009zz,Casini:2022rlv} and references therein.}. 
A pure state $\ket{\psi}$ is entangled if it cannot be written
\begin{equation}
\ket{\psi} = \ket{\phi_A}\otimes\ket{\phi_B}\otimes\ldots 
\end{equation}
as a product of states $\ket{\phi}$ of the subsystems.
A general mixed state $\rho$ of $n$ systems is entangled if it is not possible to approximate it~\cite{PhysRevA.40.4277}
\begin{equation}\label{eq:classicallycorrelated}
\rho = \sum_i^n p_i \, \rho_1^i \otimes \ldots \otimes \rho_1^n
\qquad\qquad 0<p_i\leq 1\end{equation}
as a convex combination of product states.
The states that can be expressed in the form  \eqref{eq:classicallycorrelated} are not entangled but may still be classically correlated. 

The necessary and sufficient conditions for determining entanglement of a mixed state are in general difficult (NP-hard) to calculate~\cite{arxiv.quant-ph/0303055,0810.4507}, and they are not analytically known for bipartite mixed states of dimension larger than $2\times3$~\cite{Horodecki:2009zz}. However several sufficient conditions for entanglement are known. 
One such condition is based on the observation that a state can only have non-zero \emph{concurrence} $c(\rho)$ if it is entangled~\cite{MINTERT_2005}.
The concurrence for a pure state can be defined by an operator on a two-fold copy $\ket\psi\otimes\ket\psi$ of the bipartite state~\cite{MINTERT_2005,Mintert_2007},
\[
c(\psi) = \sqrt{\bra\psi \otimes \bra\psi A \ket\psi\otimes\ket\psi}
\]
where $A$ is a self-adjoint operator which is antisymmetric with respect to exchange of both parts of the bipartite system. 
If the state is separable then it must be symmetric under this exchange, and so the expectation value of $A$ must vanish for separable states. Hence a non-zero concurrence shows that the state must be entangled. 

The concurrence $c(\rho)$ for a mixed state is defined by the method of convex roof extension. The essence of this method is that, since a mixed state can be obtained in a variety of different ways from pure states, one must find a limit in terms of the different ensembles of states $(p_i,\ket{\psi})$ that could have led to the mixed state $\rho$. 
The concurrence for a mixed state is therefore defined by
\begin{equation}
c(\rho) = {\rm inf}\sum_i p_i c(\ket{\psi}), \qquad \sum_i p_i=1, ~~p_i\geq0,
\end{equation}
where the infimum is taken over all ensembles $\{p_i, \ket{\psi_i}\}$ for which $\rho=\sum_i p_i \ket{\psi_i}\bra{\psi_i}$.
This infimum is in general difficult to evaluate. 

For bipartite systems of a pair of qubits ($d=2\times 2$) the concurrence is analytically calculable~\cite{Bennett_1996,Wootters:1997id}. It may be found using the auxiliary $4\times 4$ matrix~\cite{Wootters:1997id}
\[
R= \rho (\sigma_2 \otimes \sigma_2) \rho^* (\sigma_2 \otimes \sigma_2), 
\]
where the complex conjugate is taken in the basis in which $\sigma_z$ is diagonal. The matrix $R$, though not Hermitian, has non-negative eigenvalues. Denoting their positive square-roots $\xi_i$ with $i\in\{1,2,3,4\}$, and with $\xi_1$ the largest, the concurrence is~\cite{Wootters:1997id}
\begin{equation}
c(\rho) = {\rm max}(0,\,\xi_1-\xi_2-\xi_3-\xi_4).
\end{equation}
The maximum value of $c(\rho)$ for $2\times 2$ systems is unity, a value obtained for pure maximally-entangled states.


For bipartite systems of dimension larger than $2\times 3$ the concurrence is not analytically calculable. Nevertheless, entanglement may still be demonstrated by evaluating a calculable lower bound on the square of the concurrence.
One such lower bound for a mixed state is given by~\cite{Mintert_2007}
\begin{equation}\label{eq:concurrencebound}
\big( c(\rho) \big)^2 \geq 2\tr(\rho^2) - \tr(\rho_A^2) - \tr(\rho_B^2) \equiv \mintertboundsq.
\end{equation}
Here $\rho_A$ is the reduced density matrix 
\[
\rho_A=\tr_B(\ket{\psi}\bra{\psi})
\]
for system $A$ formed by taking the partial trace over system $B$  and $\rho_B$ has a similar meaning for system $B$. 
The key property of this bound is that $\mintertboundsq >0$ implies that $c^2(\rho) >0$ 
which in turn implies that $\rho$ is an entangled state.
If $\mintertboundsq \leq 0$ it is not possible to determine, using this measure, whether the state is entangled or not.  

For a state corresponding to the spins of a pair of qutrits, such as $WW$, $WZ$ and $ZZ$ dibison systems, 
we have parameterised the bipartite $3\times 3$ density matrix in terms of the direct products of Gell-Mann operators \eqref{eq:doubleGM}. In this basis the traces entering \eqref{eq:concurrencebound} are given by
\begin{eqnarray*}
\tr(\rho^2) &=& \tfrac{1}{9} + \tfrac{2}{3}\sum_{i=1}^8 a_i^2 + \tfrac{2}{3}\sum_{j=1}^8 b_j^2 + 4 \sum_{i,j=1}^8 c_{ij}^2,\\
\tr(\rho_A^2) &=& \tfrac{1}{3} + 2 \sum_{i=1}^8 a_i^2 ,\\
\tr(\rho_B^2) &=& \tfrac{1}{3} + 2 \sum_{j=1}^8 b_j^2.
\end{eqnarray*}
Hence the bound defined in \eqref{eq:concurrencebound} may be expressed
\begin{equation}\label{eq:mintertinGM}
\mintertboundsq = -\tfrac{4}{9} - \tfrac{2}{3}\sum_{i=1}^8 a_i^2 - \tfrac{2}{3}\sum_{j=1}^8 b_j^2 + 8 \sum_{i,j=1}^8 c_{ij}^2
\end{equation}
symmetrically in terms of the 80 GGM parameters $a_i$, $b_j$ and $c_{ij}$ of the $3\times 3$  bipartite density matrix. 

Where \eqref{eq:mintertinGM} is positive the state is determined to be entangled. Where the bound it is less than or equal to zero the $\mintertboundsq$ test is inconclusive. 

\subsection{Bell inequality tests for bipartite systems}


Classical theories are expected to obey local realism: the principles that signals cannot travel faster than light, and that subsystems have properties that are independent of the way in which they are measured.
Local realist theories must satisfy certain inequalities known as Bell inequalities, which result from the requirement that the marginal probability distribution for measurements each subsystem be non-negative. These inequalities are predicted to be violated in quantum theories.
\par
Bell~\cite{BellOnTheEPR} first showed how one could distinguish between the predictions of quantum mechanics and those of classical/local realist theories.  Thus, by experimentally testing these inequalities one could empirically discriminate between local realist and quantum theories. 

For a pair of qubits
the necessary and sufficient conditions~\cite{fine_prl} that measurements be compatible with any local realist theory is that 
they satisfy the inequality of Clauser-Horne-Shimony-Holt (CHSH)~\cite{chsh} 
\begin{equation}\label{eq:chsh}
\mathcal{I}_2 =E(a, b)-E(a, b') + E(a', b) + E(a', b') \leq 2\,.
\end{equation}
Quantum mechanics, by contrast, permits values of $\mathcal{I}_2$ larger than two,  up to the Cirel'son bound~\cite{cirelson} of $2\sqrt{2}$.

For bipartite systems of $WW$, $WZ$ or $ZZ$ bosons such tests can be performed using the Collins-Gisin-Linden-Massar-Popescu (CGLMP) inequality~\cite{PhysRevA.65.032118,Collins_2002}, the optimal Bell test for bipartite systems of pairs of qutrits. 
To construct it, one consider observers $A$ and $B$, 
each having two measurement settings,  $A_1$ and $A_2$ for $A$, and $B_1$ and $B_2$ for $B$,
with each experiment having three possible outcomes.
One denotes by $P(A_i=B_j+k)$ the probability that the outcomes $A_i$ and $B_j$
differ by $k$ modulo $3$. We then construct the linear function
 \begin{multline}\label{eq:cglmpeqn}
 \mathcal{I}_3 = P(A_1=B_1) + P(B_1=A_2+1) + P(A_2=B_2) \\ 
 + P(B_2=A_1) - P(A_1=B_1-1) - P(B_1=A_2)  \\- P(A_2=B_2-1) - P(B_2=A_1-1).
 \end{multline}
In theories admitting realism, this function is bounded by~\cite{PhysRevA.65.032118,Collins_2002}
\begin{equation}\label{eq:cglmpbound}
\mathcal{I}_3 \leq 2.
\end{equation}
\par
Quantum theories again violate this bound.
Within quantum mechanics we can calculate the expectation values 
\begin{eqnarray}
\mathcal{I}_3 &=& \langle \mathcal{B}\rangle \nonumber\\
              &=& \tr(\rho\mathcal{B})
\end{eqnarray}
of quantum Bell operators $\mathcal{B}$~\cite{PhysRevLett.68.3259}, 
and hence predict the expectation values of the corresponding experiments. 
An quantum operator for the CGLMP test \eqref{eq:cglmpbound} was defined in Ref.~\cite{Acin_2002}. A slightly modified version, suited to scalar decays, was defined in a proposal~\cite{BARR2022136866} to test qutrit Bell inequalities in Higgs boson decays to $WW^{(*)}$
\begin{equation}\label{eq:cglmpoperator}
B_\mathrm{CGLMP} =  - \tfrac{2}{\sqrt{3}} 
\left( S_x \otimes S_x + S_y \otimes S_y \right) 
        + \lambda_4\otimes\lambda_4 + \lambda_5\otimes\lambda_5 ,
\end{equation}
where $S_x = (\lambda_1+\lambda_6)/\sqrt{2}$ and $S_x = (\lambda_2+\lambda_7)/\sqrt{2}$, and is used again here. 
The largest possible value of $\langle B_\mathrm{CGLMP}\rangle$ in non-relativistic quantum mechanics for a maximally entangled 
state is $4/(6\sqrt{3}-9)\approx2.8729$~\cite{Collins_2002}.
A slightly larger extremal value of $1+\sqrt{11/3}\approx2.9149$
can be achieved for other states that are not maximally entangled~\cite{Acin_2002}.

The operator \eqref{eq:cglmpoperator} depends only on spin operators defined in the $xy$ plane~\cite{BARR2022136866}. 
It may be rotated through arbitrary angles via similarity transformations using products of the $d=3$ operators $U^\dag_{\theta,\phi}$  \eqref{eq:dmatrix}, to obtain corresponding operators in other planes. 
The expectation values of the Bell operator can then be calculated from the bipartite density matrix for arbitrary directions. 
In particular we can calculate the maximum value, 
\begin{equation}\label{eq:cglmpmax}
  \left<B_\mathrm{CGLMP}\right>_\mathrm{max} = \underset{\theta,\phi}{\mathrm{max}}\left (\tr\left( \rho\, U^\dag_{\theta,\phi}\otimes U^\dag_{\theta,\phi} \, \mathcal{B}_{\rm CGLMP} \, U_{\theta,\phi} \otimes U_{\theta,\phi} \right)\right), 
\end{equation}
of our CGLMP Bell operator evaluated in any plane\footnote{Here we apply the same rotation to both sides of the experiment. One could choose to perform independent arbitrary unitary transformations on each side of the bipartite experiment if one wished to be sure to obtain the global maximum value of the bound.}. The underlying theory is incompatible with a local realist explanation if this quantity exceeds two. 


\section{Monte Carlo Simulations}\label{sec:examples}

\subsection{\boldmath Tomography of example bipartite systems}\label{sec:tomographyWW}

To illustrate the method Monte Carlo simulations were performed of the bipartite qutrit processes listed in Table~\ref{tab:processes}.\footnote{Additional processes of other Hilbert space dimension are listed in Table~\ref{tab:additionalprocesses} in Appendix \ref{sec:qubittomography}.}
The {\texttt Madgraph v2.9.2} software~\cite{Alwall_2014} was used which includes
full spin correlation, relativistic and Breit-Wigner effects. 
Events were generated at leading order at a proton-proton centre-of-mass energy of 13\,TeV. Higgs boson events were generated using the Higgs effective-field theory model.
Higher order corrections to the shapes of the normalised angular distributions,  which for Higgs boson decays to four leptons are typically at the $\lesssim$\,5\% level~\cite{Prophecy4f2006,Boselli:2015aha,Prophecy4f2015}, are neglected. 
Care was taken not to introduce any selection cuts on the leptons on their transverse momentum or rapidity in order not to bias the expectation values of the angular observables.
In a real experiment any such biases due to trigger and acceptance effects would have to be calculated and corrected for.
When opposite-sign, same-family leptons are selected from continuum $Z^0$ processes the di-lepton invariant mass is required to be in the range [80,100]\,GeV in order to select the $Z^0$ contribution and reduce contributions from virtual photons. 

Some samples involve particles with masses far from their Standard Model values in order better to illustrate particular aspects. For example we include a process with decay of a $W$ boson to a 30\,GeV `$\tau$' lepton (and a massless neutrino) for which $v_\ell\approx 0.75$ in order to test the $W$ boson tomography for a system in which the lepton mass is important.

\begin{table}\begin{center}
\begin{tabular}{c c c}
\hline
Process & $d_1\times d_2$ \\
\hline
$pp\rightarrow H \rightarrow WW^* \rightarrow \ell^+ \nu_\ell \,\ell^-{\bar{\nu}}_\ell$ &   $3\times 3$\\
$pp \rightarrow W^+W^- \rightarrow  \ell^+ \nu_\ell \,\ell^-{\bar{\nu}}_\ell$ &   $3\times 3$ \\
$pp\rightarrow H(200) \rightarrow W^+W^- \rightarrow \ell^+ \nu_\ell \,\tau^-(30){\bar{\nu}}_\tau$ &   $3\times 3$\\
$pp\rightarrow H \rightarrow ZZ^* \rightarrow  e^+ e^- \,\mu^+ \mu^-$ &   $3\times 3$ \\
$pp \rightarrow ZZ \rightarrow  e^+ e^- \,\mu^+ \mu^-$ &   $3\times 3$ \\
$pp \rightarrow W^+Z \rightarrow  e^+ \nu_e \,\mu^+ \mu^-$ &   $3\times 3$ \\
\hline
\end{tabular}
\caption{\label{tab:processes}
Physics processes simulated for bipartite systems of qutrits. 
The final column gives the dimension of the Hilbert space of the bipartite system.
The particles $H(200)$ and $\tau(30)$ have their masses changed from their Standard Model values to those in the parentheses (in GeV) in order better to test particular aspects of the tomography method. For each process $10^6$ events were simulated. The CM energy for $pp$ collisions was $\sqrt{s}=13\,$TeV.
}
\end{center}
\end{table}

\begin{figure}\centering
\includegraphics[width=0.4\linewidth]{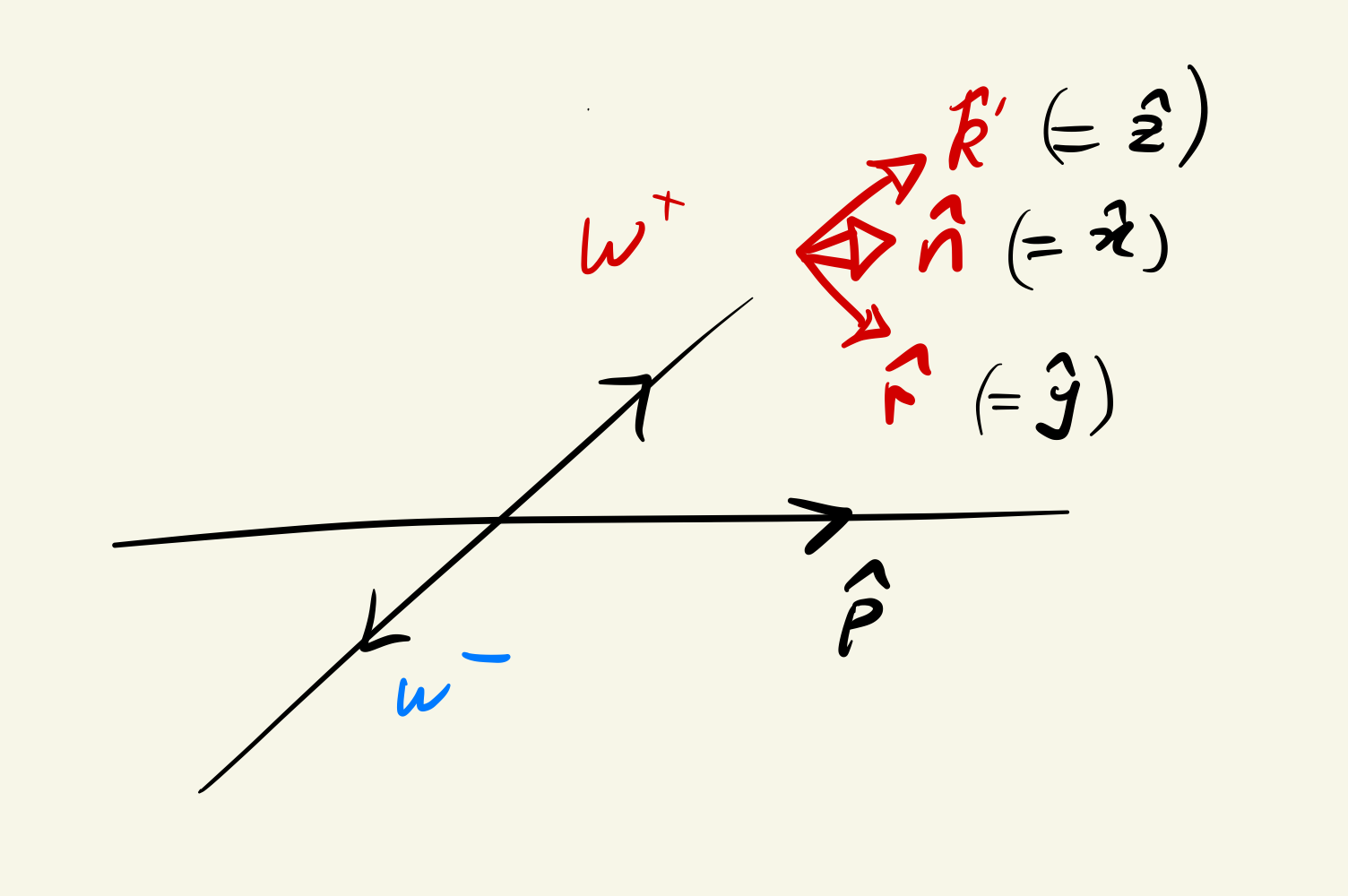}
\caption{\label{fig:axes}Cartoon of the right-handed coordinate axes used for the quantum state tomography of bipartite systems such as $W^+W^-$. The axes are aligned as indicated (with $\nhat=\hat{\bf x}$ pointing out of the page), and are each defined in the respective bosons' rest frames.
For tomography of other bipartite systems the axes are similarly defined, with the $\hat z$ axis parallel to the direction of the first of the two named particles in the bipartite centre-of-mass frame.
}
\end{figure}

A basis choice is required to perform quantum state tomography on the systems. 
The orthonormal basis chosen for this density matrix reconstruction is a modification of that
proposed for measuring spin correlation in top quarks~\cite{Bernreuther_2015}, and is described in~\cite{BARR2022136866}. 
For the purposes of these examples it is assumed that the vector boson rest frames can be identified. 
For the $W^+W^-$ case, in the centre-of-mass frame of the boson pair the direction 
of the $W^+$ is denoted $\hat{\bf k}$. The direction $\hat{\bf  p}$ of one of the beams in that frame 
is determined, and a mutually orthogonal basis constructed from them:
\[
\hat{\bf k},  \qquad 
\hat{\bf r} = \frac{1}{r}(\hat{\bf p}-y \hat{\bf{k}}), \qquad 
\hat{\bf n} = \frac{1}{r}(\hat{\bf p} \times \hat{\bf k}),
\]
where $y=\hat{\bf p} \cdot \hat{\bf k}$ and $r=\sqrt{1-y^2}$.
This provides a right-handed orthonormal basis $\{\hat{\bf n},\,\hat{\bf r},\,\hat{\bf k}\}$  
defined in the diboson CM frame. Boosts are then performed
into each of the $W^\pm$ rest frames, 
and a new basis $\{\hat{\bf x},\,\hat{\bf y},\,\hat{\bf z}\} \equiv \{\hat{\bf n},\,\hat{\bf r},\,\hat{\bf k}^\prime\}$ 
defined in each such that $\hat{\bf n}$ and $\hat{\bf r}$ are unmodified, while 
each $\hat{\bf k}^\prime$ is parallel to $\hat{\bf k}$ but has been unit-normalised after the corresponding boost. 
For other bipartite systems a similar procedure is followed, with the axes defined in the respective parents rest frames. 
A sketch of the resulting axes can be found in Figure~\ref{fig:axes}.

\par
Quantum state tomography was then performed on the simulated final states. 
For the bipartite systems of $W^+W^-$ bosons the $\hat c_{ij}$ are found from~\eqref{eq:measurecgeneral}, using the projective Wigner $P$ symbols \eqref{eq:WignerPGM}. The
parameters for the individual $W^\pm$ boson decays to leptons of negligible mass we find the GGM parameters using equation \eqref{eq:extractGMWexpt}.
For the decay of the $W$ boson to the 30 GeV `heavy $\tau$' lepton we instead use the non-projective Wigner $P$ symbols \eqref{eq:wignerPnonprojectiveW}.
For the leptonic $Z$ boson decays we find the GGM parameters from the average \eqref{eq:generalexptGM} of the generalised Wigner $P$ symbol \eqref{eq:ZP} for $Z\rightarrow \ell^+\ell^-$. For the case of a bipartite system of pair $Z$ bosons we symmetrize the density matrix parameters over exchange of labels using \eqref{eq:measureaidentical} and \eqref{eq:measurecidentical}, with the generalised Wigner $P$ symbols again being given by \eqref{eq:ZP}. Further examples illustrating the process of tomography for $2\times 2$  and $2\times 3$ dimensional bipartite systems may be found in Appendix~\ref{sec:qubittomography}.

For these initial studies we do not apply experimental selections and we assume that the parent rest frames can be reconstructed with sufficient precision to calculate the generalised Wigner $P$ symbols. 
The extent to which these approximation are valid, or can be corrected for, will depend on the details of the final state (for example whether neutrinos are created), as well as on details like the experimental selection and detector resolution.
After applying trigger and experimental selection requirement the experimentalist may wish to obtain GGM parameters from fits to data, rather than from simple averages, in order to reduce bias caused by the selection requirements, and to evaluate the corresponding uncertainties.
An advantage of our approach is that this fitting could also be performed directly on distributions of experimental interest derived from those GGM parameters, such as measures of entanglement or expectation values of Bell operators.
\par
\begin{figure}
\begin{center}
\begin{subfigure}[b]{0.48\textwidth}\centering
\includegraphics[width=\linewidth]{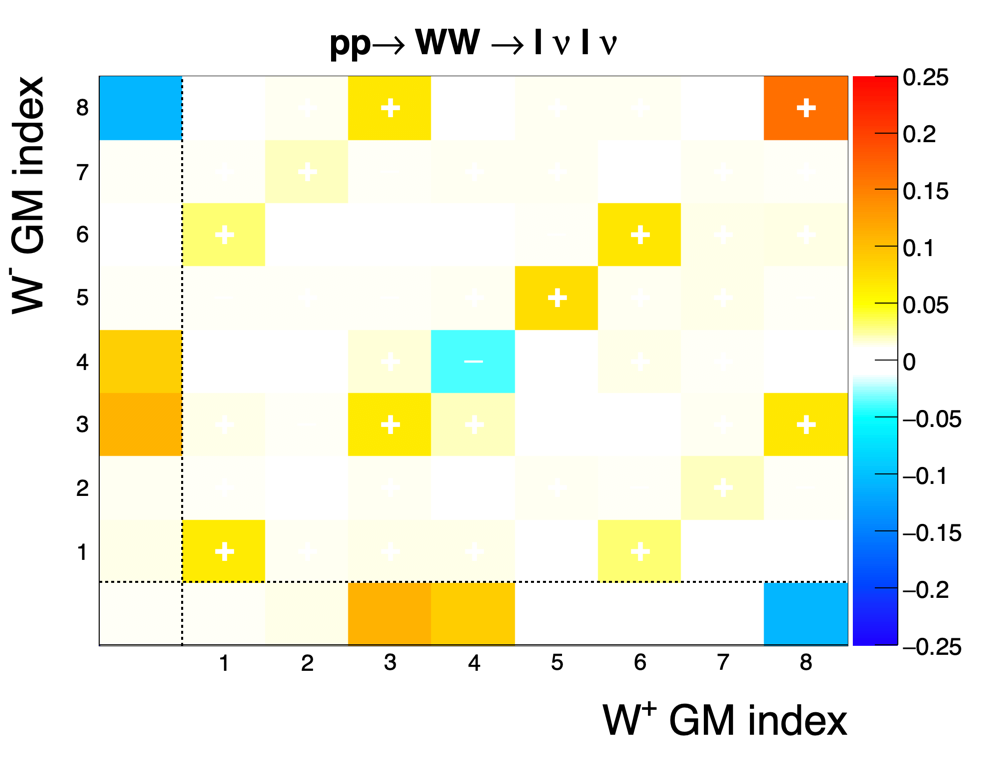} 
\vskip-4mm
\caption{\label{subfig:WW_GM_square}}
\end{subfigure}
\begin{subfigure}[b]{0.48\textwidth}\centering
\includegraphics[width=\linewidth]{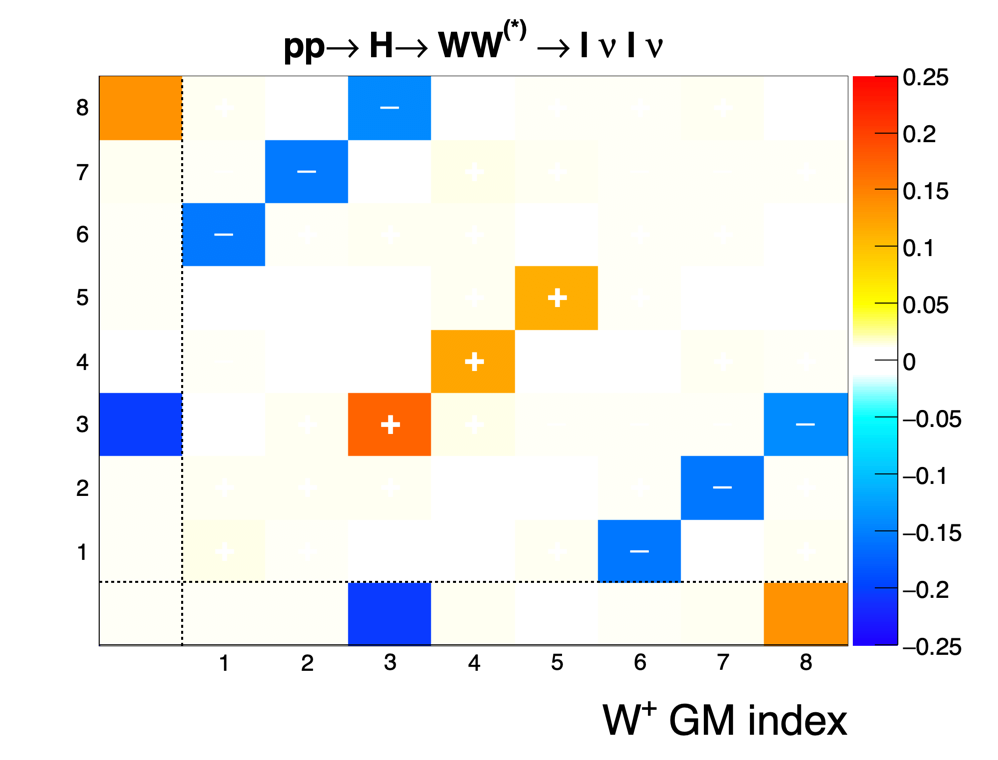} 
\vskip-4mm
\caption{\label{subfig:Hww_GM_square}}
\end{subfigure}\\
\vskip3mm
\begin{subfigure}[b]{0.48\textwidth}\centering
\includegraphics[width=\linewidth]{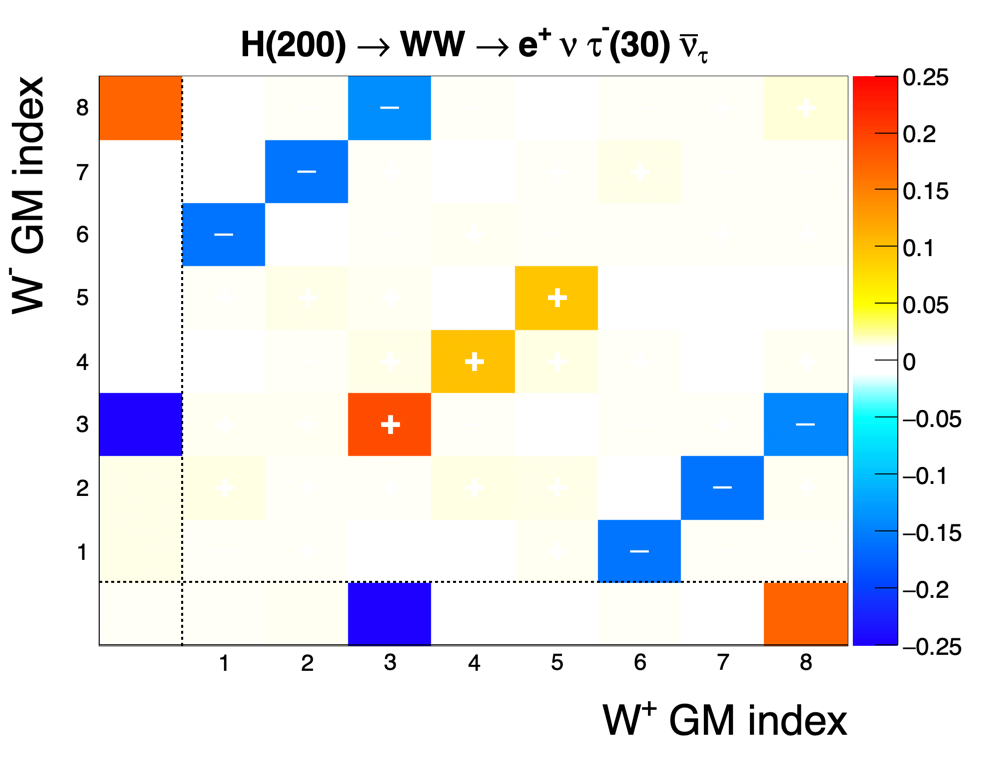} 
\vskip-4mm
\caption{\label{subfig:h200_ww_tau30_GM_square}}
\end{subfigure}
\begin{subfigure}[b]{0.48\textwidth}\centering
\includegraphics[width=\linewidth]{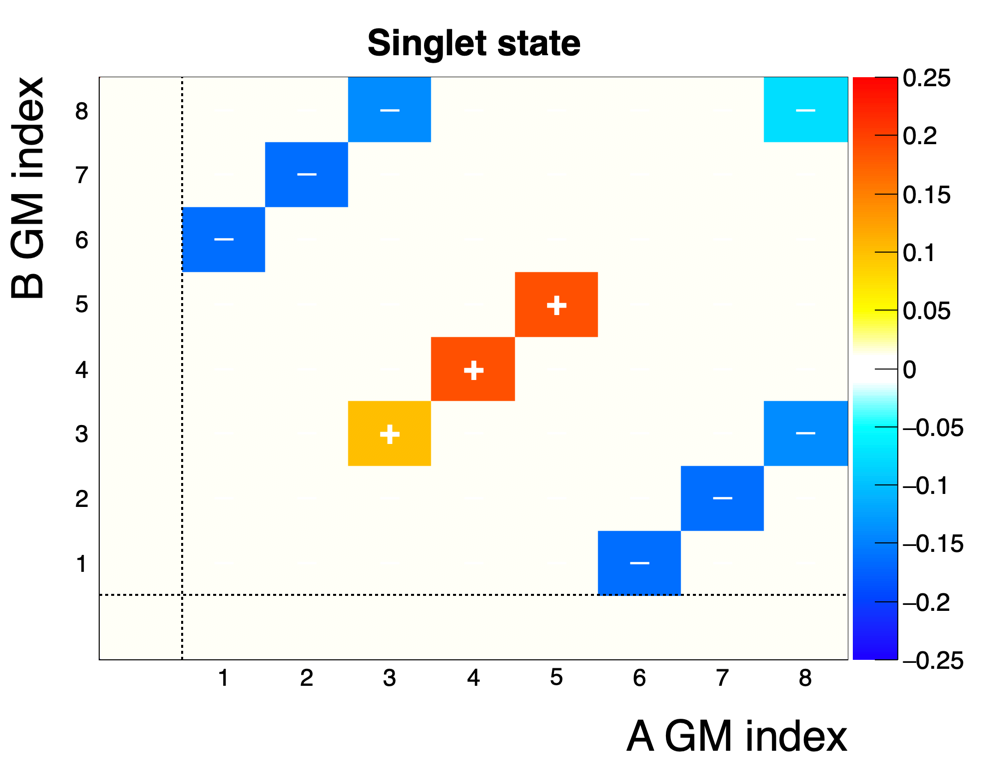}
\vskip-4mm
\caption{\label{subfig:BellState_GM_square}}
\end{subfigure}
\end{center}
\caption{\label{fig:ww_tomography_examples} 
Reconstructed Gell-Mann parameters obtained from quantum state tomography of pairs of simulated $W^\pm$ bosons obtained from (a) $W^+W^-\rightarrow \ell^+\nu \ell^- \bar\nu$ (b) $H\rightarrow WW^{(*)} \rightarrow \ell^+\nu \ell^- \bar\nu$, (c) and a 200 GeV scalar decaying to $WW$ then to a state with a 30\,GeV `$\tau$' lepton. 
We also show (d)  the results expected for an ideal singlet state \eqref{eq:singletstate} of two qutrits. 
The bottom row of each plot contains the $a_i$ parameters for a $W^+$ boson, the leftmost column the $b_j$ parameters for the $W^-$ boson and the rows and columns 1-8 the $c_{ij}$ parameters.
The (0,0) element has no meaning. 
}
\end{figure}
The results of the quantum state tomography for pairs of simulated $W^+W^-$ bosons can be found in Figure~\ref{fig:ww_tomography_examples}.
In each of the two  reconstructed samples one can observe non-zero GGM parameters corresponding to non-isotropy of, and correlations between, the spins of the two $W$ bosons. 
For comparison, Figure~\protect\ref{subfig:BellState_GM_square} shows the expected tomographic outcome for the density matrix
\begin{equation}\label{eq:singletstate}
\rho_s  = \ket{\psi_{s}}\bra{\psi_s}
\end{equation}
of a non-relativistic, narrow-width pure singlet scalar state
\begin{equation}\label{eq:singlet}
\ket{\psi_{s}} = \tfrac{1}{\sqrt{3}} \big( \ket{+}\ket{-} - \ket{0} \ket{0} + \ket{-}\ket{+} \big).
\end{equation}
This state is isotropic on measurements of each individual system but has maximal correlations between the two systems.
\par
It can be seen that as expected the parameters for the (scalar) Higgs boson decay process $gg \rightarrow H \rightarrow \ell^+ \nu_\ell \,\ell^-{\bar{\nu}}_\ell$ 
share many of the properties of the spin singlet state, a state to which it would reduce in the non-relativistic and narrow-width approximation. A difference is observed particularly in the $a_3$ and $a_8$ parameters which are related to
the longitudinal polarisation of the vector boson.
The tomographic method also performs as expected for the state with the heavy `$\tau$' lepton, provided that the generalised Wigner $P$ symbols~\eqref{eq:wignerPnonprojectiveW} for the non-projective decay are used, normalised such that only $\j=1$ states are reconstructed. 

With knowledge of the full joint density matrix one can then consider other operators. For example one may reconstruct the expectation value of various Cartesian spin operators of the individual bosons and of their correlated spin expectation values. 
For spin-half systems the spin operators are proportional to the GGM operators so their expectation values such as
\begin{eqnarray}
\langle S_i \rangle &=& \half \tr(\, \rho \, \sigma_i\, ) \nonumber\\
                    &=& a_i
\end{eqnarray}
can be read off directly from the GGM parameters.
For spin-one systems the complete set of relevant spin operators are defined from the $\j=3$ spin operators $S_i$ extended by the symmetric quadratic products, 
\[S_{\{ij\}} \equiv S_iS_j+S_jS_i.\]
The various spin operators are related to the $d=3$ GGM operators and parameters as described in  Appendix~\ref{sec:spinmatrices}.

\begin{figure}
\begin{center}
\begin{subfigure}[b]{0.48\textwidth}\centering
\includegraphics[width=\linewidth]{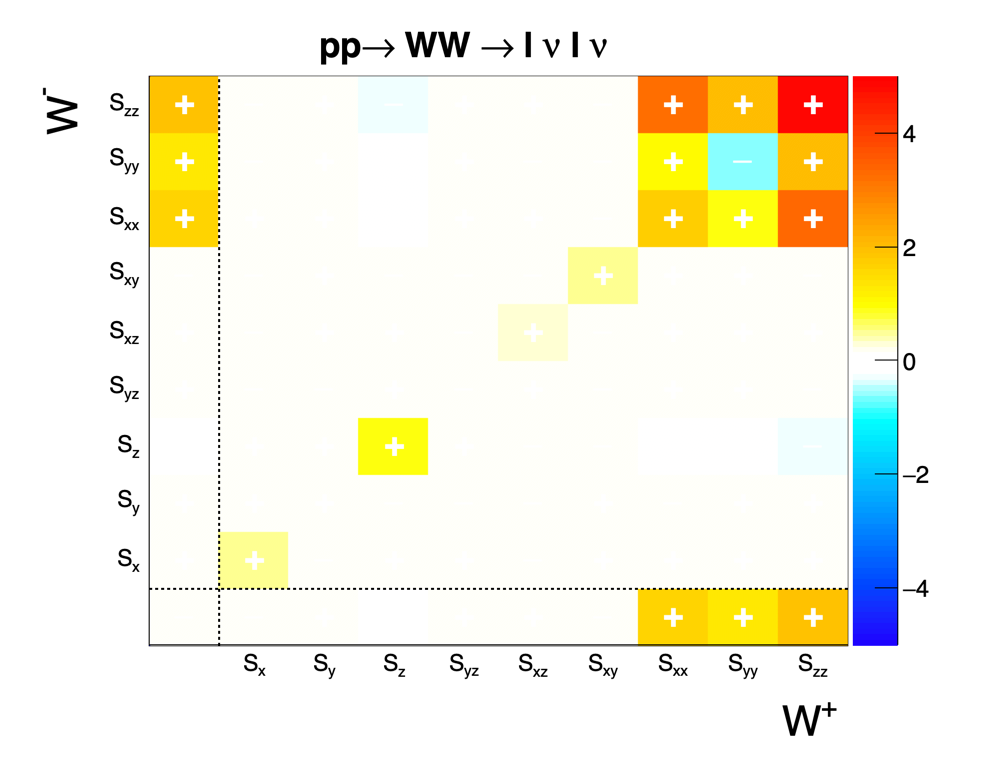}
\vskip-4mm
\caption{\label{subfig:WW_SP_square}} 
\end{subfigure}
\begin{subfigure}[b]{0.48\textwidth}\centering
\includegraphics[width=\linewidth]{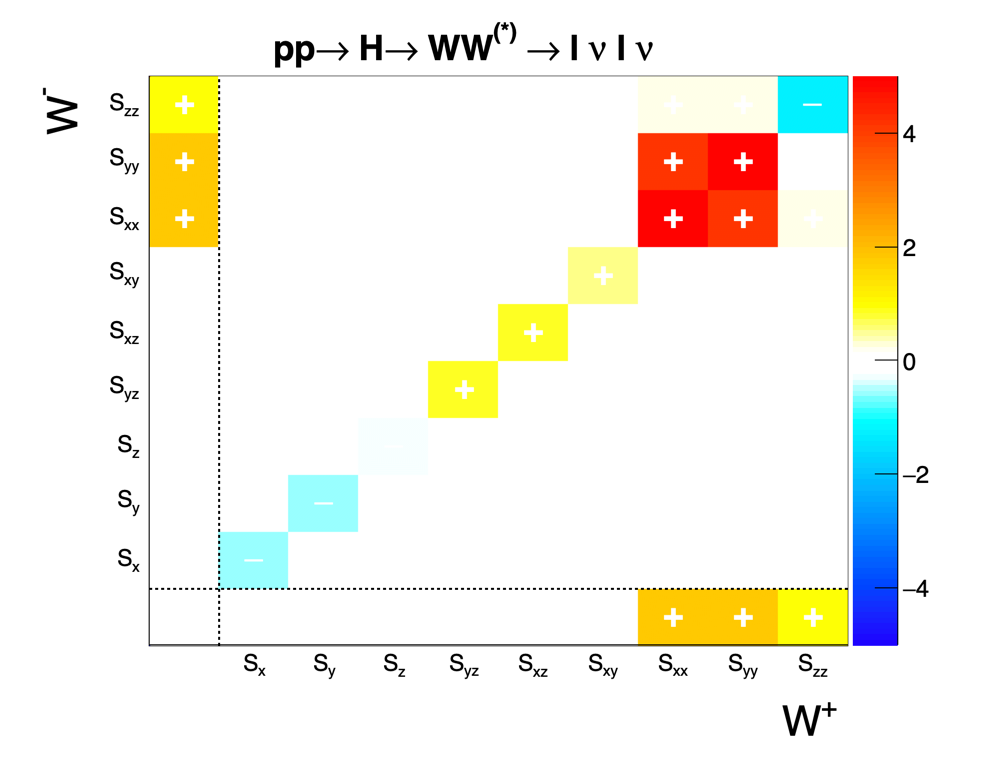} 
\vskip-4mm
\caption{\label{subfig:Hww_SP_square}}
\end{subfigure}\\
\vskip4mm
\begin{subfigure}[b]{0.48\textwidth}\centering
\includegraphics[width=\linewidth]{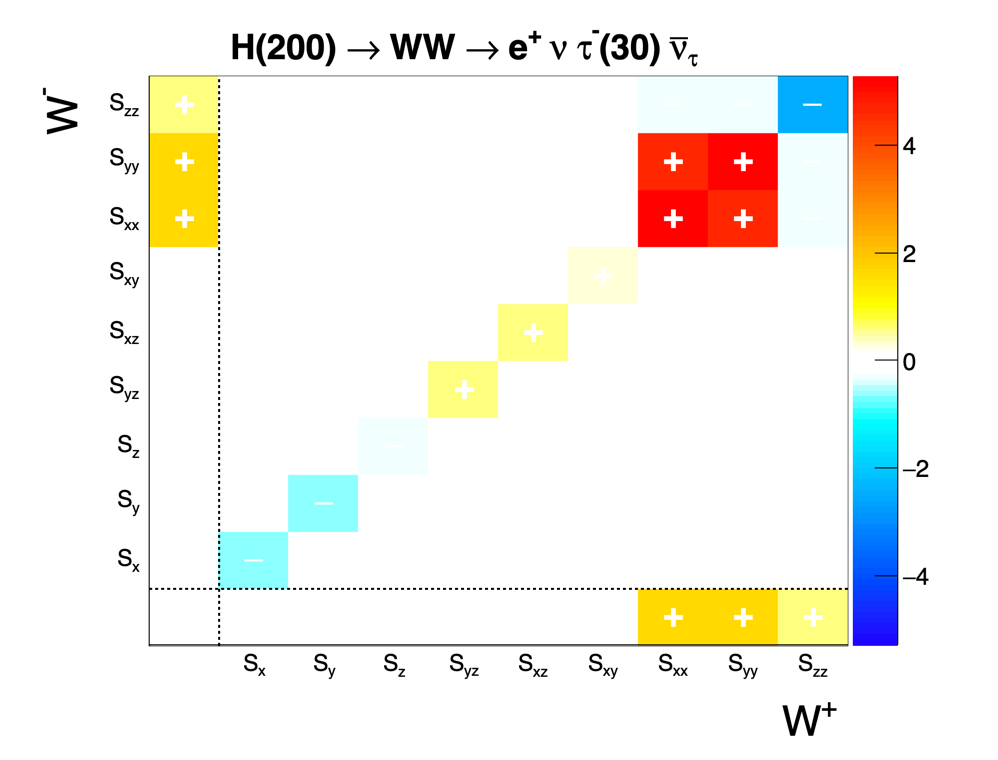} 
\vskip-4mm
\caption{\label{subfig:h200_ww_tau30_SP_square}}
\end{subfigure}
\begin{subfigure}[b]{0.48\textwidth}\centering
\includegraphics[width=\linewidth]{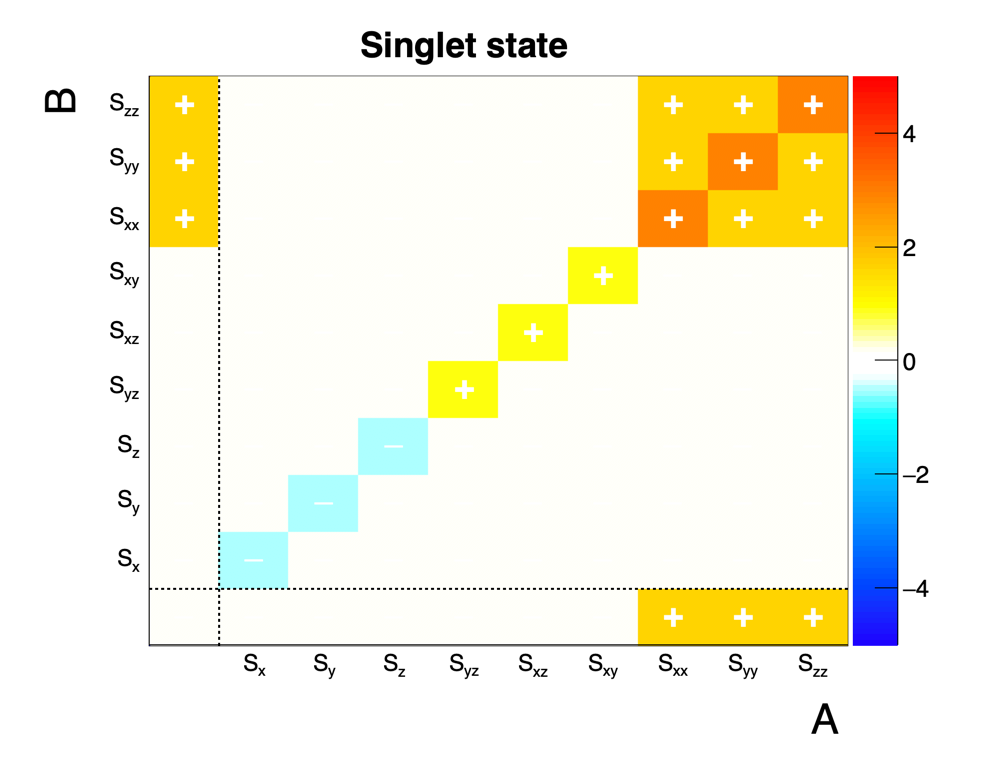} 
\vskip-4mm
\caption{}
\end{subfigure}
\end{center}
\caption{\label{fig:ww_tomography_examples_spin} 
Reconstructed expectation values of the complete set of $\j=1$ spin operators for pairs of $W^\pm$ bosons obtained from (a) $pp\rightarrow W^+W^-\rightarrow  \ell^+\nu \ell^- \bar\nu$ 
(b) $pp\rightarrow H\rightarrow WW^{(*)} \rightarrow \ell^+\nu \ell^- \bar\nu$, 
(c) a 200 GeV Higgs boson decaying to $WW$ involving a 30 GeV `$\tau$' lepton, and (d) an ideal singlet state \eqref{eq:singletstate} of two qutrits. 
In (a)--(c) the bottom row contains the operators for the $W^+$ boson, the leftmost column the $b_j$ parameters for the $W^-$ boson and others rows and columns the product operators for those bosons. The bottom-left-hand bin of each plot has no meaning.
}\end{figure}

The resulting spin-operator expectation values (after tomographic reconstruction of our Monte Carlo simulations in the Gell-Mann basis)
are shown for the two tested $W^+W^-$ physics processes in Figure~\ref{fig:ww_tomography_examples_spin}, 
along with the corresponding values for the spin-singlet state \eqref{eq:singletstate}. 
Again, one observes that the tomographically reconstructed states for $gg \rightarrow H \rightarrow \ell^+ \nu_\ell \,\ell^-{\bar{\nu}}_\ell$ events
have the expected correspondence with the idealised spin singlet state. 
However, unlike the case for the spin singlet, the expectation values both for the single $W$ boson and for the $WW$ system,
have different values in the $W$-boson momentum direction, $z$, compared to the transverse ($x,y$) directions. 

\begin{figure}
\begin{center}
\includegraphics[width=0.48\linewidth]{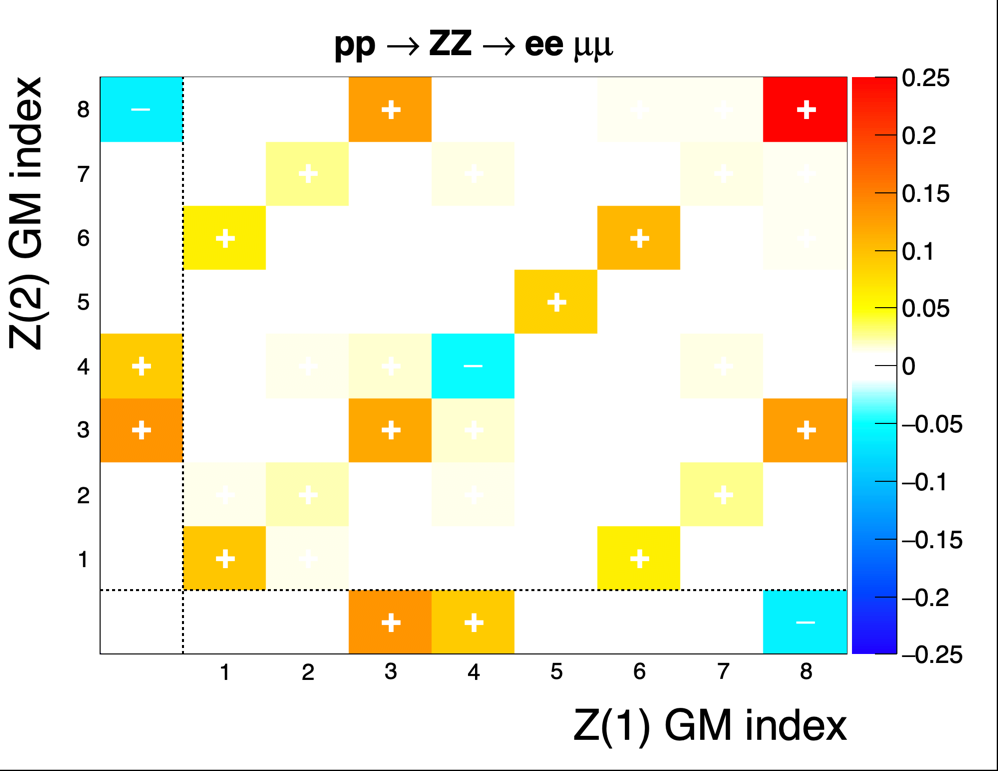} 
\includegraphics[width=0.48\linewidth]{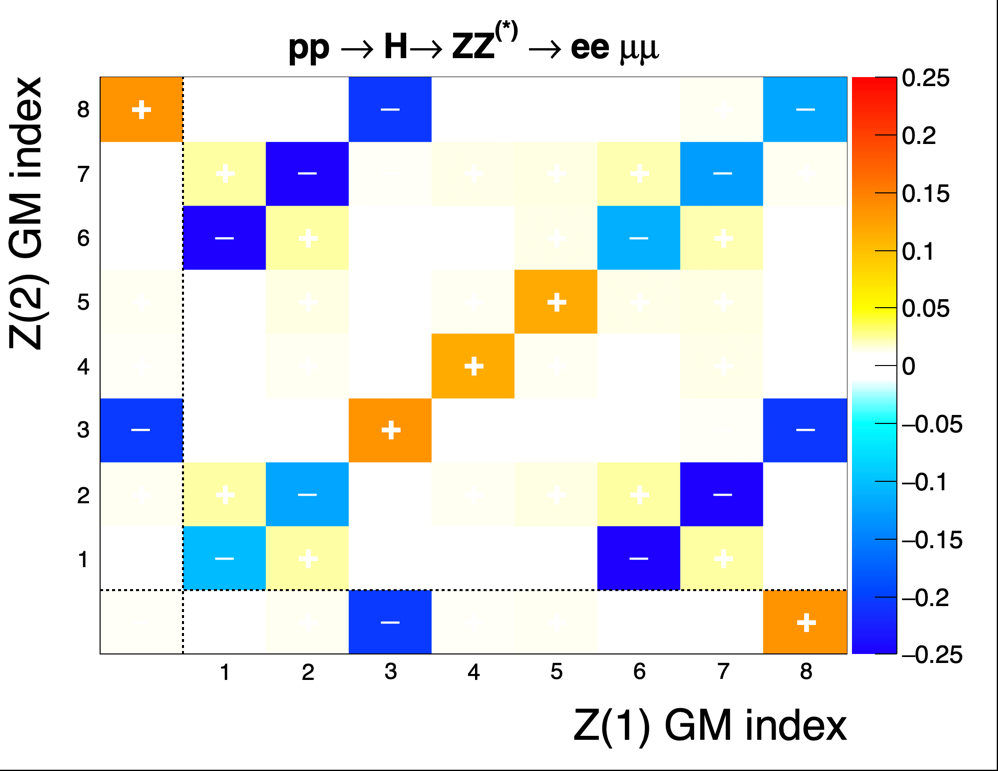} 
\includegraphics[width=0.48\linewidth]{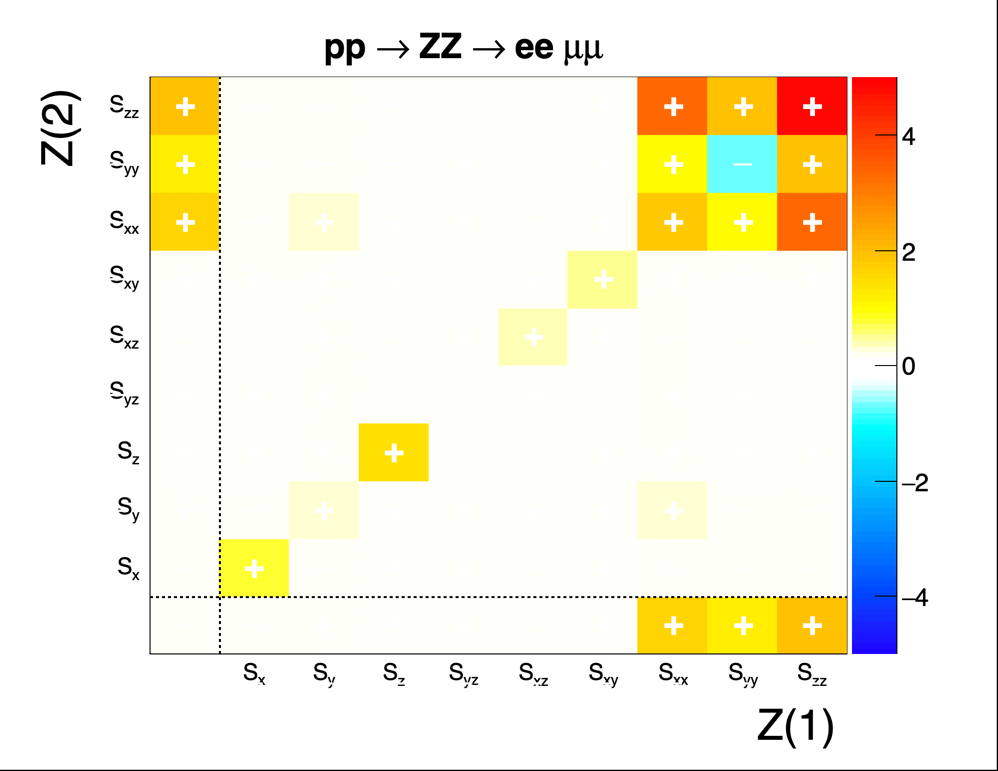} 
\includegraphics[width=0.48\linewidth]{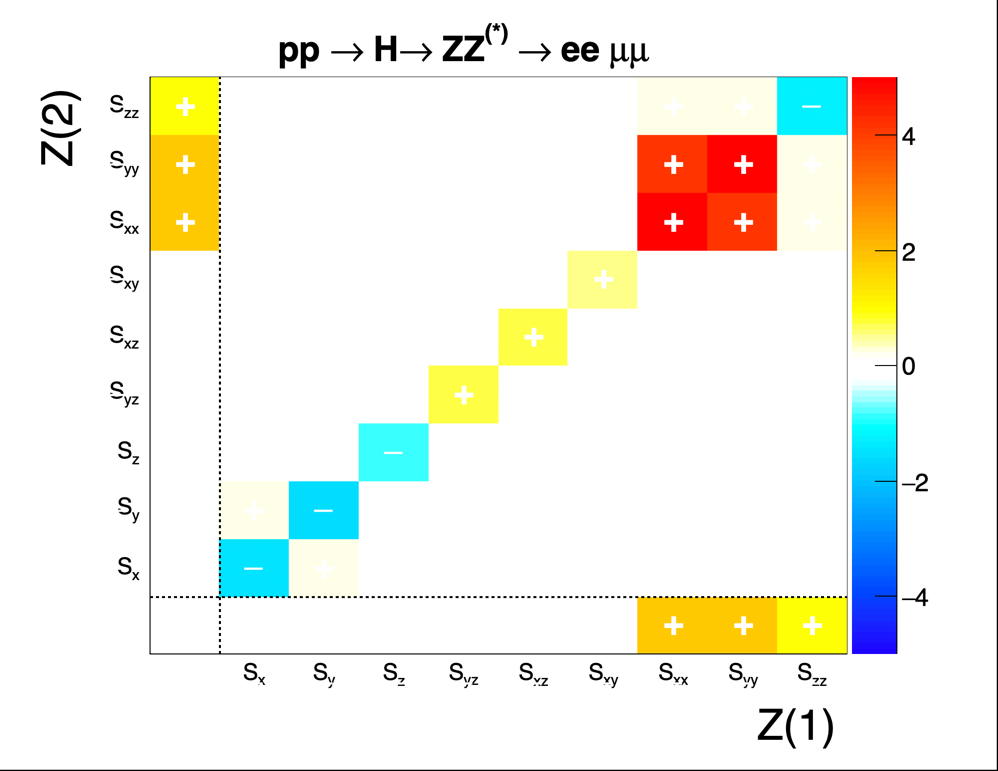} 
\end{center}
\caption{\label{fig:zz_tomography_examples} 
Reconstructed Gell-Mann parameters (top) and spin expectation values (bottom) obtained from quantum state tomography of pairs of $Z$ bosons obtained from (left) $pp \rightarrow ZZ\rightarrow e^+e^-\mu^+\mu^-$ and (right) $pp\rightarrow H\rightarrow ZZ^{(*)} \rightarrow e^+e^-\mu^+\mu^-$ simulations in Madgraph. The plots are symmetric under the interchange of labels $1\leftrightarrow 2$ by exchange symmetry.
}
\end{figure}

Figure~\ref{fig:zz_tomography_examples} shows the results of performing tomography on pairs of simulated $Z$ bosons using \eqref{eq:measureaidentical} and \eqref{eq:measurecidentical}.
The density matrix has been constructed to be symmetrical under interchange of their labels. 
The GM parameters and the expectation values of the various Cartesian tensor operators are shown. The high degree of correlation between the GM parameters is again visible.
The reconstructed GM parameters and spin expectation values have a similar but not identical structure to that for the $WW$ case, as expected for decays with the same spin structure but different masses.

\subsection{\boldmath Entanglement measures in diboson systems}\label{sec:entanglement}

Having found a method to obtain analytically the full density matrix we are in a position to address other questions about the reconstructed quantum state. 
One of the most interesting is whether a multipartite state is separable or entangled.  

\begin{table}
\centering
\begin{tabular}{l c c}
\hline\hline
Process              & & $\mintertboundsq$  \\
\hline\hline
\multicolumn{2}{l}
{$pp \rightarrow W^+W^-\rightarrow \ell^+\nu \ell^- \bar\nu$} 
 & -0.147 \\ 
\multicolumn{2}{l}{$H(125) \rightarrow WW^{(*)} \rightarrow \ell^+ \nu_\ell \,\ell^-{\bar{\nu}}_\ell$}   & 0.973  \\ 
\multicolumn{2}{l}{$H(200) \rightarrow WW \rightarrow \ell^+ \nu_\ell \,\tau^-(30){\bar{\nu}}_\tau$}   &  0.946  \\ 
\multicolumn{2}{l}
{$pp \rightarrow ZZ\rightarrow e^+e^-\mu^+\mu^-$} 
 & 0.50 \\ 
\multicolumn{2}{l}{$H(125) \rightarrow ZZ^{(*)}\rightarrow e^+e^-\mu^+\mu^-$}  & $3.51^\dagger$  \\ 
\multicolumn{2}{l}{$pp \rightarrow W^+Z\rightarrow e^+\nu_e\mu^+\mu^-$} 
& 0.10 \\ 
\hline\hline
$\ket{\psi_s}\bra{\psi_s}$ & \eqref{eq:singlet} & $\tfrac{4}{3}$ \\
$\ket{\Phi^+_{\rm Bell2}}\bra{\Phi^+_{\rm Bell2}}$ & \eqref{eq:effectivebellqubit} & 1 \\
$\ket{\psi_{\rm sep}}\bra{\psi_{\rm sep}}$ & \eqref{eq:separableproduct} & 0\\
$\rho_{{\rm I}_9}$ & \eqref{eq:maxmixed} & $-\tfrac{4}{9}$ \\
\hline\hline 
\end{tabular}
\caption{\label{tab:concurrence}
Lower bound $\mintertboundsq$ \eqref{eq:concurrencebound} on the square of the concurrence for the spin density matrices obtained by quantum state tomography from simulated diboson Monte Carlo event samples. For entanglement to be detected it is sufficient that $\mintertboundsq>0$. For comparison the values for particular example states are also given. Values indicated with a dagger $\dagger$ show unphysically large values ($>\frac{4}{3}$) for the reasons discussed in the text.}
\end{table}

To test entanglement in our simulations of bipartite systems the bound \eqref{eq:mintertinGM} on the square of the concurrence was calculated. The results may be found in Table~\ref{tab:concurrence}.
It is observed that inclusive $pp\rightarrow W^+W^-$ production (i.e. with no selection requirements) yield a negative $\mintertboundsq$, so entanglement can not be inferred for this ensembles using this method.\footnote{Numerical approximations to the  concurrence~\cite{agi} do however suggest that these states still may be entangled but not in a manner that is not detectable using $\mintertboundsq$.} 
By contrast, for the decays of Higgs bosons to $WW^{(*)}$ pairs the $\mintertboundsq$ bounds both exceed zero, meaning that entanglement of the vector bosons pairs could indeed be inferred in each case. 
\par
Inclusive $pp\rightarrow ZZ$ production yields a positive $\mintertboundsq$, suggesting that entanglement should be able to be determined. However care is required in this interpretation, as can be seen from the unphysically large value obtained for $\mintertboundsq$ for the case of Higgs boson decays to $ZZ^*$. The reason for the unphysical bound is that the reconstructed density matrix for the $ZZ^*$ system contains unphysical eigenvalues, ranging from -0.27 to 1.36.
The source of the unphysical values is expected to be that described (after the initial submission of this article) in Refs.~\cite{Aguilar-Saavedra:2024jkj,DelGratta:2025qyp,Goncalves:2025qem,DelGratta:2025xjp} where effects of higher-order perturbative corrections, sensitivity to the electroweak mixing angle, sensitivity to the kinematical selection and identical-particle symmetry are analyzed.
\par
For comparison the results for various idealised pure and mixed 
qutrit pair states are also calculated. Results are tabulated for two highly entangled states, the singlet state \eqref{eq:singletstate}, and the qubit-pair-like state:
\begin{equation}\label{eq:effectivebellqubit}
\ket{\Phi^+_{\rm Bell2}} = \tfrac{1}{\sqrt{2}}\big( \ket{-}\ket{-} + \ket{+}\ket{+}  \big)\,, 
\end{equation}
as are those for two states known not to be entangled: a separable state
\begin{equation}\label{eq:separableproduct}
\ket{\psi_{\rm sep}} = \ket{+}\ket{+},
\end{equation}
and the maximally mixed state in a bipartite system of two qutrits
\begin{equation}\label{eq:maxmixed}
\rho_{{\rm I}_9} = \inine.
\end{equation}
Comparing the magnitudes of the values we observe that the reconstructed bounds $\mintertboundsq$ for the Higgs decays to $WW^*$ are of the same order as the values expected for maximally-entangled qubit or qutrit states (1 and $\tfrac{4}{3}$ respectively). 

\begin{figure}
    \centering
\begin{subfigure}[b]{0.7\textwidth}\centering
\includegraphics[width=\linewidth]{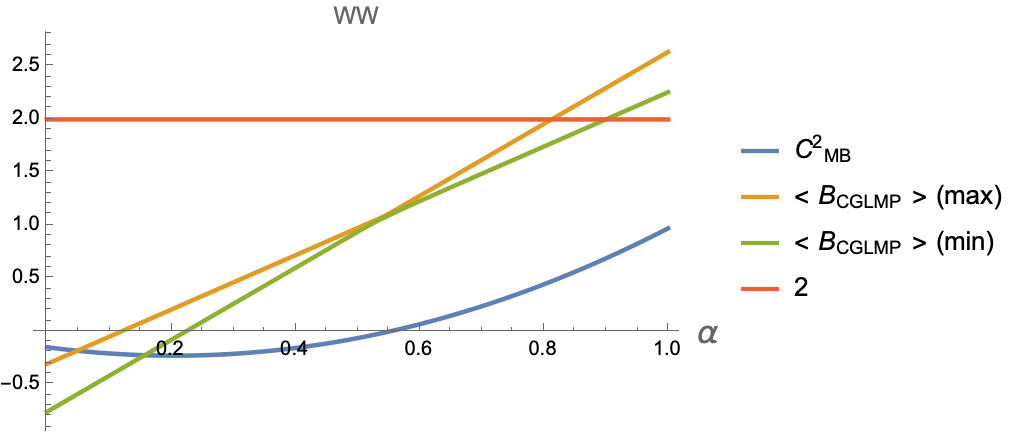}
\vskip-2mm
\caption{\label{subfig:mixalphaplotWW}$\rhomixWW$}
\end{subfigure}
\begin{subfigure}[b]{0.7\textwidth}\centering
\includegraphics[width=\linewidth]{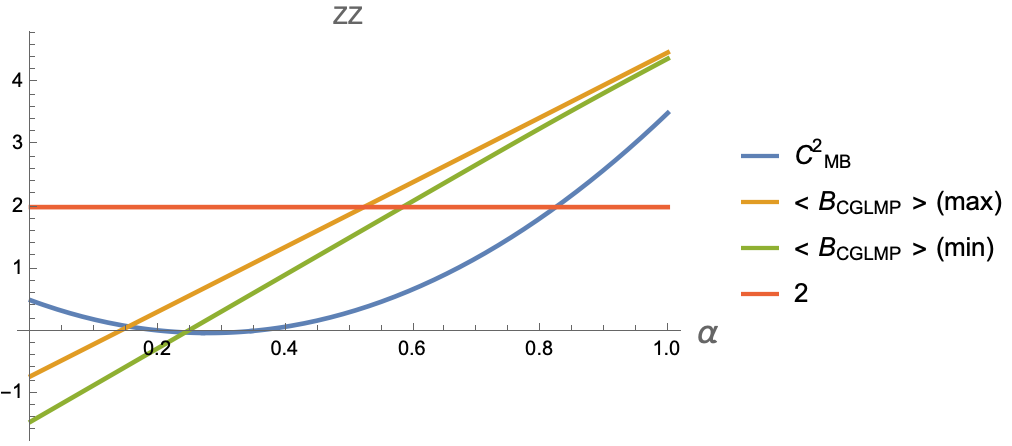}
\vskip-2mm
\caption{\label{subfig:mixalphaplotZZ}$\rhomixZZ$}
\end{subfigure}
\caption{Bell violation and entanglement detection in Higgs boson decays to vector bosons. Results are shown for the `signal+background' states $\rhomixWW$ \eqref{eq:rho_HWW_WW} 
and $\rhomixZZ$ \eqref{eq:rho_HZZ_ZZ} 
as a function of the signal fraction $\alpha$. 
Where the bound \eqref{eq:concurrencebound} on the square of the concurrence exceeds zero the simulated state is entangled. Where it is less than zero the test is inconclusive.
For the Bell inequality test the simulated state violates the qutrit Bell inequality \eqref{eq:cglmpbound} if either expectation value of the CGLMP operator \eqref{eq:cglmpmax} exceeds two.
}
    \label{fig:mixalpha}
\end{figure}

To investigate the effect of background processes mixed signal and background state were constructed from the linear combinations
\begin{equation}
\rhomixWW = \alpha \, \rho_\mathrm{H\rightarrow WW^{*}} + (1-\alpha)\, \rho_\mathrm{WW} \label{eq:rho_HWW_WW}
\end{equation}
that would be obtained from a fraction $\alpha$ of $H\rightarrow WW^*$ signal events and fraction $(1-\alpha)$ of $W^+W^-$ continuum background events, assuming no selection bias. 
A corresponding density matrix
\begin{equation}
\rhomixZZ = \alpha\, \rho_\mathrm{H\rightarrow ZZ^{*}} + (1-\alpha)\, \rho_\mathrm{ZZ} \label{eq:rho_HZZ_ZZ}
\end{equation}
is defined from the states obtained from the $H\rightarrow{ZZ^{(*)}}$ and $pp\rightarrow ZZ$ processes.
\par
As we vary the signal fraction for $H\rightarrow WW^{(*)}$ we find that the bound \eqref{eq:concurrencebound} on the square of the concurrence is greater than zero for values of $\alpha\gtrsim 0.55$, as shown in Figure~\ref{fig:mixalpha}. 
Hence one could determine using this method that the state is non-separable over the range of $0.55 \lesssim \alpha\leq 1$. 
\par
For the $ZZ$ state $\rhomixZZ$ the corresponding range is larger.
However, both the the bound on the concurrence and the Bell expectation values yield unphysically high values (above $\tfrac{4}{3}$ and $1+\sqrt{11/3}$ respectively) as $\alpha$ approaches unity, indicating a need to account for the effects described Refs.~\cite{Aguilar-Saavedra:2024jkj,DelGratta:2025qyp,Goncalves:2025qem,DelGratta:2025xjp}.
Nevertheless the $ZZ$ case might be  more experimentally accessible since, unlike the case of the leptonic $W$ boson decays, there are no neutrinos in the final state so the signal may be cleanly identified from events in which the four-lepton invariant mass is close to $m_H$.

\begin{figure}
    \centering
    \begin{subfigure}[b]{0.48\textwidth}\centering
         \includegraphics[width=\linewidth]{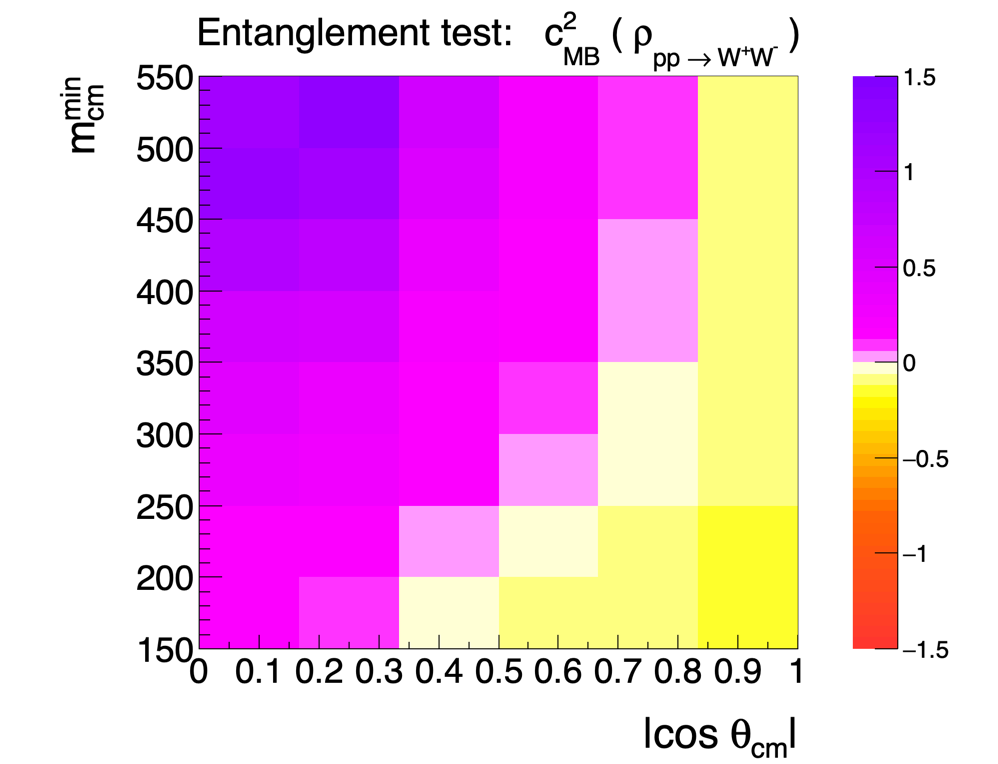} 
        \caption{$pp\rightarrow W^+W^-$}
     \end{subfigure}
    \begin{subfigure}[b]{0.48\textwidth}\centering
    \includegraphics[width=\linewidth]{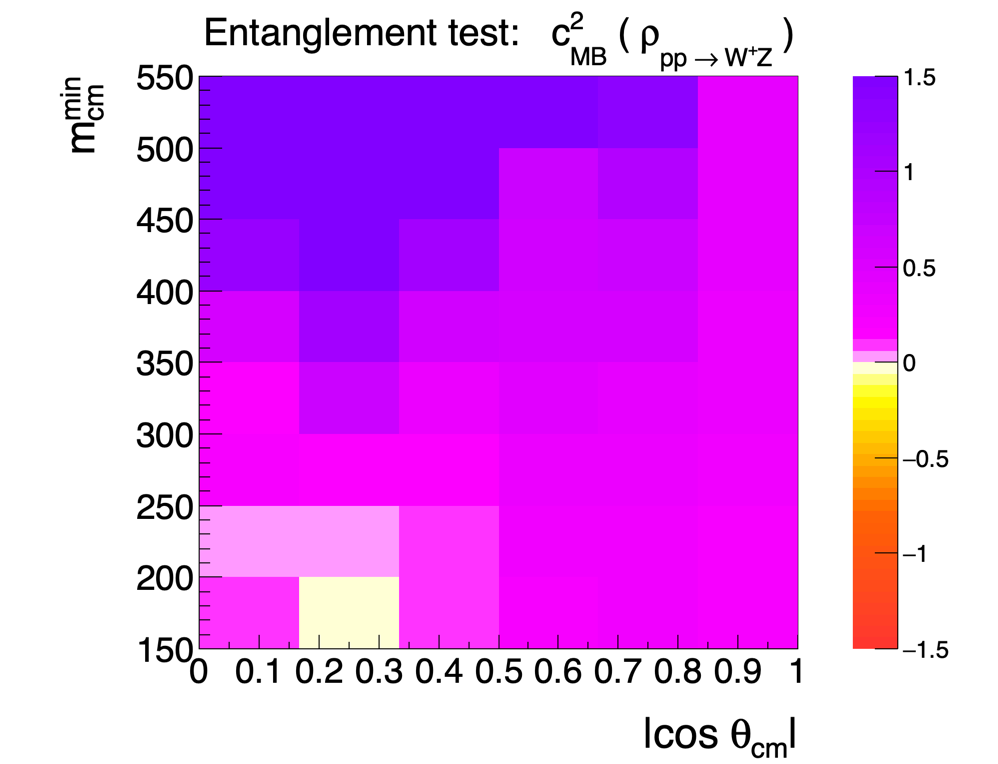} 
        \caption{$pp\rightarrow W^+Z$}    
    \end{subfigure}\\
    \begin{subfigure}[b]{0.48\textwidth}\centering
    \includegraphics[width=\linewidth,trim=0 6 6 0, clip]{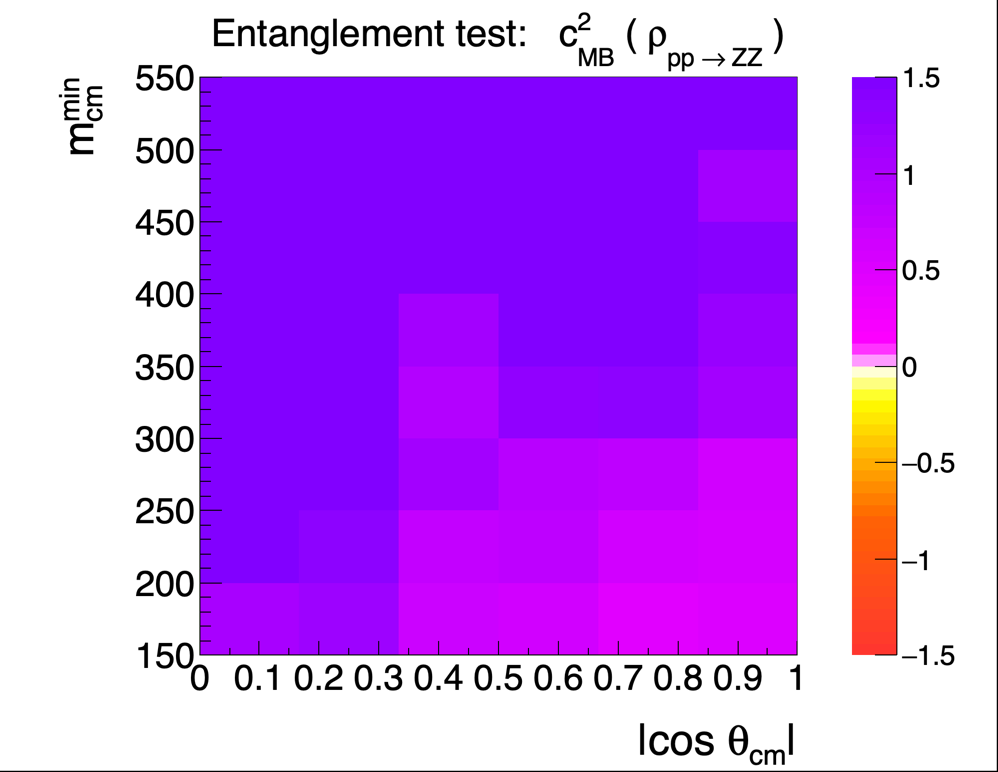} 
        \caption{$pp\rightarrow ZZ$}
    \end{subfigure}
\caption{
Entanglement detection in diboson samples. The color scale shows $\mintertboundsq(\rho)$, the bound on the square of the concurrence of the bipartite density matrix for diboson production $pp\rightarrow VV$. 
The density matrices have been obtained from tomography of the simulated data.
Each bin is plotted deferentially in $|\cos\theta_{\rm cm}|$ (horizontal axis) and with a lower bound $m_{CM}^{\min}$ on the invariant mass of the diboson pair given by the value on the vertical axis.
Values of $\mintertboundsq(\rho)>0$ (pink/purple bins) indicate non-separable, entangled, states. Values larger than that obtained for a singlet state ($4/3$) are due to statistical fluctuations.
}
\label{fig:c_mb_dibosons}
\end{figure}

When selection cuts are applied to the diboson processes the contributions of different Feynman diagrams with different spin structures differ, and so the density matrix varies. 
To investigate the effect of these effects we calculate $\mintertboundsq$ separately 
for different ensembles of $W^+W^-$, $W^+Z$ and $ZZ$ events, each defined by a coarse binning on the angle $\theta_{\rm cm}$ of one of the parent bosons relative to the beam in the CM frame, and a threshold requirement $m_{\rm cm}^{\min}$ on the invariant mass of the diboson pair. 

We observe in these simulated diboson systems  that the bound $\mintertboundsq$ on the square of the concurrence typically increases with increasing diboson invariant mass (Figure~\ref{fig:c_mb_dibosons}). The bound is typically larger than zero, indicating an entangled systems, for invariant mass larger than about a few hundred GeV, and has an angular dependence which depends on which bosons are considered. 
These results suggest that entanglement might be measurable in these systems provided that sufficiently large ensembles can be collected with the appropriate kinematic selections. 

\subsection{Bell inequality tests in diboson systems}

We next explore Bell inequality violation using the simulated bipartite vector boson systems $W^+W^-$, $ZZ$ and $W^+Z$. 
After performing quantum state tomography we calculate the expectation value of the CGLMP Bell operator \eqref{eq:cglmpoperator}. 
\par
Our background-free simulations of Higgs boson decays to $WW^*$ have expectation values which violate the maximum value (two) allowed in local-realist theories. Unphysically large values for $H\rightarrow ZZ^*$ are observed for reasons previously indicated.
These results confirm and extend previous findings for projective $H\rightarrow WW^{(*)}$ decays~\cite{BARR2022136866} and suggest that, if higher-order corrections can be managed, they might be extended to $H\rightarrow ZZ^{(*)}\rightarrow 4\ell$ decays which could be easier to investigate experimentally due to the absence of invisible neutrinos in the final state. 

The effect of generic diboson backgrounds on the Bell operator expectation value was checked by considering the mixed Higgs signal+background states \eqref{eq:rho_HWW_WW} and \eqref{eq:rho_HZZ_ZZ}. 
The results are plotted in Figure~\ref{fig:mixalpha}. 
The maximum expectation values (over all test planes) of the tested CGLMP operator exceeds the local realist limit of two for signal fraction $\alpha\gtrsim 0.81$ for $WW^{(*)}$ 
and $\alpha\gtrsim 0.5$ for $ZZ^{(*)}$. 
Thus observing Bell violation in Higgs boson decays is possible, but the caveats are that it making the measurement is likely to require samples with similarly large signal purity, at least using the operator, and that the higher-order effects in EW perturbation theory described in Refs~\cite{Aguilar-Saavedra:2024jkj,DelGratta:2025qyp,Goncalves:2025qem,DelGratta:2025xjp} will need to be considered. 


\par
For the continuum $pp\rightarrow W^+W^-$, $pp\rightarrow W^+Z$, and $pp\rightarrow ZZ$ simulations we also calculate the CGLMP Bell operator expectation values for the same kinematic selections on $m_{cm}^{\min}$ and $\cos\theta_{\rm cm}$ as were used for entanglement detection.
The expectation values for operators defined in the $xy$, $xz$ and $yz$ planes are calculated according to the axes defined in Figure~\ref{fig:axes} and the largest of the three determined for each ensemble of events. 

The $WW$ and $WZ$ samples do not show Bell violation for our (scalar-optimised) CGLMP operator \eqref{eq:cglmpoperator}, nor for its $xz$- or $yz$-plane equivalents.\footnote{An apparent violation reported in $WZ$ events in an earlier version of this report was found to be due to selection requirements made by the Monte Carlo calculation, and should be disregarded.} 
The $ZZ$ diboson continuum sample show a region at large diboson invariant mass and small $\cos\theta_{\rm CM}$ in which the $xz$- and $yz$-plane versions of the operator violates the local-realist limits in high-statistics samples. If higher-order corrections can be managed then CGLMP inequlity violation might be measurable in this region. 
None of this excludes the possibility of observable Bell violation for other Bell operators optimised for this particular process.\footnote{During the review of this paper analytical calculations were performed by others for the $ZZ$ process and tested with sets of Bell operators following various unitary transformations $(U\otimes V)^\dag\cdot\mathcal{B}_{\rm CGLMP}\cdot (U\otimes V)$ where $U$ and $V$ are independent three-dimensional unitary matrices. Those results show violation of these optimised Bell operators in both $ZZ$ and $W^+W^-$ continuum diboson events~\cite{Fabbrichesi:2023cev}.}

\section{Conclusions}
We have outlined a broad programme for performing experimental tests of the foundations of quantum theory using weak vector bosons at colliders. 

The enabling technology developed for this purpose is a rather general method of quantum state tomography.
We have described the simplifications that result from using a generalised Gell-Mann parameterisation of the density matrix, which allows its reconstruction for arbitrary numbers of parents of arbitrary spin, provided only that their decays are spin-dependent. 
The method links to quantum information theory by making use of Kraus and measurement operators and by exploiting the Wigner-Weyl formalism for spin. 
We have generalised the Wigner-Weyl formalism to deal with non-projective decays, and provided step-by-step examples for the most relevant spin-half and spin-one cases, including of the top quark, and the $W^\pm$ and $Z$ bosons.

Monte Carlo simulations were performed and analysed of various bipartite systems.  
These illustrate how this procedure of quantum state tomography can experimentally reconstruct the full bipartite spin-density matrix for Hilbert spaces of dimensions $2\times2$, $2\times 3$ and $3\times 3$.
\par
We have explored quantum separability/entanglement criteria for $W^+W^-$ $W^+Z$ and $ZZ$ pairs. 
The $W$ and $Z$ bosons resulting from the SM Higgs boson decay processes $H\rightarrow WW^{(*)}$ and $H\rightarrow ZZ^{(*)}$ are highly entangled, and could be measured to be so provided that sufficiently pure samples can be selected.
In simulations of $pp$ collisions we have explored general $W^+W^-$ $ZZ$ and $W^+Z$ production, all of which show potential for measurements of entanglement for particular kinematic selections. 

Bell inequality tests for gauge boson pairs were also investigated in simulations. The expectation value of the Bell operators for the decays $H\rightarrow WW^{(*)}$ and $H\rightarrow ZZ^{(*)}$ each violate the qutrit CGLMP Bell inequality provided that the signal-to-background ratios are sufficiently large. 


For any collider experiments the effects of statistical and systematic uncertainties, and higher-order corrections in perturbation theory, will need to be determined and accounted for before definitive statements can be made. 
The calculation of these uncertainties requires multiple numerical experiments which go beyond the scope of this work. However in general one can say that with increasing LHC luminosity these measurements should become increasingly accessible. The methods are also general and could later be used at future $e^+e^-$, $pp$ or $\mu^+\mu^-$ colliders. Applications may also be found in other spin-dependent decays, such as the weak decays of hadrons. 

\newpage\appendix


\renewcommand\arraystretch{1.1}
\section{\boldmath Generalised Gell-Mann matrix representations, density matrix representations, and Wigner $Q$ and $P$ symbols}\label{sec:ggm}

The $d^2-1$ generalised Gell-Mann matrices $\lambdad$ for a Hilbert space of dimension $d$ are representations of the generators of the group $SU(d)$. 
They are characterized by the commutator structure constants $f_{ijk}$ and anticommutator structure constants
$g_{ijk}$:
\begin{eqnarray}
[\lambdad_i,\,\lambdad_j] &=& 2\i f_{ijk} \lambdad_k \\
\{\lambdad_i,\,\lambdad_j\} &=& \frac{4}{d} \delta_{ij} I_d + 2g_{ijk}\lambdad_k.
\end{eqnarray}

The generators can be constructed systematically for any $d$; an orthogonal set is given by~\cite{kimura2003bloch}
\begin{equation}\label{eq:GGM_all}
\{\lambdad_i\}_{i=1}^{d^2-1} = \{\lambdad_{S,jk},\,\lambdad_{A,jk},\,\lambdad_{D,l} \}
\end{equation}
comprising a set of symmetric
\begin{equation}\label{eq:GGM_symmetric}
\lambdad_{S,jk} = \ket{j}\bra{k} + \ket{k}\bra{j},
\end{equation}
antisymmetric 
\begin{equation}\label{eq:GGM_antisymmetric}
\lambdad_{A,jk} = -\i(\ket{j}\bra{k} - \ket{k}\bra{j}),
\end{equation}
and diagonal 
\begin{equation}\label{eq:GGM_diagonal}
\lambdad_{D,l} = \sqrt{\frac{2}{l(l+1)}}
\left( \sum_{j=1}^l  \ket{j}\bra{j} - l\ket{l+1}\bra{l+1} \right)
\end{equation}
matrices where $1\leq j < k\leq d$, and $1\leq l\leq d-1$.

\subsection{\boldmath $\j=1/2$, $d=2$ (qubits)}

For $d=2$ the GGM matrices are the Pauli matrices  
\begin{align}\label{eq:Paulimatrices}
\sigma_1 = \lambda^{(2)}_{S,12} &= \left(
\begin{array}{cc}
 0 & 1 \\
 1 & 0 \\
\end{array}
\right), &
\sigma_2 = \lambda^{(2)}_{A,12} &= \left(
\begin{array}{cc}
 0 & -\i \\
 \i & 0 \\
\end{array}
\right), &
\sigma_3 &= \lambda^{(2)}_{D,1} = \left(
\begin{array}{cc}
 1 & 0 \\
 0 & -1 \\
\end{array}
\right),
\end{align}
where for each of the three matrices we label them above both by  the Pauli-matrix index (e.g. $\sigma_1$ and the GGM label e.g. $\lambda^{(2)}_{S,12}$.

The single-particle density matrix~\eqref{eq:rhosingleGM} written in terms of these three parameters for $d=2$ is 
\begin{eqnarray}\label{eq:rho2GM}
\rho^{(2)} &=& \half\itwo + \sum_{i=1}^3 a_i\sigma_i \nonumber\\
&=& 
\left(
\begin{array}{cc}
 \half + a_3 & a_1 - \i a_2 \\
 a_1 + \i a_2 & \half - a_3 \\
\end{array}
\right).
\end{eqnarray}
The Wigner $Q$ symbols for $d=2$ are given by \eqref{eq:fermionQ} with $\kappa=1$. The corresponding Wigner $P$ symbols are given by \eqref{eq:fermionP} with $\kappa=1$.

\subsection{\boldmath $\j=1$, $d=3$ (qutrits)}

For spin-1 particles the $d=3$ Gell-Mann matrices are
\begin{align}\label{eq:GMmatrices}
\lambdathree_1 = \lambda^{(3)}_{S,12} &= \left(
\begin{array}{ccc}
 0 & 1 & 0 \\
 1 & 0 & 0 \\
 0 & 0 & 0 \\
\end{array}
\right), &
\lambdathree_2 = \lambda^{(3)}_{A,12} &=\left(
\begin{array}{ccc}
 0 & -\i & 0 \\
 \i & 0 & 0 \\
 0 & 0 & 0 \\
\end{array}
\right), &
\lambdathree_3 = \lambda^{(3)}_{D,1} &=\left(
\begin{array}{ccc}
 1 & 0 & 0 \\
 0 & -1 & 0 \\
 0 & 0 & 0 \\
\end{array}
\right), \nonumber\\
\lambdathree_4 = \lambda^{(3)}_{S,13} &= \left(
\begin{array}{ccc}
 0 & 0 & 1 \\
 0 & 0 & 0 \\
 1 & 0 & 0 \\
\end{array}
\right), &
\lambdathree_5 = \lambda^{(3)}_{A,13}  &=\left(
\begin{array}{ccc}
 0 & 0 & -\i \\
 0 & 0 & 0 \\
 \i & 0 & 0 \\
\end{array}
\right), &
\lambdathree_6 = \lambda^{(3)}_{S,23}  &=\left(
\begin{array}{ccc}
 0 & 0 & 0 \\
 0 & 0 & 1 \\
 0 & 1 & 0 \\
\end{array}
\right), \nonumber\\ 
\lambdathree_7 = \lambda^{(3)}_{A,23} &=\left(
\begin{array}{ccc}
 0 & 0 & 0 \\
 0 & 0 & -\i \\
 0 & \i & 0 \\
\end{array}
\right), &
\lambdathree_8 = \lambda^{(3)}_{D,2} &=\frac{1}{\sqrt{3}}\left(
\begin{array}{ccc}
 1 & 0 & 0 \\
 0 & 1 & 0 \\
 0 & 0 & -2 \\
\end{array}
\right),
\end{align}
where for each $d=3$ matrix both the conventional Gell-Mann notation (e.g. $\lambda^{(3)}_1$) and the GGM label (e.g. $\lambda^{(3)}_{S,12}$) are given. 

The density matrix for $d=3$ is 
\begin{eqnarray} 
\rho^{(3)}
&=& \tfrac{1}{3} \ithree + \sum_{i=1}^8 a_i \lambdathree_i \nonumber\\
&=&\left(
\begin{array}{ccc}
\frac{1}{3} + {a_3}+\frac{1}{\sqrt{3}}{a_8} & {a_1}-\i {a_2} & {a_4}-\i {a_5} \\
 {a_1}+\i {a_2} & \frac{1}{3}-{a_3}+\frac{1}{\sqrt{3}}{a_8} & {a_6}-\i {a_7} \\
 {a_4}+\i {a_5} & {a_6}+\i {a_7} & \frac{1}{3}-\frac{2}{\sqrt{3}} {a_8}
\end{array}
\right).\label{eq:rhothree}
\end{eqnarray}

The $d=3$ Wigner $Q$ and $P$ symbols are the $\WignerQp$ and $\WignerPp$ given in equations \eqref{eq:WignerQGM} and \eqref{eq:WignerPGM} respectively.

\subsection{\boldmath $\j=3/2$, $d=4$}
In $d=4$ there are 15 GGM matrices calculable from \eqref{eq:GGM_all}--\eqref{eq:GGM_diagonal}.
The Wigner $Q$ symbols for the six symmetric matrices \eqref{eq:GGM_symmetric} are
\begin{eqnarray}
\renewcommand\arraystretch{1.4}
\WignerQ_{\lambda^{(4)}_{S,12}} &=& 2 \sqrt{3} \sin \left(\tfrac{\theta }{2}\right) \cos ^5\left(\tfrac{\theta }{2}\right) \cos \phi  \label{eq:threehalfs_ggm_Q_1} \nonumber\\
\WignerQ_{\lambda^{(4)}_{S,13}} &=& 2 \sqrt{3} \sin ^2\left(\tfrac{\theta }{2}\right) \cos ^4\left(\tfrac{\theta }{2}\right) \cos 2 \phi  \nonumber\\
\WignerQ_{\lambda^{(4)}_{S,23}} &=& \tfrac{3}{4} \sin ^3 \theta  \cos \phi  \nonumber\\
\WignerQ_{\lambda^{(4)}_{S,14}} &=& \tfrac{1}{4} \sin ^3\theta  \cos  3 \phi  \nonumber\\
\WignerQ_{\lambda^{(4)}_{S,24}} &=& 2 \sqrt{3} \sin ^4\left(\tfrac{\theta }{2}\right) \cos ^2\left(\tfrac{\theta }{2}\right) \cos 2 \phi  \nonumber\\
\WignerQ_{\lambda^{(4)}_{S,34}} &=& 2 \sqrt{3} \sin ^5\left(\tfrac{\theta }{2}\right) \cos \left(\tfrac{\theta }{2}\right) \cos \phi .
\end{eqnarray}
Those for the six $d=4$ anti-symmetric GGM matrices \eqref{eq:GGM_antisymmetric} are
\begin{eqnarray}
\WignerQ_{\lambda^{(4)}_{A,12}} &=& -2 \sqrt{3} \sin \left(\tfrac{\theta }{2}\right) \cos ^5\left(\tfrac{\theta }{2}\right) \sin \phi  \nonumber\\
\WignerQ_{\lambda^{(4)}_{A,13}} &=& -2 \sqrt{3} \sin^2\left(\tfrac{\theta}{2}\right) \cos ^4\left(\tfrac{\theta }{2}\right) \sin  2\phi  \nonumber\\
\WignerQ_{\lambda^{(4)}_{A,23}} &=& -\tfrac{3}{4} \sin ^3\theta  \sin \phi  \nonumber\\
\WignerQ_{\lambda^{(4)}_{A,14}} &=& -\tfrac{1}{4} \sin ^3\theta  \sin 3 \phi  \nonumber\\
\WignerQ_{\lambda^{(4)}_{A,24}} &=& -2\sqrt{3} \sin ^4\left(\tfrac{\theta }{2}\right) \cos^2\left(\tfrac{\theta }{2}\right) \sin 2\phi  \nonumber\\
\WignerQ_{\lambda^{(4)}_{A,34}} &=& -2\sqrt{3} \sin ^5\left(\tfrac{\theta }{2}\right) \cos\left(\tfrac{\theta }{2}\right) \sin \phi .
\end{eqnarray}
Those for the three $d=4$ diagonal GGM matrices \eqref{eq:GGM_diagonal} are
\begin{eqnarray}
\WignerQ_{\lambda^{(4)}_{D,1}} &=& \cos ^4\left(\tfrac{\theta }{2}\right) (2 \cos \theta-1) \nonumber\\
\WignerQ_{\lambda^{(4)}_{D,2}} &=& \tfrac{1}{8 \sqrt{3}} \left( 6 \cos \theta+3 \cos 2 \theta -2 \cos 3 \theta +1 \right) \nonumber\\
\WignerQ_{\lambda^{(4)}_{D,3}} &=& \tfrac{1}{8 \sqrt{6}} \left( 15 \cos \theta-6 \cos 2 \theta +\cos 3 \theta -2 \right).\label{eq:threehalfs_ggm_Q_15}
\end{eqnarray}
These $Q$ symbols are shown graphically in Figure~\ref{fig:threehalfs_15Q}.

\newcommand\threehalfswidth{0.16\textwidth}
\begin{figure}[htb]
    \centering
    \includegraphics[width=\threehalfswidth]{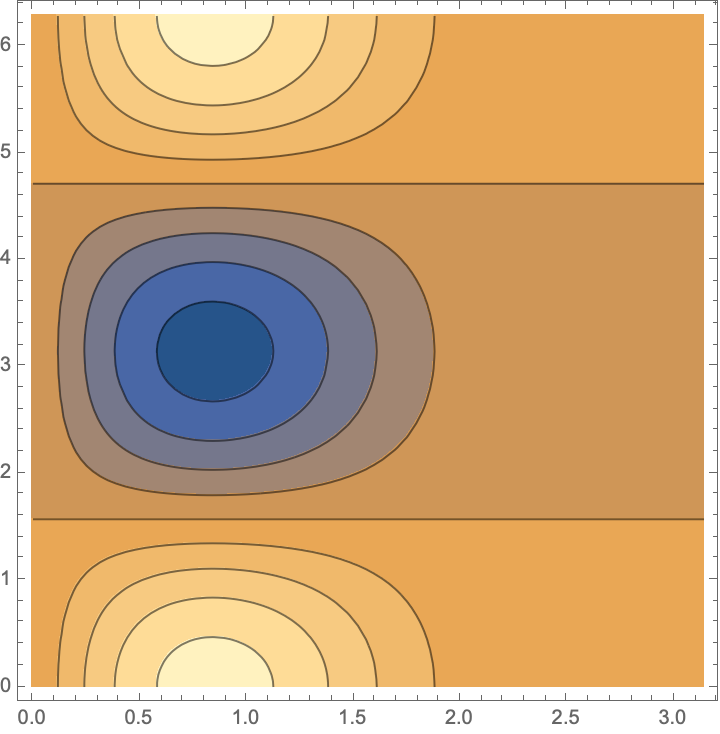}
    \includegraphics[width=\threehalfswidth]{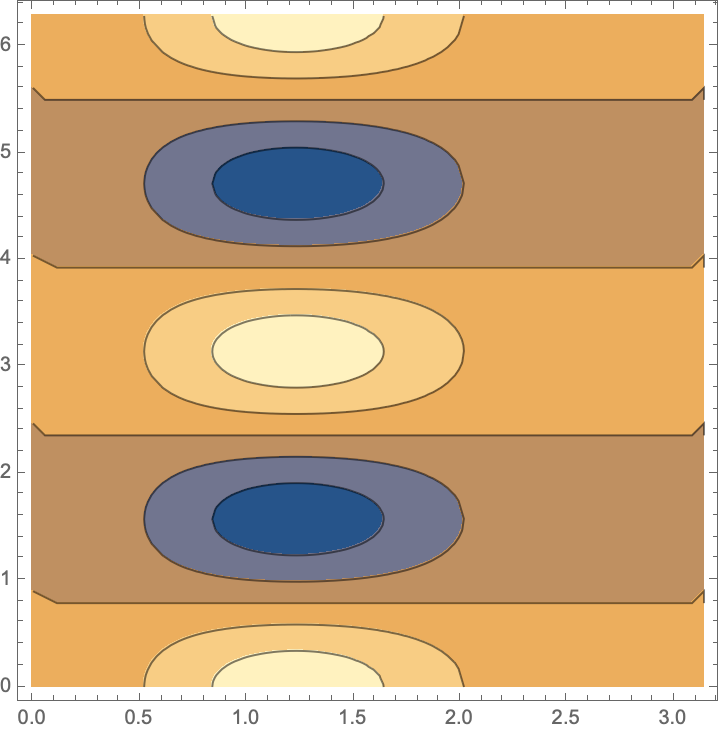}
    \includegraphics[width=\threehalfswidth]{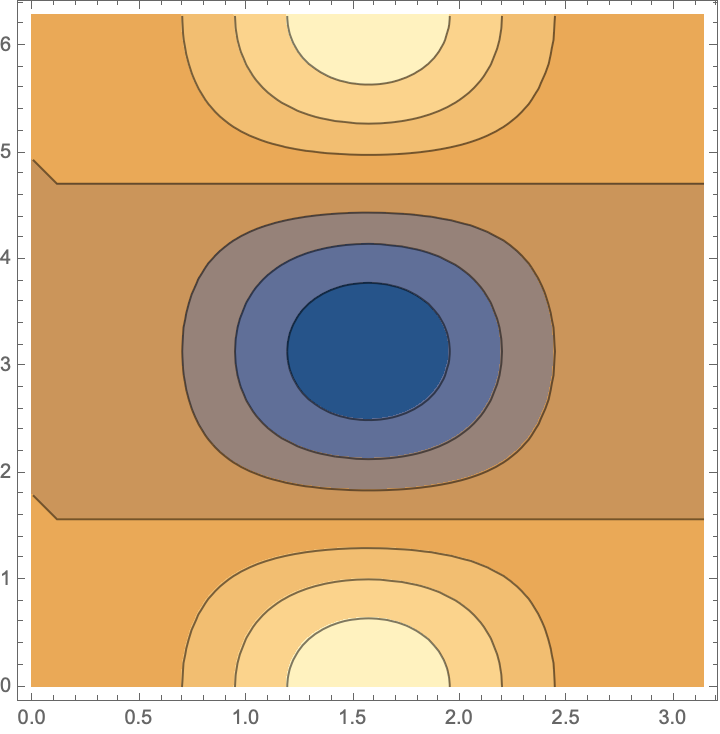}
    \includegraphics[width=\threehalfswidth]{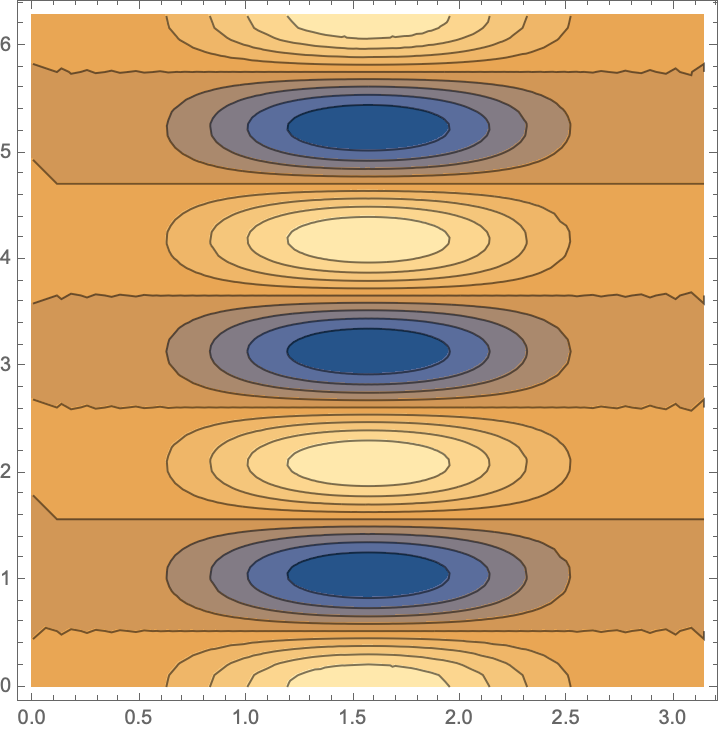}
    \includegraphics[width=\threehalfswidth]{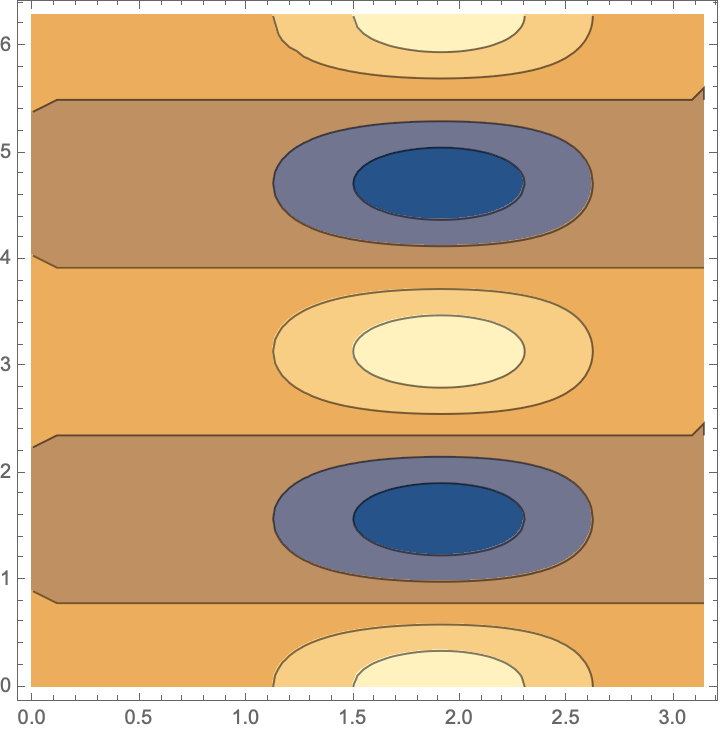}        \includegraphics[width=\threehalfswidth]{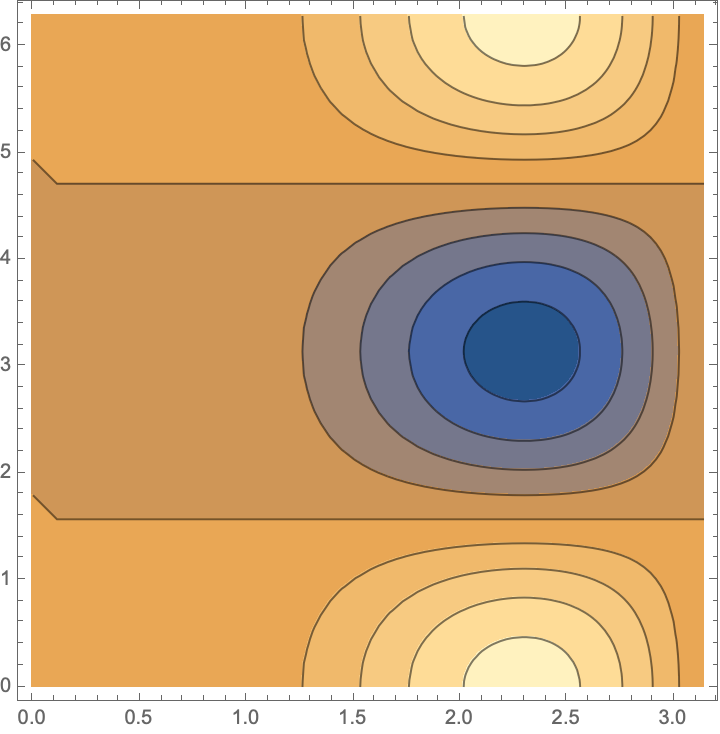} \\
    \includegraphics[width=\threehalfswidth]{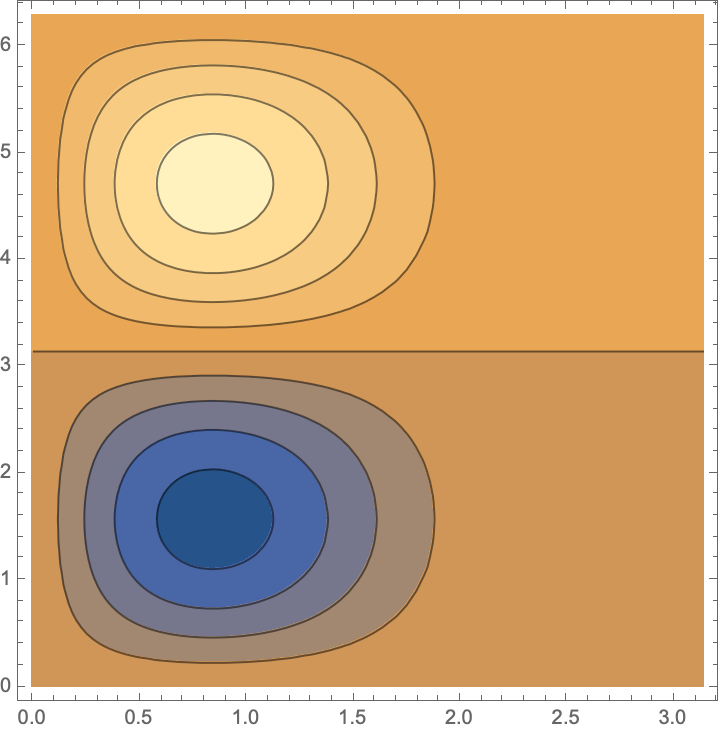}  \includegraphics[width=\threehalfswidth]{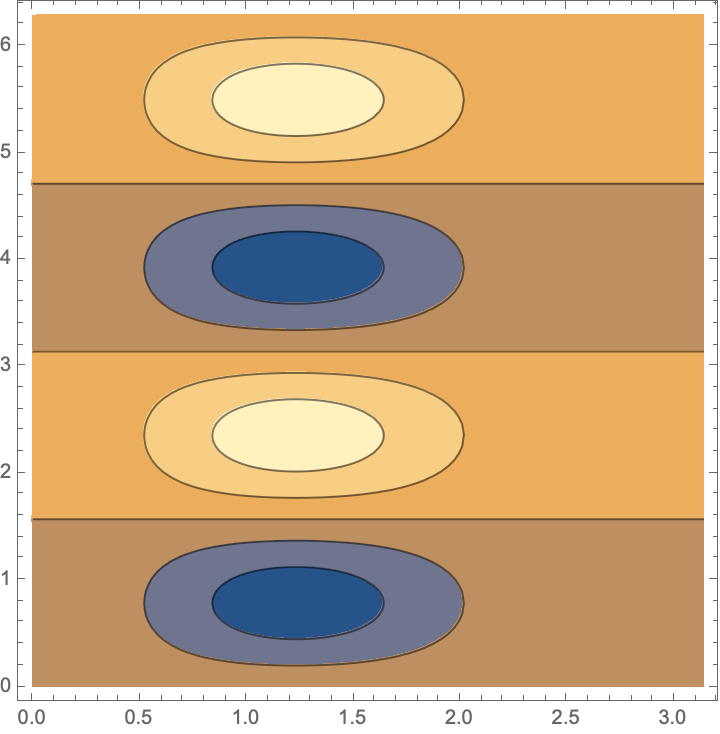}
    \includegraphics[width=\threehalfswidth]{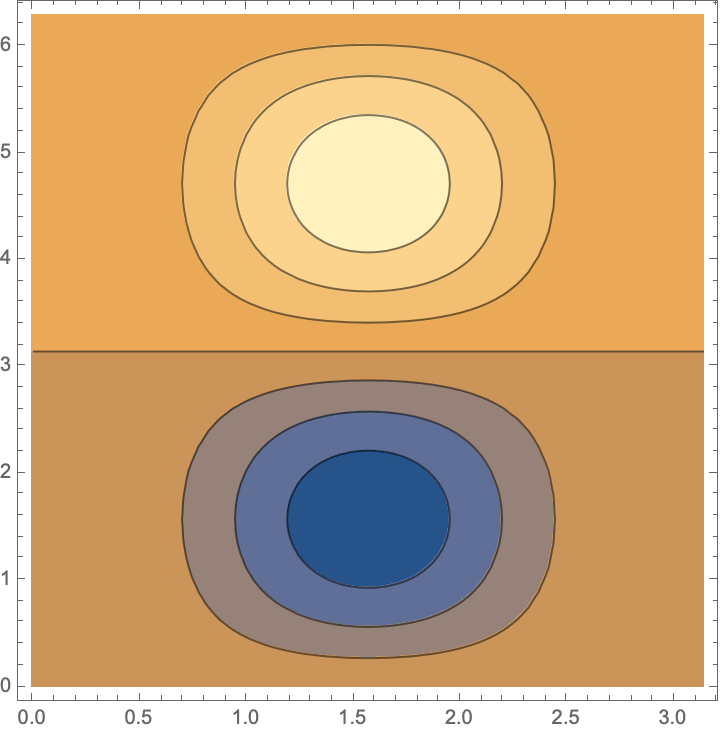}
    \includegraphics[width=\threehalfswidth]{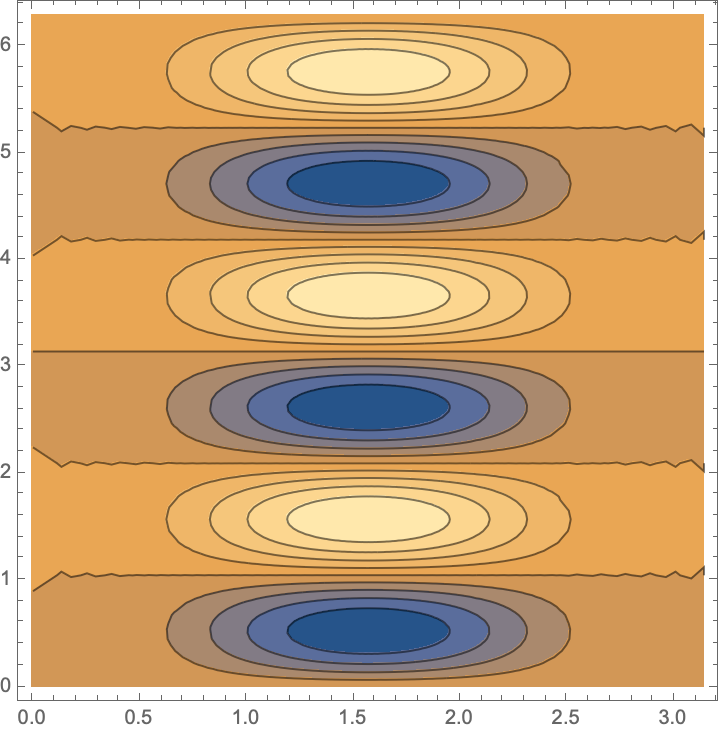}
    \includegraphics[width=\threehalfswidth]{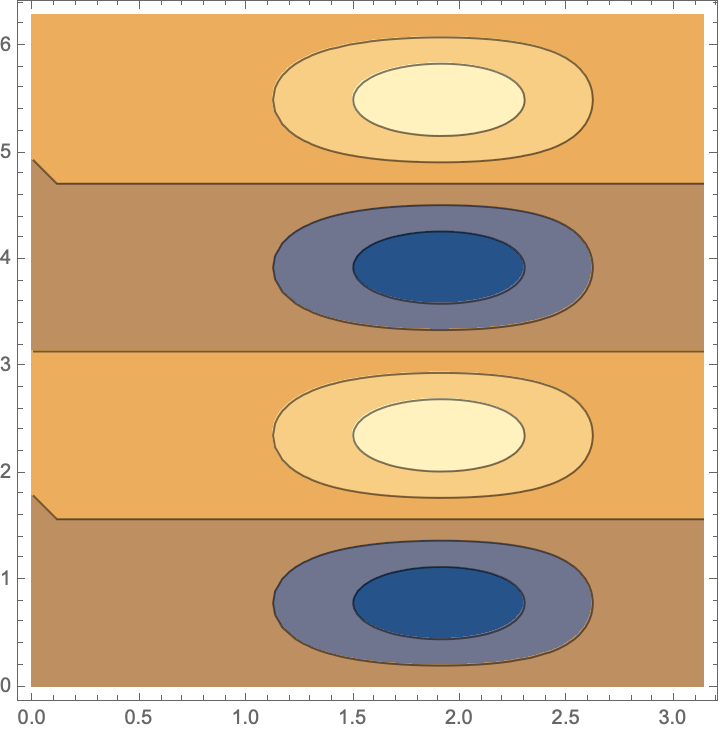}
    \includegraphics[width=\threehalfswidth]{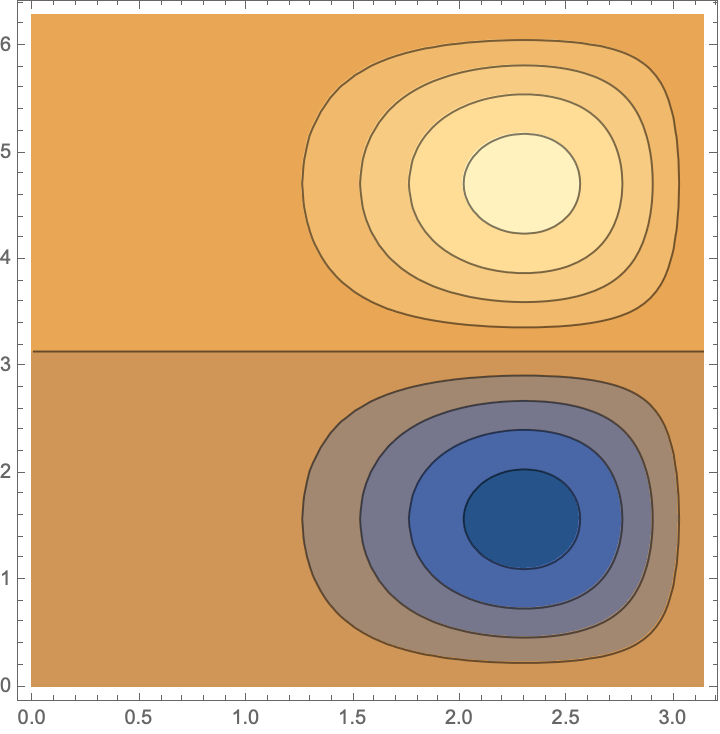} \\
    \includegraphics[width=\threehalfswidth]{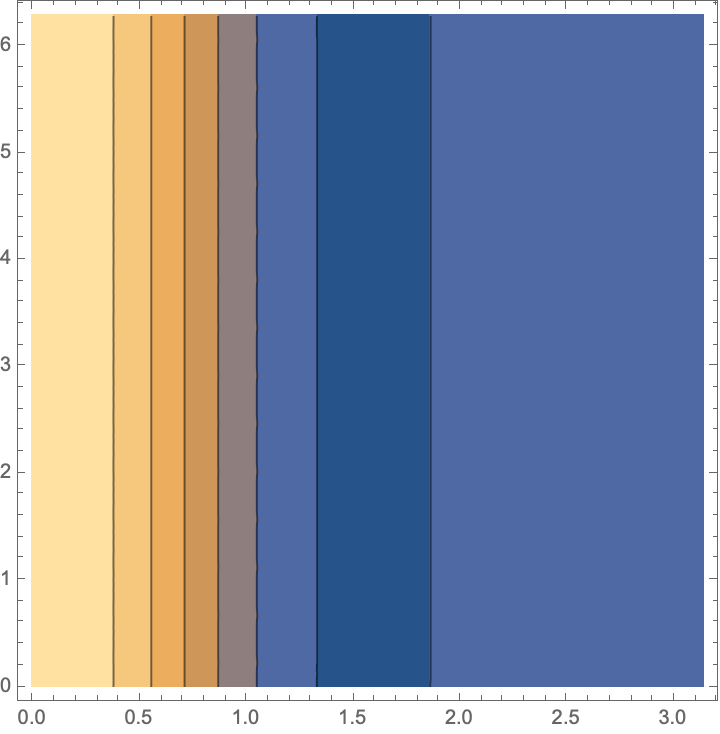}~        \includegraphics[width=\threehalfswidth]{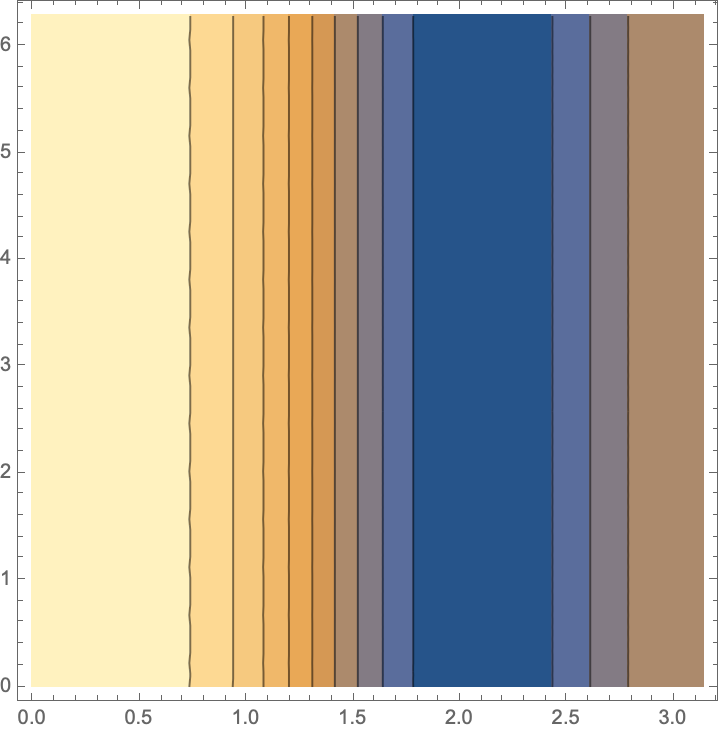}~
    \includegraphics[width=\threehalfswidth]{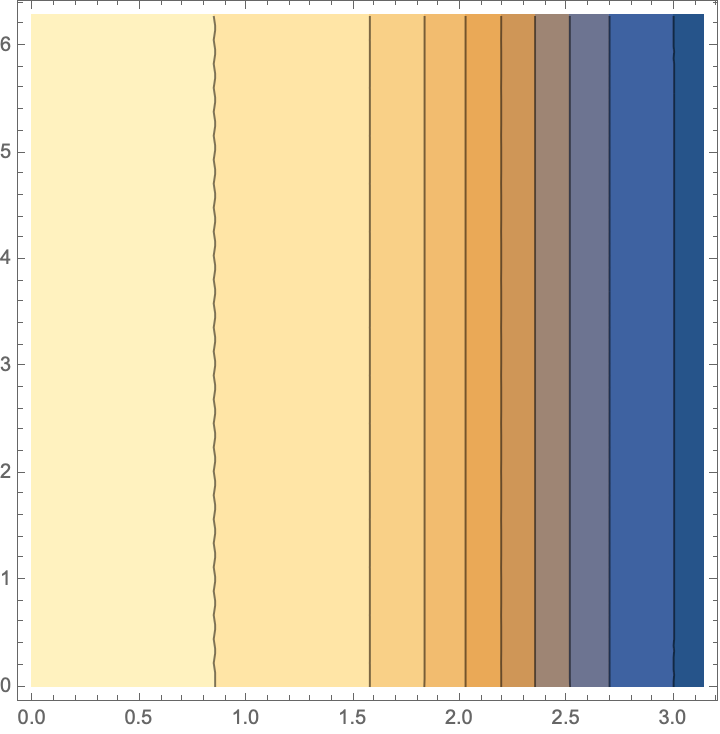}    
    \caption{Graphical display of the Wigner $Q$ symbols for the $d=4$ GGM matrices. In each case $\theta\in[0,\pi]$ is shown on the horizontal axis and $\phi\in[0,2\pi)$  on the vertical axis. The plots are arranged in rows following the order of the symbols in Eqns \eqref{eq:threehalfs_ggm_Q_1}--\eqref{eq:threehalfs_ggm_Q_15}.
    }
    \label{fig:threehalfs_15Q}
\end{figure}

The corresponding $d=4$ Wigner $P$ functions are, for the symmetric GGM matrices:
\begin{eqnarray}
\WignerP_{\lambda^{(4)}_{S,12}} &=&  \tfrac{5\sqrt{3}}{16}  \big(3 \sin \theta+4 \sin 2 \theta +7 \sin 3 \theta  \big) \cos \phi  \nonumber\\
\WignerP_{\lambda^{(4)}_{S,13}} &=&  \tfrac{5\sqrt{3}}{4}  \sin ^2\theta  \big(7 \cos \theta+1 \big) \cos 2 \phi  \nonumber\\
\WignerP_{\lambda^{(4)}_{S,23}} &=&  -\tfrac{5}{16}  \big(\sin \theta+21 \sin 3 \theta  \big) \cos \phi  \nonumber\\
\WignerP_{\lambda^{(4)}_{S,14}} &=&  \tfrac{35}{4} \sin ^3\theta  \cos 3 \phi  \nonumber\\
\WignerP_{\lambda^{(4)}_{S,24}} &=&  \tfrac{5\sqrt{3}}{4}  \sin ^2\theta  (1-7 \cos \theta) \cos 2 \phi  \nonumber\\
\WignerP_{\lambda^{(4)}_{S,34}} &=&  \tfrac{5\sqrt{3}}{16}  \big(3 \sin \theta-4 \sin 2 \theta +7 \sin 3 \theta  \big) \cos \phi \,,
\end{eqnarray}
for the antisymmetric matrices:
\begin{eqnarray}\WignerP_{\lambda^{(4)}_{A,12}} &=&  - \tfrac{5 \sqrt{3}}{16}  \big(3 \sin \theta+4 \sin 2 \theta +7 \sin 3 \theta  \big) \sin \phi  \nonumber\\
\WignerP_{\lambda^{(4)}_{A,13}} &=&  - \tfrac{5\sqrt{3}}{4}   \sin ^2\theta  (7 \cos \theta + 1 ) \sin 2 \phi  \nonumber\\
\WignerP_{\lambda^{(4)}_{A,23}} &=&  \tfrac{5}{16} \big(\sin \theta + 21 \sin 3 \theta  \big) \sin \phi  \nonumber\\
\WignerP_{\lambda^{(4)}_{A,14}} &=&  - \tfrac{35}{4}  \sin ^3(\theta ) \sin 3 \phi  \nonumber\\
\WignerP_{\lambda^{(4)}_{A,24}} &=&  - \tfrac{5\sqrt{3} }{4} \sin ^2\theta  \big(1 - 7 \cos \theta\big) \sin 2 \phi  \nonumber\\
\WignerP_{\lambda^{(4)}_{A,34}} &=&  -\tfrac{5\sqrt{3}}{16}  \big(3 \sin \theta-4 \sin 2 \theta +7 \sin 3 \theta  \big) \sin \phi 
\end{eqnarray}
and for the diagonal matrices:
\begin{eqnarray}\WignerP_{\lambda^{(4)}_{D,1}} &=&  \tfrac{5}{8} \big(5 \cos \theta+3 \cos 2 \theta +7 \cos 3 \theta +1 \big) \nonumber\\
\WignerP_{\lambda^{(4)}_{D,2}} &=&  -\tfrac{5}{8 \sqrt{3}} \big(6 \cos \theta-3 \cos 2 \theta +14 \cos 3 \theta -1\big) \nonumber\\
\WignerP_{\lambda^{(4)}_{D,3}} &=&  \frac{5}{8 \sqrt{6}} \big(9 \cos \theta-6 \cos 2 \theta +7 \cos 3 \theta -2\big)\,.
\end{eqnarray}



\section{\boldmath Spin matrix representation for $d=3$}\label{sec:spinmatrices}

\par
The $\j=1$ spin operators are represented by the $3$-dimensional  traceless Hermitian matrices
\begin{equation} \label{eq:spinoperators}
S_x = \frac{1}{\sqrt{2}}\left(
\begin{array}{ccc}
 0 & 1 & 0 \\
 1 & 0 & 1 \\
 0 & 1 & 0 \\
\end{array}
\right),
\quad
S_y = 
\frac{1}{\sqrt{2}}\left(
\begin{array}{ccc}
 0 & -\i & 0 \\
 \i & 0 & -\i \\
 0 & \i & 0 \\
\end{array}
\right),
\quad
S_z = 
\left(
\begin{array}{ccc}
 1 & 0 & 0 \\
 0 & 0 & 0 \\
 0 & 0 & -1 \\
\end{array}
\right).
\end{equation}
In this basis the states $\{\ket{+},\ket{0},\ket{-}\}$ label the eigenstates of the $S_z$ operator. 
Under an active rotation of angle $\theta$ about an axis in direction $\nhat$ the state transforms as
\begin{equation}
\ket\psi \rightarrow \ket{\psi^\prime} = \exp\left(-\i \nhat \cdot \mathbf{S}\, \theta\right)\ket{\psi},
\end{equation}
where we make use of the Cartesian inner product
\[
\nhat \cdot \mathbf{S} = \sum_{j=1}^3 \hat{n}_j S_j .
\]

By the Cayley-Hamilton theorem the spin operators for spin $\j$, matrices of order $2\j+1$, can be written as polynomials of degree 2$\j$ or less in the spin operator. 
It follows that a possible parameterisation for $\rho$ is in terms of the $S_i$ together with the $S_{\{ij\}}$ --
the pairwise symmetric products of those operators:
\begin{equation}
S_{\{ij\}} \equiv S_i S_j + S_j S_i.
\end{equation}
Their pairwise symmetric products take the form
\renewcommand\arraystretch{1}
\begin{align} %
S_{\{xy\}} &= 
\left(
\begin{array}{ccc}
 0 & 0 & -\i \\
 0 & 0 & 0 \\
 \i & 0 & 0 \\
\end{array} \right), & 
S_{{\{xz\}}}&=\frac{1}{\sqrt{2}}\left(
\begin{array}{ccc}
 0 & 1 & 0 \\
1 & 0 & -1 \\
 0 & -1 & 0 \\
\end{array} \right), &
S_{\{yz\}}  &=
\frac{1}{\sqrt{2}}\left(
\begin{array}{ccc}
 0 & -\i & 0 \\
 \i & 0 & \i \\
 0 & -\i & 0 \\
\end{array} \right), \nonumber\\
S_{\{xx\}} &= \left(
\begin{array}{ccc}
 1 & 0 & 1 \\
 0 & 2 & 0 \\
 1 & 0 & 1 \\
\end{array} \right), & 
S_{\{yy\}} &= \left(
\begin{array}{ccc}
 1 & 0 & -1 \\
 0 & 2 & 0 \\
 -1 & 0 & 1 \\
\end{array} \right), & 
S_{\{zz\}} &= 
\left(
\begin{array}{ccc}
 2 & 0 & 0 \\
 0 & 0 & 0 \\
 0 & 0 & 2 \\
\end{array}
\right). \nonumber
\end{align}
\par
The density matrix therefore may be written in terms of the $S_s$ and $S_{\{ij\}}$ as
\begin{equation}\label{eq:rhospin}
\rho = \tfrac{1}{3}\ithree + \sum_{i=1}^3 \alpha_j S_i + \sum_{i,j=1}^3 \beta_{ij} S_{\{ij\}}
\end{equation}
where the three real parameters $\alpha_i$ form a vector, and the $\beta_{ij}$ form a traceless real symmetric matrix.
\par
For the purposes of quantum state tomography it is inconvenient that the set of nine operators
$ \left\{ S_j,\,S_{\{jk\}}  \right\}$ is not linearly independent of the identity $\ithree$. The dependence can be seen from the relation
\begin{equation}
S_{\{xx\}} + S_{\{yy\}} + S_{\{zz\}} = 2\left(S_x^2 + S_y^2 + S_z^2\right) = 4 \ithree,
\end{equation}
which makes it necessary that $\sum_{j} c_{jj}=0$ in order that $\rho$ retains unit trace.

These spin-based operators are given in terms of the Gell-Mann operators by
\begin{align}
 {S_x} &= \tfrac{1}{\sqrt{2}} (\lambda_1 + \lambda_6) &
 {S_{\{yz\}}} &= \tfrac{1}{\sqrt{2}} (\lambda_2 - \lambda_7 )  & 
 {S_{\{xx\}}} &= \sqrt{\tfrac{8}{3}} \lambda_0 - \tfrac{1}{2} \lambda_3  + \lambda_4 + \tfrac{1}{2\sqrt{3}} \lambda_8 \nonumber\\
 {S_y} &= \tfrac{1}{\sqrt{2}} (\lambda_2 + \lambda_7 ) &
 {S_{\{xz\}}} &= \tfrac{1}{\sqrt{2}} (\lambda_1 - \lambda_6)  &
 {S_{\{yy\}}} &= \sqrt{\tfrac{8}{3}}\lambda_0 - \tfrac{1}{2} \lambda_3 - \lambda_4 + \tfrac{1}{2\sqrt{3}} \lambda_8 \nonumber\\
 {S_z} &= \tfrac{1}{2} (\lambda_3 + \sqrt{3} \lambda_8)  &
 {S_{\{xy\}}} &= \lambda_5 &
 {S_{\{zz\}}} &= \sqrt{\tfrac{8}{3}}  \lambda_0 + \lambda_3 - \tfrac{1}{\sqrt{3}} \lambda_8, \nonumber
\end{align}
where $\lambda_0$ was defined in \eqref{eq:lambda_0}.
Hence the expectation values are given in terms of the Gell-Mann parameters:
\begin{align}
 \braket{S_x} &= \sqrt{2} ({a_1}+{a_6}) &
 \braket{S_{\{yz\}}} &= \sqrt{2} ({a_2}-{a_7}) & 
 \braket{S_{\{xx\}}} &= -{a_3}+2 {a_4}+\frac{{a_8}}{\sqrt{3}}+\frac{4}{3} \nonumber\\
 \braket{S_y} &= \sqrt{2} ({a_2}+{a_7}) &
 \braket{S_{\{xz\}}} &= \sqrt{2} ({a_1}-{a_6}) & 
 \braket{S_{\{{yy\}}}} &= -{a_3}-2 {a_4}+\frac{{a_8}}{\sqrt{3}}+\frac{4}{3} \nonumber\\
 \braket{S_z} &= {a_3}+\sqrt{3} {a_8} & 
 \braket{S_{\{xy\}}} &= 2 {a_5} &
 \braket{S_{\{zz\}}} &= 2 {a_3}-\frac{2 {a_8}}{\sqrt{3}}+\frac{4}{3}.
\end{align}


\section{Generalised Wigner $Q$ and $P$ symbols \boldmath $W\rightarrow \ell^+\nu$ decay with massive lepton} \label{sec:Wnonprojective}

In Section~\ref{sec:singleW} we examined $W\rightarrow\ell\nu$ decay under the assumption of a massless charged lepton. Here we calculate the Wigner symbols allowing that lepton to be of non-negligible mass $m_\ell$, of the same order as $m_W$. 

The vertex factor for $W^+\rightarrow \ell^+\nu_\ell$ is~\cite{thomson_2013}
\begin{equation}
-\i  \frac{g_{\rm W}}{\sqrt{2}}  \gamma^\mu \half (1-\gamma^5)
\end{equation}
The positive and negative helicity Dirac spinors for the $\ell^+$ may be written
\begin{eqnarray}
\mathsf{v}_+ &\propto& \half (1+\chi) \mathsf{v}_R + \half (1-\chi) \mathsf{v}_L \\
\mathsf{v}_- &\propto& \half (1-\chi) \mathsf{v}_R + \half (1+\chi) \mathsf{v}_L
\end{eqnarray}
where $\mathsf{v}_{L,R}$ are the left- and right-chiral projected components and $\chi=\frac{p}{E+m}$, where $p$, $E$ and $m$ are the momentum, energy and mass respectively of the heavy lepton $\ell^+$, as measured in its parent $W$ boson's rest frame. 


Allowing for a scalar component of the $W$ boson, neglecting the neutrino mass, and including a Clebsch-Gordan coefficient coefficients 
\begin{eqnarray}
\langle{1,0}\ket{\uparrow\downarrow} &=& 2^{-1/2} \nonumber\\
\langle{0,0}\ket{\uparrow\downarrow} &=& 2^{-1/2}
\end{eqnarray}
for the $\ket{1,0}$ and $\ket{0,0}$ states, the $\j=1$ components of the Kraus operator are
\begin{equation}\label{eq:krausWheavylepton}
K_{W^+\rightarrow\ell^+\nu} = \frac{\i}{\sqrt{2(1+\chi^2)}}\,\mathrm{diag}\big((1+\chi),\,(1-\chi)/\sqrt{2},\,0\big)
\end{equation}
in the order of $m_\j$: $(+,0,-)$. The $\j=0$ component has magnitude equal to that for the $\ket{1,0}$ state if it's assumed that its coupling is equal. 
The $\j=1$ components of the diagonal measurement operator are then
\begin{eqnarray}\label{eq:measurementoperatorWlnu}
F_{W^+\rightarrow\ell^+\nu} &=& K_{W^+\rightarrow\ell^+\nu}^\dag \, K_{W^+\rightarrow\ell^+\nu} \nonumber\\
 &=& \mathrm{diag}\,\left( \tfrac{1}{2}(1+{v}),\,\tfrac{1}{4}(1-{v}),\, 0\right),
\end{eqnarray}
where $v$ is the speed of the heavy $\ell^+$ in the $W^+$ boson rest frame.
The generalised $Q$ functions for the $j=1$ components can be calculated from \eqref{eq:nonprojectiveQ} and \eqref{eq:measurementoperatorWlnu}. 
\par
The $\j=0$ component of $F_{W^+\rightarrow\ell^+\nu}$, under the assumption of uniform couplings, is $\tfrac{1}{4}(1-v)$. That component has a generalised Wigner Q symbol $\tfrac{1}{4}(1-v)$ proportional to the identity. It contributes to the Kraus and measurement operators of the $\j=1$ components by contributing to the normalisation terms in their denominators.
\par
The 9$\times$9 matrix (eight components corresponding to $\j=1$ one for $\j=0$) of inner products \eqref{eq:QQinner} of generalised Wigner $Q$ functions is then calculated. The $\j=0$ functions are orthogonal to the $\j=1$ functions, so the matrix breaks into two blocks. 
The $\j=1$ block (of dimension 8$\times$8) is invertible except when $v=0$. The $\j=0$ block (1$\times$1) is invertible except when $v=1$. 
\par
Other than at these singular values the generalised $P$ functions for can be calculated for each block according to \eqref{eq:pfromqgen}. For $\j=1$ they are given in terms of the projective $P$ symbols by
\begin{equation}\label{eq:wignerPnonprojectiveW}
\WignerNonProjectiveP_{F_{W^+\rightarrow\ell^+\nu},i} = A_{ij} \WignerPp_j\,,
\end{equation}
where the matrix $A$ is 
\begin{equation}\renewcommand\arraystretch{1}\label{eq:heavyleptonMatrixA}
A=
\left(
\begin{array}{cccccccc}
 \frac{1}{v+1}+\frac{1}{2 v} & 0 & 0 & 0 & 0 & \frac{1}{v+1}-\frac{1}{2 v} & 0 & 0 \\
 0 & \frac{1}{v+1}+\frac{1}{2 v} & 0 & 0 & 0 & 0 & \frac{1}{v+1}-\frac{1}{2 v} & 0 \\
 0 & 0 & \frac{1}{2(v+1)}+\frac{3}{4 v} & 0 & 0 & 0 & 0 & \frac{\sqrt{3} (v-1)}{4 v (v+1)} \\
 0 & 0 & 0 & ~~\frac{1}{v}~~ & 0 & 0 & 0 & 0 \\
 0 & 0 & 0 & 0 & ~~\frac{1}{v}~~ & 0 & 0 & 0 \\
 \frac{1}{v+1}-\frac{1}{2 v} & 0 & 0 & 0 & 0 & \frac{1}{v+1}+\frac{1}{2 v} & 0 & 0 \\
 0 & \frac{1}{v+1}-\frac{1}{2 v} & 0 & 0 & 0 & 0 & \frac{1}{v+1}+\frac{1}{2 v} & 0 \\
 0 & 0 & \frac{\sqrt{3} (v-1)}{4 v (v+1)} & 0 & 0 & 0 & 0 & \frac{3}{2(v+1)}+\frac{1}{4 v} \\
\end{array}
\right)
\end{equation}
The corresponding Wigner $P$ symbols for the $W^-$ can be obtained from those for the $W^+$ from by the simultaneous replacements $\phi\rightarrow\phi+\pi$, $\theta\rightarrow\pi-\theta$. 
\par
If scalar components for the vector boson are forbidden, for example due to the argument that the boson must be on-shell, then the measurement operator \eqref{eq:measurementoperatorWlnu} should be normalised such that its has unit trace without any scalar component. The $\j=1$ components of the measurement operator are then multiplied by a factor $f_v=4/(3+v)$; the generalised $\j=1$ Wigner $Q$ functions are multiplied by $f_v$; and the generalised $\j=1$ Wigner $P$ functions \eqref{eq:wignerPnonprojectiveW} are multiplied by  $f_v^{-1}$. 
\par
In the limit $v\rightarrow 1$ 
the normalisation factor $f\rightarrow1$ and 
the question as to whether one permits a scalar component for the $W^\pm$ becomes irrelevant. We have that the product of Kraus operators $F_{W^+\rightarrow \ell^+\nu}F^*_{W^+\rightarrow \ell^+\nu} \rightarrow \Pi_+$, the matrix $A=I_8$ and for the $\j=1$ block we recover the projective Wigner $Q$ \eqref{eq:WignerQGM} and $P$ \eqref{eq:WignerPGM} symbols for $d=3$.


\section{\boldmath Examples of quantum state tomography in bipartite systems of dimension $3\times 2$ and $2\times 2$}\label{sec:qubittomography}

\begin{table}\begin{center}
\renewcommand\arraystretch{1.2}
\begin{tabular}{c c c}
\hline
Process & $d_1\times d_2$ \\
\hline
$pp\rightarrow H(400) \rightarrow t \bar{t} 
\rightarrow (b e^+\nu_e) (\bar b \mu^- \bar\nu_\mu)$ &  $2\times 2$ \\
$pp \rightarrow W^+ \bar t  b \rightarrow (e^+\nu_e) (\bar b \mu^- \bar\nu_\mu) b $ &   $3\times 2$\\
\hline
\end{tabular}
\caption{\label{tab:additionalprocesses}
Additional physics processes simulated of Hilbert space dimension different from $3 \times 3$. 
The final column gives the dimension of the Hilbert space of the bipartite system.
For each process $10^6$ events were simulated. The CM energy for $pp$ collisions was $\sqrt{s}=13\,$TeV.
}
\end{center}
\end{table}

\begin{figure}
    \centering
\begin{subfigure}[b]{0.48\textwidth}\centering
\includegraphics[width=\textwidth]{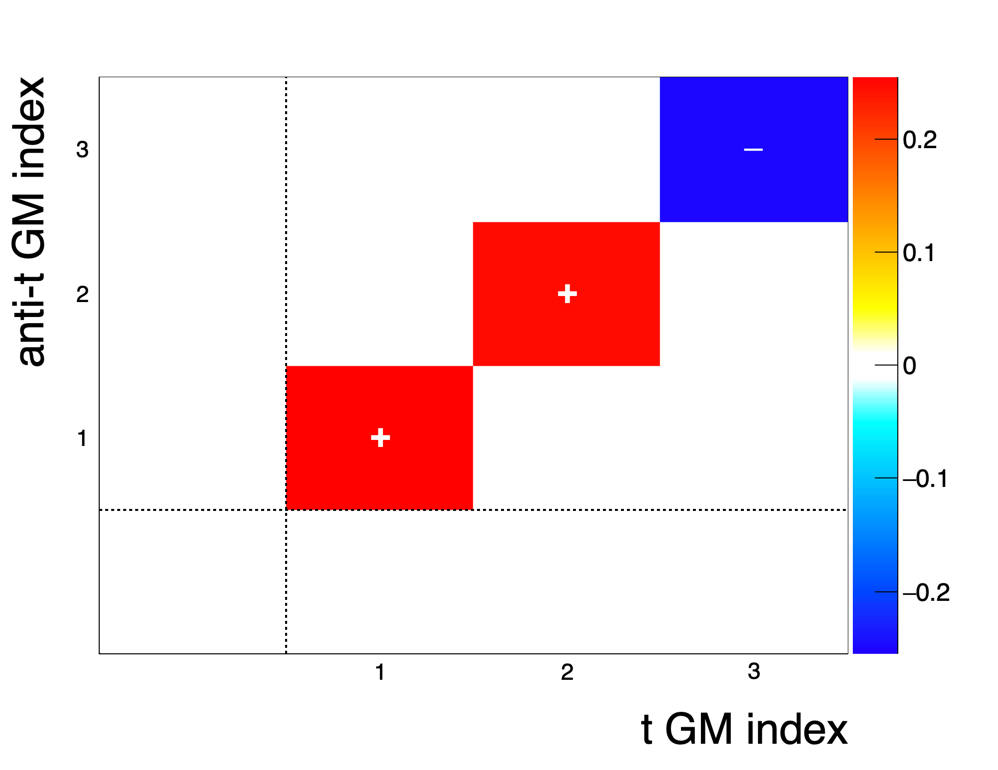}
\vskip-1mm
\caption{\label{subfig:tt_GM_square}$H(400)\rightarrow t\bar{t}$}
\end{subfigure}
\begin{subfigure}[b]{0.48\textwidth}\centering
\includegraphics[width=\textwidth]{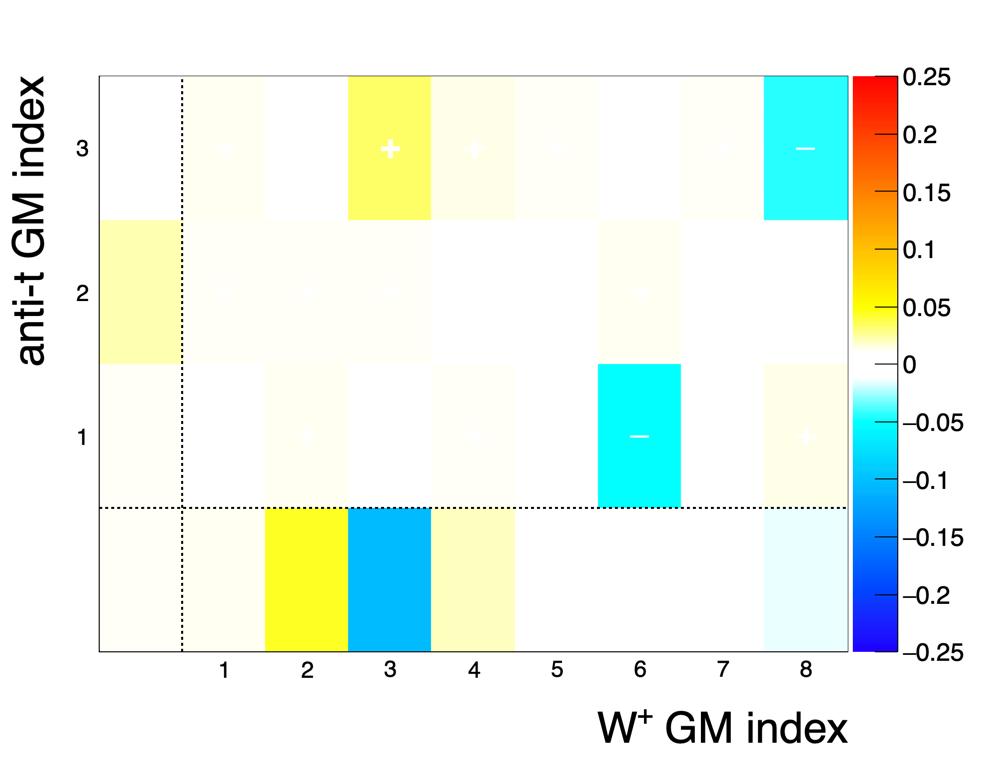}~
\vskip-1mm
\caption{\label{subfig:Wt_GM_square}$pp\rightarrow W^+\bar{t}$ }
\end{subfigure}
\caption{Generalised Gell-Mann parameters obtained by quantum state tomography of simulated decays for bipartite systems of Hilbert space $d_1\times d_2=3\times 2$ (qutrit-qubit) and $2\times 2$ (qubit pair).}
\label{fig:rhogm_fermionic}
\end{figure}

We have concentrated in the main body of the text on $3\times 3$ bipartite systems, which have not previously been investigated in detail in particle physics. 
To illustrate the method also test examples of bipartite systems of Hilbert space dimension $d_1\times d_2=3\times 2$ and $2\times 2$. 
We follow the procedures described in Section~\ref{sec:tomographyWW}, but now also performing qubit tomography of leptonic top and anti-top quark decays using the generalised Wigner $P$ symbols \eqref{eq:fermionP} with spin analysing power $\kappa=+1.0$ for $\ell^+$ from $t$ and $\kappa=-1.0$ for $\ell^-$ from $\bar{t}$.
\par
The physics processes simulated are shown in Table~\ref{tab:additionalprocesses}. 
Our $2\times 2$ test system is the hypothetical decay $H(400)\rightarrow t\bar{t}$, of a Higgs boson with mass of 400\,GeV to a pair of top quarks. The GGM density matrix parameters obtained after simulation by quantum state tomography are displayed in Figure~\ref{subfig:tt_GM_square}. 
The $a_i$ and $b_j$ parameters are consistent with zero while the $c_{ij}$ parameters of the reconstructed bipartite density matrix are, within the statistical uncertainties of the simulation, consistent with those  
\[\renewcommand\arraystretch{1.0}
c_{ij} = \frac{1}{4}\left(
 \begin{array}{ccc} 
 1 & 0 & 0 \\
 0 & 1 & 0 \\
 0 & 0 & -1 \\
\end{array}
\right)
\]
of a maximally entangled state a pair of qubits.\footnote{They are also consistent with those calculated in Ref.~\cite{Fabbrichesi:2022ovb} for the similar process of $H(125)\rightarrow \tau\tau$. } 
Such a state has the largest concurrence possible (unity) for a $2\times 2$ system.
The reconstructed concurrence is 0.97, confirming that the tomographic process has reconstructed an almost maximally-entangled state of a pair of qubits.

To test a  bipartite system of Hilbert space dimension $3\times 2$ we reconstruct the spin density matrix of a $W^+$ boson together with a leptonically decaying anti-top quark, for the process indicated in \eqref{tab:additionalprocesses}. The lepton is used as the probe daughter in each decay. The reconstructed GGM parameters are plotted in Figure~\ref{subfig:Wt_GM_square}. Correlations are visible, but they are not as strong as in the Higgs boson decay cases.
\par

After reconstructing its spin density matrix we find that our test decay $H(400)\rightarrow t\bar{t}$ has an expectation value of the Clauser-Horne-Shimony-Holt (CHSH) operator~\cite{chsh} that is consistent with the value $2\sqrt{2}$ within statistical uncertainties. This is the maximum value for a $2\times 2$ bipartite system~\cite{cirelson}, and in violation of the upper bound of two required of local realist theories~\cite{chsh}. Thus if nature permitted such a process it would be an ideal laboratory for a Bell test. 
Further exploration of possible qubit-pair entanglement and Bell tests in particle decays can be found in previous works~\cite{Tornqvist,Baranov_2008,10.1093/ptep/ptt032,Ac_n_2001,Afik_2021,Gong:2021bcp,Severi:2021cnj,Afik:2022kwm,Aoude:2022imd,Fabbrichesi:2022ovb,Afik:2022dgh}.


\acknowledgments

We are very grateful to Bryan Webber for his careful reading of an early version of this work, and for his very helpful suggestions for its extension to a wider range of cases. 
We are indebted to Hussain Anwar who provided valuable suggestions and references regarding numerical entanglement detection methods for $d>2\times3$. 
We are also grateful for helpful comments provided by an anonymous referee. 
Further helpful comments and suggestions were given by Ben Allanach, Clelia Altomonte, George Barker, Artur Ekert, James Newton, and from the members of the Oxford ATLAS `SM \& beyond' group. 
AJB is grateful to Marcel Vos who encouraged the further study of entanglement in diboson systems. 
AJB is funded through STFC grants ST/R002444/1 and ST/S000933/1, by the University of Oxford and by Merton College, Oxford. AW is funded by an undergraduate summer project grant from Merton College, Oxford. 


\bibliographystyle{JHEP}
\bibliography{tomography}

\end{document}